\documentclass[useAMS,usedcolumn,usenatbib]{mn2e}
\usepackage{graphicx}
\usepackage{epsfig}

\begin{document}
\title{Starburst and AGN activity in ultraluminous infrared galaxies}
\author[D.~Farrah et al]{D.~Farrah$^{1}$, J.~Afonso$^{2,3}$, A.~Efstathiou$^4$, M.~Rowan-Robinson$^5$, M.~Fox$^5$, and 
\newauthor D.~Clements$^5$\\
 $^1$SIRTF Science Centre, California Institute of Technology, Jet Propulsion Laboratory, Pasadena, CA 91125, USA\\
 $^2$CAAUL, Observat\'{o}rio Astron\'{o}mico de Lisboa, Tapada da Ajuda, 1349-018 Lisboa, Portugal\\
 $^3$ Onsala Space Observatory, S-43992 Onsala, Sweden\\
 $^4$Department of Computer Science and Engineering, Cyprus College, 6 Diogenous Street, PO Box 22006, \\
     1516 Nicosia, Cyprus\\
 $^5$Astrophysics Group, Blackett Laboratory, Imperial College, Prince Consort Road, London SW7 2BW, UK\\
}
\date{2002 September 30}
\pubyear{2001} \volume{000} \pagerange{1} \twocolumn

\maketitle \label{firstpage}
 
\begin{abstract}
We examine the power source of 41 local Ultraluminous Infrared Galaxies (ULIRGs) using archival 
infrared (IR) and optical photometry. We fit the observed Spectral Energy Distributions (SEDs) 
with starburst and AGN components; each component being drawn from a family of 
templates. We find all of the sample require a starburst, whereas only half require an AGN. In 
$90\%$ of the sample the starburst provides over half the IR emission, with a mean 
fractional luminosity of $82\%$. When combined with other galaxy samples we find that starburst 
and AGN luminosities correlate over 6 decades in IR luminosity suggesting that a common 
factor governs both luminosities, plausibly the gas masses in the nuclear regions. We 
find no trend for increasing fractional AGN luminosity with increasing total luminosity, contrary 
to previous claims. We find that the mid-IR $F_{7.7}/C_{7.7}$ line-continuum ratio is no 
indication of the starburst luminosity, or the fractional AGN luminosity, and 
therefore that $F_{7.7}/C_{7.7}$ is not a reliable diagnostic of the power source in ULIRGs. 
The radio flux correlates with the starburst luminosity, but shows no correlation with the AGN 
luminosity, in line with previous results. We propose that the scatter in this correlation is due 
to a skewed starburst IMF and/or relic relativistic electrons from a previous 
starburst, rather than contamination from an obscured AGN. We 
show that most ULIRGs undergo multiple starbursts during their lifetime, and by inference 
that mergers between more than two galaxies may be common amongst ULIRGs. Our results 
support the evolutionary model for ULIRGs proposed by \citet{far1}, where they can follow many different 
evolutionary paths of starburst and AGN activity in transforming merging spiral galaxies into elliptical galaxies, but that most 
do not go through an optical QSO phase. The lower level of AGN activity in our local sample than in $z\sim1$ HLIRGs implies 
that the two samples are distinct populations. We postulate that different galaxy formation 
processes at high-$z$ are responsible for this difference.  
\end{abstract}

\begin{keywords}
 infrared: galaxies -- galaxies: active -- galaxies: Seyfert -- galaxies: starburst -- Quasars: general 
\end{keywords}

\section{Introduction}\label{sect:intro}

\begin{table*}
\begin{minipage}{170mm}
\caption{All Ultraluminous Infrared Galaxies at $z<0.1$ \label{ulirgsample}}
\begin{tabular}{@{}lcccccc}
\hline
IRAS Name  & Other names & RA (2000)  & Dec                      & $z$     & Optical spectrum & $F_{25}/F_{60}^{a}$ \\ 
           &             & hh:mm:ss   & \degr\ \arcmin\ \arcsec\ &         &                  &                 \\ 
\hline
00198-7926 &             & 00 21 52.9 & -79 10 08.0              & 0.0728  & Sy2              & 0.370 \\ 
00199-7426 &             & 00 22 07.0 & -74 09 41.6              & 0.0964  & --               & 0.078 \\ 
00262+4251 &             & 00 28 54.2 & +43 08 15.3              & 0.0927  & LINER            & 0.112 \\ 
00335-2732 &             & 00 36 00.5 & -27 15 34.5              & 0.0693  & Starburst        & 0.147 \\ 
01388-4618 &             & 01 40 55.9 & -46 02 53.3              & 0.0903  & HII              & 0.121 \\ 
02364-4751 &             & 02 38 13.1 & -47 38 10.5              & 0.0983  & --               & 0.063 \\ 
03068-5346 &             & 03 08 21.3 & -53 35 12.0              & 0.0745  & --               & 0.057 \\ 
04232+1436 &             & 04 26 05.0 & +14 43 38.0              & 0.0796  & LINER            & 0.113 \\ 
05189-2524 &             & 05 21 01.5 & -25 21 45.4              & 0.0426  & Sy2              & 0.252 \\ 
06035-7102 &             & 06 02 54.0 & -71 03 10.3              & 0.0795  & HII              & 0.112 \\ 
06206-6315 &             & 06 21 01.2 & -63 17 23.2              & 0.0924  & Sy2              & 0.074 \\ 
08572+3915 &             & 09 00 25.4 & +39 03 54.4              & 0.0584  & LINER            & 0.229 \\ 
09111-1007 &             & 09 13 38.8 & -10 19 20.3              & 0.0541  & HII+Sy2          & 0.067 \\ 
09320+6134 & UGC 05101   & 09 35 51.7 & +61 21 11.3              & 0.0394  & LINER+Sy2        & 0.090 \\ 
09583+4714 &             & 10 01 31.2 & +46 59 44.0              & 0.0859  & Sy1+Sy2          & 0.182 \\ 
10035+4852 &             & 10 06 46.1 & +48 37 44.1              & 0.0648  & Starburst        & 0.062 \\ 
10190+1322 &             & 10 21 42.5 & +13 06 53.9              & 0.0766  & HII              & 0.114 \\ 
10494+4424 &             & 10 52 23.6 & +44 08 47.3              & 0.0921  & LINER            & 0.047 \\ 
10565+2448 &             & 10 59 18.1 & +24 32 34.3              & 0.0431  & HII+LINER        & 0.094 \\ 
12112+0305 &             & 12 13 45.7 & +02 48 40.3              & 0.0733  & LINER            & 0.060 \\ 
12540+5708 & Mrk 231     & 12 56 14.2 & +56 52 25.2              & 0.0422  & Sy1              & 0.271 \\ 
13428+5608 & Mrk 273     & 13 44 42.1 & +55 53 12.6              & 0.0378  & Sy2              & 0.105 \\ 
14348-1447 &             & 14 37 38.4 & -15 00 22.8              & 0.0827  & LINER            & 0.072 \\ 
14378-3651 &             & 14 40 58.9 & -37 04 33.0              & 0.0676  & LINER+Sy2        & 0.085 \\ 
15250+3609 &             & 15 26 59.4 & +35 58 37.5              & 0.0552  & LINER            & 0.182 \\ 
15327+2340 & Arp220      & 15 34 57.1 & +23 30 11.5              & 0.0181  & LINER            & 0.076 \\ 
17132+5313 &             & 17 14 20.5 & +53 10 30.4              & 0.0509  & HII+AGN          & 0.109 \\ 
17208-0014 &             & 17 23 21.9 & -00 17 00.4              & 0.0428  & HII              & 0.053 \\ 
18470+3233 &             & 18 48 54.2 & +32 37 31.0              & 0.0784  & HII              & 0.104 \\ 
19254-7245 & Super-Antena& 19 31 21.6 & -72 39 21.7              & 0.0617  & Sy2              & 0.226 \\ 
19297-0406 &             & 19 32 20.7 & -04 00 06.0              & 0.0857  & HII              & 0.084 \\ 
19458+0944 &             & 19 48 15.7 & +09 52 05.0              & 0.0999  & --               & 0.066 \\ 
20046-0623 &             & 20 07 19.4 & -06 14 26.0              & 0.0844  & Starburst        & 0.141 \\ 
20414-1651 &             & 20 44 18.2 & -16 40 16.2              & 0.0871  & HII              & 0.079 \\ 
20551-4250 &             & 20 58 26.9 & -42 39 06.2              & 0.0428  & HII              & 0.149 \\ 
21130-4446 &             & 21 16 18.5 & -44 33 37.7              & 0.0926  & HII              & 0.048 \\ 
21504-0628 &             & 21 53 05.5 & -06 14 49.9              & 0.0776  & --               & 0.111 \\ 
22491-1808 &             & 22 51 49.3 & -17 52 23.4              & 0.0778  & HII              & 0.101 \\ 
23128-5919 &             & 23 15 47.0 & -59 03 16.9              & 0.0446  & HII+Sy2          & 0.147  \\ 
23365+3604 &             & 23 39 01.3 & +36 21 08.7              & 0.0645  & HII+LINER        & 0.114  \\ 
23389-6139 &             & 23 41 43.6 & -61 22 50.9              & 0.0928  & HII              & 0.067 \\ 
\hline
\end{tabular}

\medskip
Coordinates, optical spectral classifications and redshifts were taken from the NASA Extragalactic Database. 
$^{a}$Ratio of the IRAS $25\mu$m flux to the IRAS $60\mu$m flux.

\end{minipage}
\end{table*}

One of the most important results from the Infrared Astronomical Satellite ({\em IRAS}) all 
sky surveys was the detection of a new class of galaxy where most of the bolometric emission 
lies in the infrared waveband \citep{soi1,san}. These 'Luminous Infrared Galaxies' (LIRGs), become 
the dominant extragalactic population at IR luminosities above 
$10^{11}L_{\sun}$, with a higher space density than all other classes of galaxy of comparable 
bolometric luminosity. At the most luminous end of the {\em IRAS} galaxy population lie the 
Ultraluminous Infrared Galaxies (ULIRGs), those with $L_{ir}>10^{12}L_{\odot}$. Although 
ULIRGs are rare in the local Universe, their luminosity function shows strong evolution with 
redshift \citep{vei2} and deep sub-mm surveys \citep{bar,hug,eal,scot,fox} have found 
that systems with ULIRG-like luminosities are very numerous at $z\geq1$. ULIRGs are thus an 
important population in understanding the cosmic history of star formation. 
ULIRGs are also thought to play a role in the evolution of spiral and elliptical galaxies. 
Nearly all ULIRGs are observed to be ongoing mergers between two or more spirals 
\citep{san2,far1}, and it is thought that such a merger will form an elliptical \citep{barn}.

Despite this, the evolution of ULIRGs, their power source and the trigger behind the 
IR emission are poorly understood. Although it is now accepted that a mixture of star 
formation and AGN activity powers the IR emission, the {\em dominant} power source, and how 
ULIRGs evolve, are unknown. There are similarities between ULIRGs and starburst galaxies 
\citep{jow,rrc,con1}. Conversely, many ULIRGs display emission lines characteristic of 
Seyferts \citep{san2}. An early evolutionary model for ULIRGs was that of \citet{san2} 
who suggested, based on the similar space density, bolometric emission and luminosity function 
of ULIRGs and QSOs in the local Universe, that ULIRGs are the dust enshrouded precursors 
to optical QSOs and that all QSOs emerge from a luminous infrared phase. Conversely, 
a more recent evolutionary model \citep{far1} proposes that ULIRGs are not a simple dust shrouded 
precursor to optical QSOs but instead follow multiple evolutionary paths.  

In this paper we examine the power source in ULIRGs, their evolution, and their relationship to high-$z$ 
IR luminous galaxies using archival photometry for a sample of 41 local ULIRGs, and advanced 
radiative transfer models for starbursts and AGN. We also examine the origin of the radio-IR correlation 
in ULIRGs and the power of mid-IR spectroscopy as a diagnostic of the active power source. Sample selection 
and data analysis are described in \S\ref{sect:sample} and \S\ref{sect:models}. Results are presented in 
\S\ref{sect:res} and notes on individual objects are given in \S\ref{sect:ind}. Discussion is 
presented in \S\ref{sect:discuss} and conclusions are summarized in \S\ref{sect:conc}. We adopt 
$H_{0}=65$ km s$^{-1}$ Mpc$^{-1}$, $\Omega_{0}=1.0$ and $\Lambda=0.0$ and quote all luminosities in this system. 
Unless otherwise stated, the terms 'IR luminosity', 'starburst luminosity', and 'AGN luminosity' refer to 
the luminosity over the wavelength range $1-1000\mu$m. Luminosities are quoted in units of bolometric 
solar luminosities, where $L_{\odot} = 3.826\times10^{26}$ Watts.

\section{The sample}\label{sect:sample}
We assembled from the literature a sample of objects comprising all ULIRGs with $z\leq0.1$. This was complicated
by the fact that different authors use different minimum luminosities for ULIRGs (e.g. \citet{con1}: 
$L_{40-120}\geq10^{11.02}L_{\odot}$, \citet{cle1}: $L_{60}\geq10^{11.77}L_{\odot}$). We first adopted our own 
minimum IR luminosity for a ULIRG, namely a rest-frame $1-1000\mu$m luminosity (hereafter referred to as $L_{ir}$) 
of greater than $10^{12}L_{\odot}$. Using a luminosity derived across a broad wavelength range is to be 
preferred, as luminosities derived across smaller wavelength ranges may 
erroneously exclude or include objects due to their continuum shape. We then assembled a parent sample 
of all objects claimed to be ULIRGs, or close to ULIRG luminosity, at $z\leq0.1$, from the IRAS Point Source Catalogue, 
the IRAS Faint Source Catalogue, and the sample presented by \citet{cle1}. We then determined their 
$1-1000\mu$m luminosities using the methods described in sections \ref{sect:models} and 
\ref{sect:res}, and excluded those objects that did not satisfy our luminosity criterion. From a parent 
sample of 73 objects, 32 were excluded. The final sample of 41 objects, presented in Table \ref{ulirgsample},
are all contained within the IRAS Point Source Catalogue. This final sample thus comprises a complete sample 
of ULIRGs at $z<0.1$ with $L_{ir}>10^{12}L_{\odot}$, and is thus free from selection effects arising from 
infrared colours, or luminosities measured across smaller wavelength ranges.  

Fluxes were assembled from online catalogues and from the literature. Optical and near-IR fluxes were taken from 
the APM and 2MASS databases and also from \citet{car0} \& \citet{spi}. IRAS fluxes were taken from the IRAS 
Faint Source Survey. The XSCANPI software was used to derive IRAS fluxes where only upper limits were present in the 
catalogues. Other infrared data were taken from \citet{car0,mai,kla0,rig,dal,zld,tra2} \& \citet{kla}. 
Sub-millimetre data were taken from \citet{rlr,dun} \& \citet{dun2}. The assembled optical and near-infrared 
photometry is listed in Table \ref{ulirgoptnearir}\footnote{A table containing the assembled infrared and sub-millimetre 
photometry can be obtained from http://spider.ipac.caltech.edu/staff/duncan/irphotom.tex}.

\begin{table*}
\begin{minipage}{170mm}
\caption{Ultraluminous Infrared Galaxies: Optical and near-IR data \label{ulirgoptnearir}}
\begin{tabular}{@{}lcccccccc}
\hline
Name       & U            &  B            &  V            &  R            &  J           &  H           &  K           & L       \\        
\hline
00198-7926 & 	          &               &               &               & $619\pm41$   & $764\pm51$   & $565\pm38$   & $232\pm15$ \\
00199-7426 &              &               &               &               & $1.0\pm0.1$  & $5.4\pm0.6$  & $6.4\pm0.7$  &  \\
00262+4251 &              &               &               &               & $5.4\pm1.6$  &              & $5.5\pm1.6$  &  \\
00335-2732 &              &               &               & $0.94\pm0.1$  & $4.2\pm0.4$  & $3.3\pm0.3$  & $5.0\pm0.5$  &  \\
01388-4618 &              & $0.31\pm0.04$ &               &               &              &              &              &  \\
02364-4751 &              & $0.44\pm0.05$ &               &               &              &              &              &  \\
03068-5346 &              &               &               &               &              &              &              & \\  
04232+1436 &              &               &               &               & $5.5\pm1.6$  &              & $10.6\pm3.2$ & \\ 
05189-2524 &              & $0.66\pm0.07$ & $0.65\pm0.07$ & $0.76\pm0.08$ & $14.7\pm0.9$ & $30.8\pm2.1$ & $57.1\pm3.8$ & $136\pm30$ \\
06035-7102 &              &               &               &               & $2.1\pm0.2$  & $2.8\pm0.3$  & $4.2\pm0.4$  &  \\
06206-6315 &              &               &               &               & $1.3\pm0.1$  & $1.7\pm0.2$  & $2.8\pm0.3$  &  \\
08572+3915 &              & $0.51\pm0.05$ & $0.65\pm0.07$ & $0.90\pm0.1$  & $1.7\pm0.1$  & $3.0\pm0.2$  & $3.9\pm0.3$  & $50\pm10$ \\
09111-1007 &              &               &               &               & $3.4\pm0.3$  & $5.5\pm0.6$  & $7.4\pm0.7$  & \\
09320+6134 &              & $0.96\pm0.1$  & $10.91\pm1.2$ & $1.11\pm0.1$  & $12.9\pm0.9$ & $21.2\pm1.4$ & $29.2\pm1.9$ & $33\pm7$     \\  
09583+4714 &              &               &               &               & $1.2\pm0.1$  & $1.5\pm0.2$  & $2.0\pm0.2$  &  \\ 
10035+4852 &              &               &               &               &              &              &              & \\
10190+1322 &              &               &               &               & $2.3\pm0.2$  & $3.0\pm0.3$  & $2.1\pm0.2$  &  \\
10494+4424 &              &               &               &               & $1.0\pm0.1$  & $2.0\pm0.2$  & $2.8\pm0.3$  & \\
10565+2448 &  	          &               &               &               & $14.0\pm1.4$ & $17.0\pm1.7$ & $18.0\pm1.8$ & \\
12112+0305 &              & $0.16\pm0.02$ & $0.25\pm0.03$ & $0.32\pm0.03$ & $2.0\pm0.2$  & $2.9\pm0.3$  & $3.5\pm0.4$  & $2.2\pm0.5$  \\
12540+5708 & $2.77\pm0.5$ & $7.34\pm1.24$ & $13.1\pm2.2$  & $11.5\pm0.1$  & $49\pm3$     & $107\pm7$    & $186\pm12$   & $368\pm72$   \\
13428+5608 &              & $0.74\pm0.07$ & $1.03\pm0.1$  & $1.76\pm0.2$  & $9.0\pm0.9$  & $13.5\pm1.4$ & $16.1\pm1.6$ & $17.4\pm1.7$   \\
14348-1447 &              & $0.22\pm0.02$ & $0.24\pm0.02$ & $0.43\pm0.04$ & $1.7\pm0.2$  & $2.5\pm0.3$  & $3.3\pm0.3$  & \\
14378-3651 &              &               &               &               & $2.5\pm0.3$  & $2.8\pm0.3$  & $5.0\pm0.5$  & \\
15250+3609 &              & $0.26\pm0.03$ & $0.27\pm0.03$ & $0.42\pm0.04$ & $3.1\pm0.3$  & $4.0\pm0.4$  & $4.1\pm0.4$  & $4.0\pm0.4$  \\
15327+2340 &              & $0.96\pm0.1$  & $3.16\pm0.3$  & $4.07\pm0.4$  & $9.6\pm1.0$  & $17.9\pm1.8$ & $23.0\pm2.3$ & $22.7\pm2.3$ \\ 
17132+5313 &              &               &               &               & $5.6\pm0.6$  & $8.2\pm0.8$  & $8.9\pm0.9$  & $10.5\pm1.6$ \\
17208-0014 &              &               &               &               &              &              &              & \\ 
18470+3233 &              &               &               &               & $2.7\pm0.4$  &              & $3.6\pm0.5$  & \\
19254-7245 &              &               &               &               &              &              &              & \\
19297-0406 &              &               &               &               &              &              &              & \\
19458+0944 &              &               &               &               &              &              &              & \\
20046-0623 &              &               &               &               & $3.9\pm0.6$  &              & $4.0\pm0.6$  & \\  
20414-1651 &              &               &               &               &              &              &              & \\
20551-4250 &              & $3.93\pm0.34$ &               & $5.04\pm0.34$ & $22.9\pm5.9$ & $30.2\pm7.8$ & $25.7\pm6.7$ & \\ 
21130-4446 &              &               &               & $0.89\pm0.09$ &              &              &              &\\ 
21504-0628 &              &               &               &               & $5.5\pm0.8$  &              & $5.7\pm0.9$  & \\ 
22491-1808 &              & $0.22\pm0.02$ & $0.20\pm0.02$ & $0.26\pm0.03$ & $2.1\pm0.2$  & $2.7\pm0.3$  & $2.6\pm0.3$  & $2.4\pm0.2$ \\   
23128-5919 &              & $4.09\pm0.4$  &               & $7.69\pm0.7$  & $9.6\pm0.6$  & $12.6\pm0.8$ & $11.9\pm.08$ & \\   
23365+3604 &              &               &               &               & $7.5\pm0.4$  &              & $6.5\pm0.3$  & \\ 
23389-6139 &              & $0.35\pm0.04$ &               &               &              &              &              & \\
\hline
\end{tabular}

\medskip

All fluxes are given in mJy and are in the observed frame. $1\sigma$ errors are quoted.

\end{minipage}
\end{table*}

\section{Infrared emission models}\label{sect:models}

\subsection{Starburst Models}

To model the IR emission due to starburst activity we used the starburst models of \citet{ef1}. These models consider 
an ensemble of evolving HII regions containing hot young stars, embedded within Giant Molecular Clouds (GMCs) of gas 
and dust. The composition of the dust is given by the dust grain model of \citet{sie}, which includes the Polycyclic 
Aromatic 
Hydrocarbons (PAHs) thought to be responsible for the $7.7\mu$m emission feature. The stellar populations 
within the GMCs evolve according to the stellar synthesis codes of \citet{bru}. 
The star formation rate is assumed to decay exponentially with an $e$-folding timescale of $2\times10^{7}$ years. The models 
vary in starburst age from zero years up to $7.2\times10^{7}$ years (11 discrete values) and in effective visual optical depth from
$\tau_{V}=50$ to $\tau_{V}=200$ (4 discrete values). A plot showing the range in starburst SED template shapes can be 
found in figure 3 of \citet{ef1}, with additional information in figure 1 of the same paper.

\subsection{AGN Torus Models}

To model the IR emission due to AGN, we used the AGN models of \citet{ef0}.
These models incorporate accurate solutions to the axially symmetric radiative-transfer problem in dust clouds to model 
the IR emission from dust in active galactic nuclei. Dust composition is given by the multigrain dust model of \citet{rra}. 
We have used the thick tapered disk models following an $r^{-1}$ density distribution, as this subset of models has been 
found to be most successful in satisfying the observational constraints of AGN. The AGN models vary in torus opening angle 
from $0\degr$ to $90\degr$ (15 discrete values) and in UV equatorial optical depth, from $\tau_{UV}=1000$ to $\tau_{UV}=1500$ 
(3 discrete values). A plot showing the range in AGN torus SED template shapes can be found in figure 5 of \citet{ef0}.

\section{Results}\label{sect:res}

\subsection{Spectral Energy Distributions}
We combined the archival photometry from the optical to the millimetre to fit Spectral Energy Distributions (SEDs) 
for each object. Goodness of fit was examined by using the reduced $\chi^{2}$ statistic. Fits for all sources were good, 
with $\chi^{2}_{best}\leq5$ in all cases, and $\chi^{2}_{best}\leq4$ in all but one case (this object, IRAS 17208-0014, is 
discussed further in Section \ref{sect:ind}). To determine the errors on the model parameters we explored 
reduced $\chi^{2}$ space between $\chi^{2}_{best}$ and $\chi^{2}_{best}+2$. The resulting luminosities, model parameters and 
errors are given in Table \ref{ulirgparams}. The SEDs are presented in Figure \ref{ulirg_seds}.

Emission from unobscured population II stars from the merger progenitors is not included in either the starburst or AGN 
models, and care must be taken in accounting for this, as a recent Hubble Space Telescope ({\it HST}) study of 
ULIRGs \citep{far1} found that the optical emission was in most cases dominated by old stellar populations and not by light 
from a starburst or AGN. The range in possible optical/near-IR SEDs for evolved stellar populations is very large, and 
in addition an obscured starburst or AGN can in principle contribute significantly in the optical or near-IR. Constraining 
the properties of the evolved stellar component in any of our sample with the limited available optical photometry therefore 
proved impossible. We have therefore assumed that all emission shortwards of $3.5\mu$m in the rest frame of the objects 
contains a significant but unquantified contribution from old stellar populations. As this contribution could lie between 
$0\%$ and $100\%$, any measured flux at a rest-frame wavelength shorter than $3.5\mu$m is treated as a  $3\sigma$ upper 
limit in the fitting.

\subsection{Star Formation Rates}
Estimating obscured star formation rates from IR data is based on silicate and graphite dust grains absorbing 
the optical and UV light from young stars and reradiating in the IR and sub-mm. A recent estimate for deriving 
star formation rates from IR luminosities has been made by \citet{rr0}:

\begin{equation}
\stackrel{.}{M}_{*,all} = 2.6\times10^{-10}\frac{\phi}{\epsilon}\frac{L_{60}}{L_{\odot}}
\label{equnsfr}
\end{equation}

\noindent where $\stackrel{.}{M}_{*,all}$ is the star formation rate, $\epsilon$ is the fraction of 
optical/UV light from the starburst that is absorbed by dust and re-emitted in the IR, and $L_{60}$ is the 
$60\mu$m starburst luminosity. The factor $\phi$ incorporates the correction between a Salpeter IMF and the 
true IMF ($\times1.0$ for a Salpeter IMF, $\times3.3$ for a Miller-Scalo IMF), and a correction for the 
assumed upper and lower stellar mass bounds for the stars forming in the starburst ($\times1.0$ if the mass 
range is $0.1 < M_{\odot} < 100$, $\times 0.323$ if the mass range spans only OBA type stars, i.e.  
$1.6 < M_{\odot} < 100$). We have assumed that all of the optical/UV light from the starburst is absorbed 
($\epsilon\sim1.0$), and that the IMF of the starburst is a Salpeter IMF forming stars across the mass range 
$0.1 < M_{\odot} < 100$, ($\phi=1.0$). The star formation rates for our sample calculated using Equation 
\ref{equnsfr} are given in Table \ref{ulirgparams}.

\subsection{Dust Masses and Temperatures}
The starburst and AGN models do not assume a monolithic dust temperature, but instead invoke a range of dust temperatures 
spanning 10K to 1000K. An estimate of the mass-weighted mean dust temperature can be obtained by fitting an optically thin 
greybody function of the form:

\begin{equation}
S_{\nu} = \nu^{\beta}B(\nu, T_{dust})
\label{greybody}
\end{equation}

\noindent to the FIR SED over the wavelength range $200 - 1000\mu$m, which measures the colour temperature of the system. 
In Equation \ref{greybody}, $S_{\nu}$ is the source flux at a frequency $\nu$, $T_{dust}$ is the dust temperature, 
$B(\nu, T_{dust})$ is the Planck function, and $\beta$ is the frequency dependence of the grain emissivity. Monolithic 
dust temperatures derived in this way are unphysical simplifications in AGN, but can serve as a useful comparison with 
previous work. Mass-weighted mean dust temperatures for our sample are presented in Table \ref{ulirgparams}.

Gas and dust masses can be estimated in two ways. The first way is to calculate the gas mass directly from the models using 
equation 4 of \citet{far3}. This however gives the total gas mass at the current age of the starburst without accounting 
for gas that is blown out of the nuclear regions by supernovae or superwinds, and thus may overestimate the amount of 
gas and dust in the system. The second approach is to estimate the dust mass directly from the sub-mm SED. An approach for 
estimating dust and gas masses in galaxies using sub-mm data is to assume the system is optically thin at these wavelengths, 
and to follow the prescription of \citet{hil0}:

\begin{equation}
M_{dust}=\frac{1}{1+z}\frac{S_{\nu_{o}}D_{L}^{2}}{\kappa(\nu_{r})B(\nu_{r},T_{dust})}
\label{eqn:mdust}
\end{equation} 

\noindent where $\nu_{o}$ and $\nu_{r}$ are the observed and rest frame frequencies respectively, $S_{\nu_{o}}$ is the 
flux in the observed frame, $B(\nu_{r},T_{dust})$ is the Planck function in the rest frame and $T_{dust}$ is the dust 
temperature. The gas mass is then obtained by assuming a fixed gas to dust ratio. For the most extreme {\it IRAS} galaxies, 
the best current estimate of the gas to dust ratio is $540\pm290$ \citep{sss}. For comparison, the gas to dust ratio 
in spiral galaxies is thought to be $\sim500$ \citep{dev}, and $\sim700$ in ellipticals \citep{wik}. The mass absorption 
coefficient in the rest frame, $\kappa_{r}$, is taken to be:

\begin{equation}
\kappa_{r}= 0.067\left(\frac{\nu_{r}}{2.5\times10^{11}}\right)^{\beta}
\label{eqn:mabsorb}
\end{equation} 

\noindent in units of m$^{2}$kg$^{-1}$. Dust masses and temperatures calculated using Equations 
\ref{greybody} and \ref{eqn:mdust} are given in Table \ref{ulirgparams}.

\begin{figure*}
\begin{minipage}{180mm}
\epsfig{figure=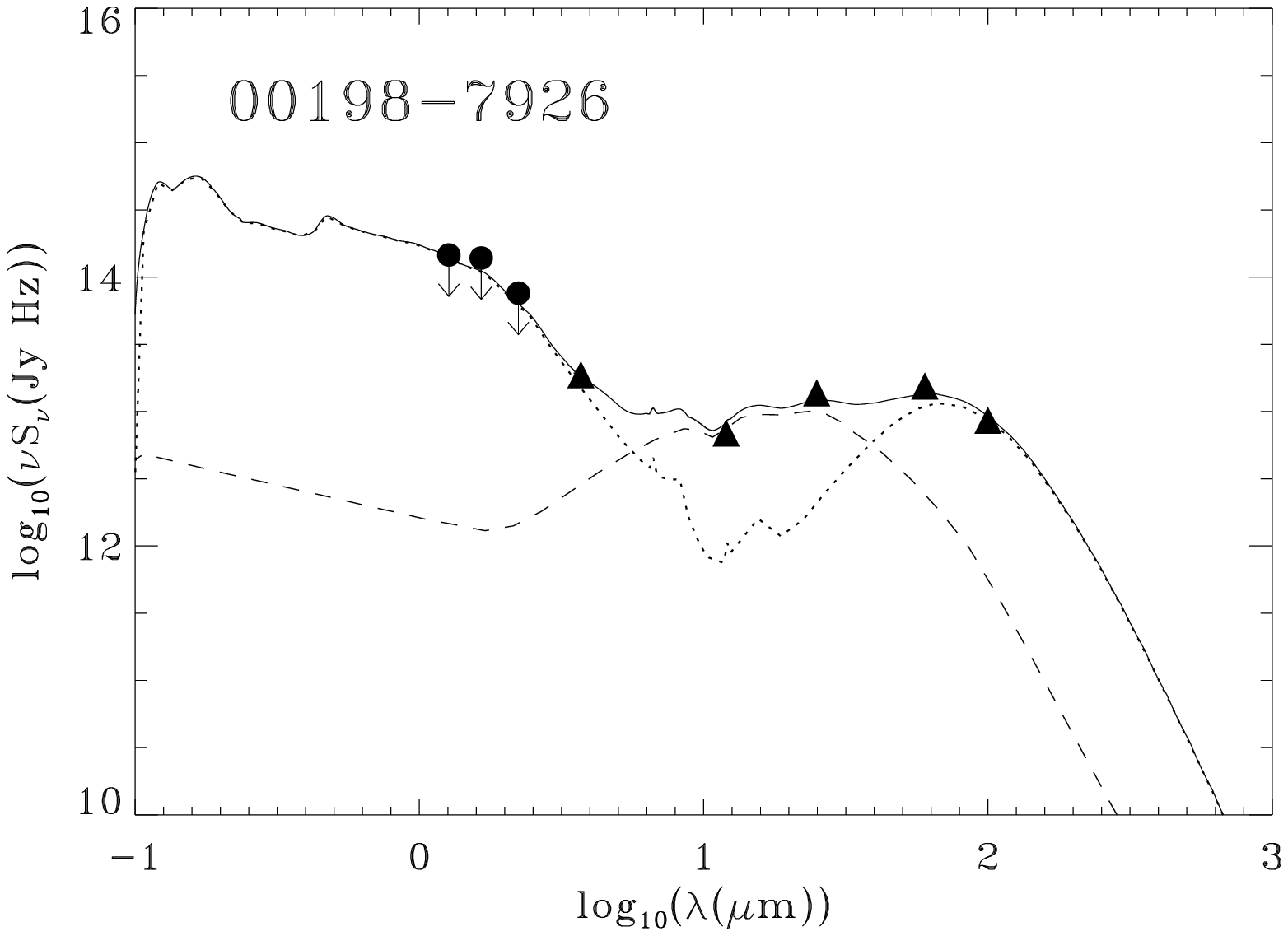,width=89mm}
\epsfig{figure=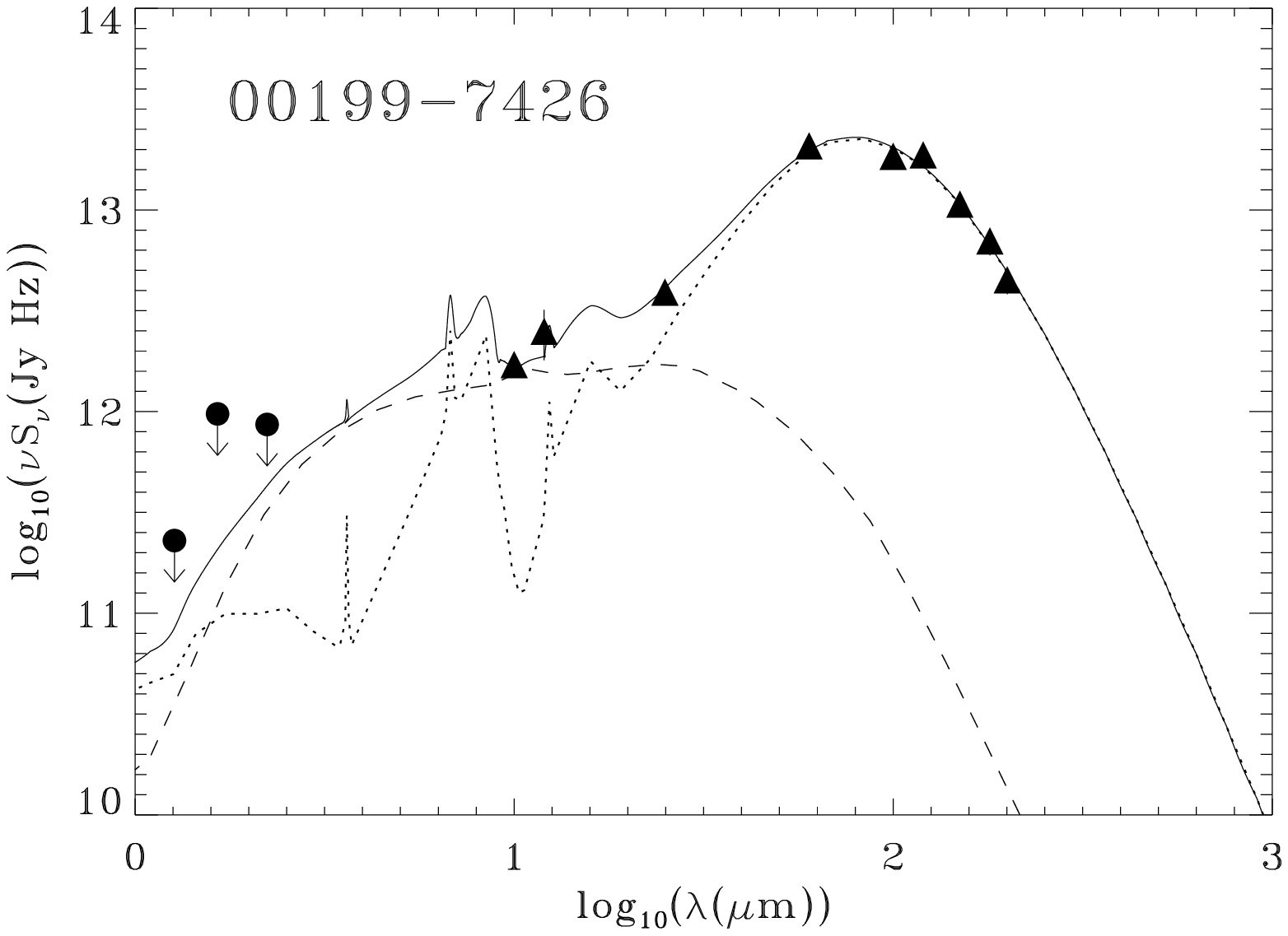,width=89mm}
\end{minipage}
\begin{minipage}{180mm}
\epsfig{figure=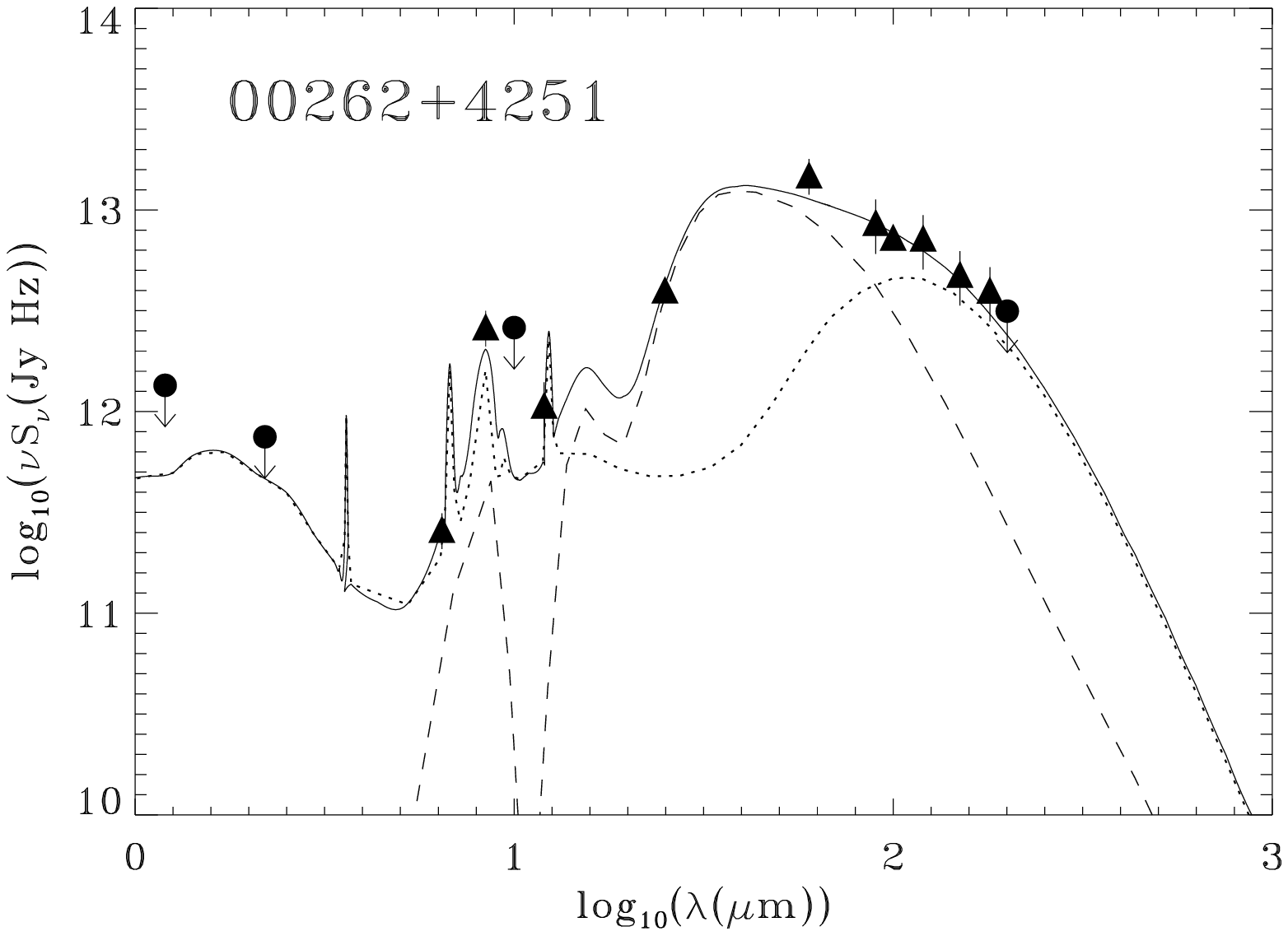,width=89mm}
\epsfig{figure=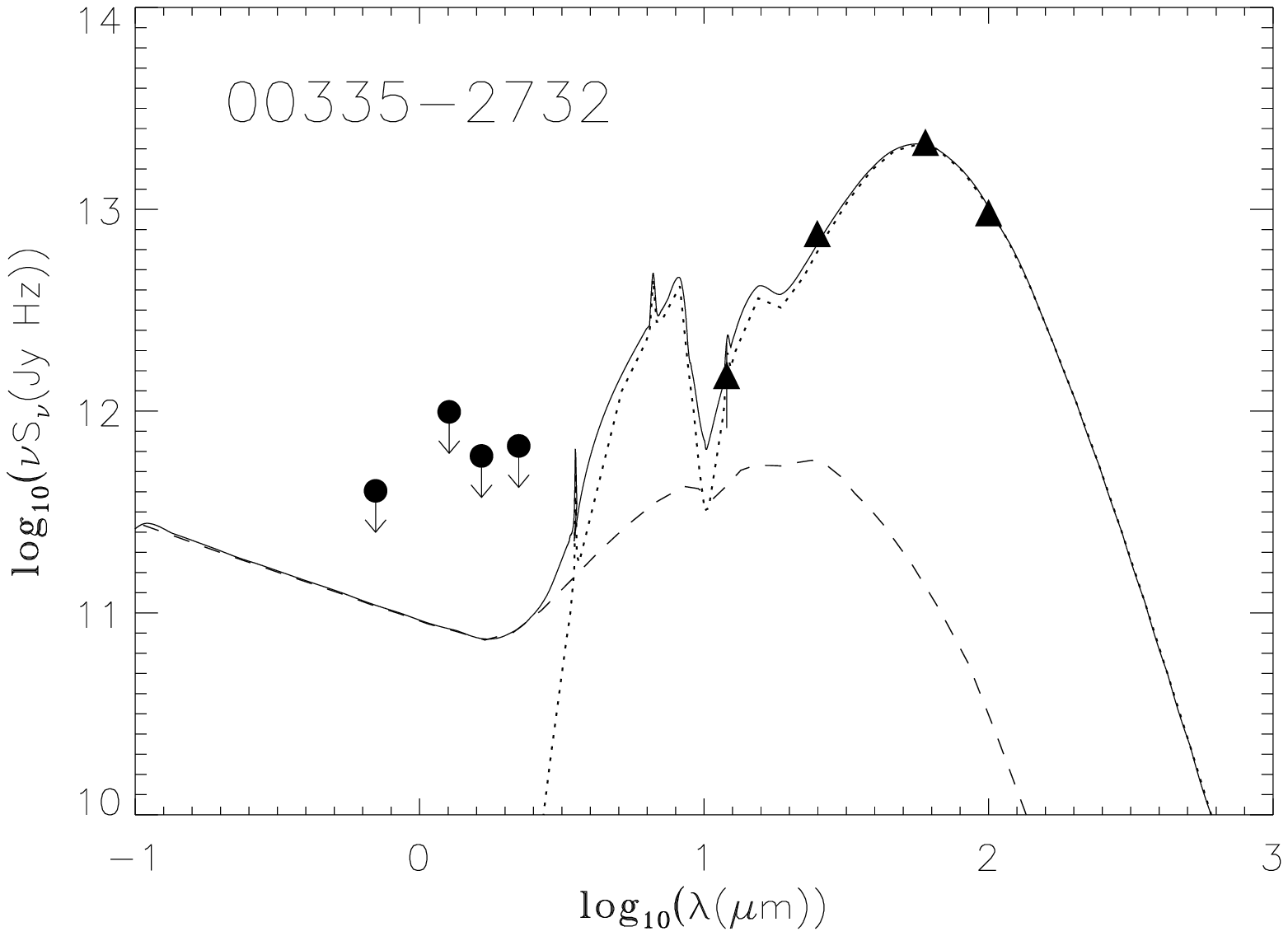,width=89mm}
\end{minipage}
\begin{minipage}{180mm}
\epsfig{figure=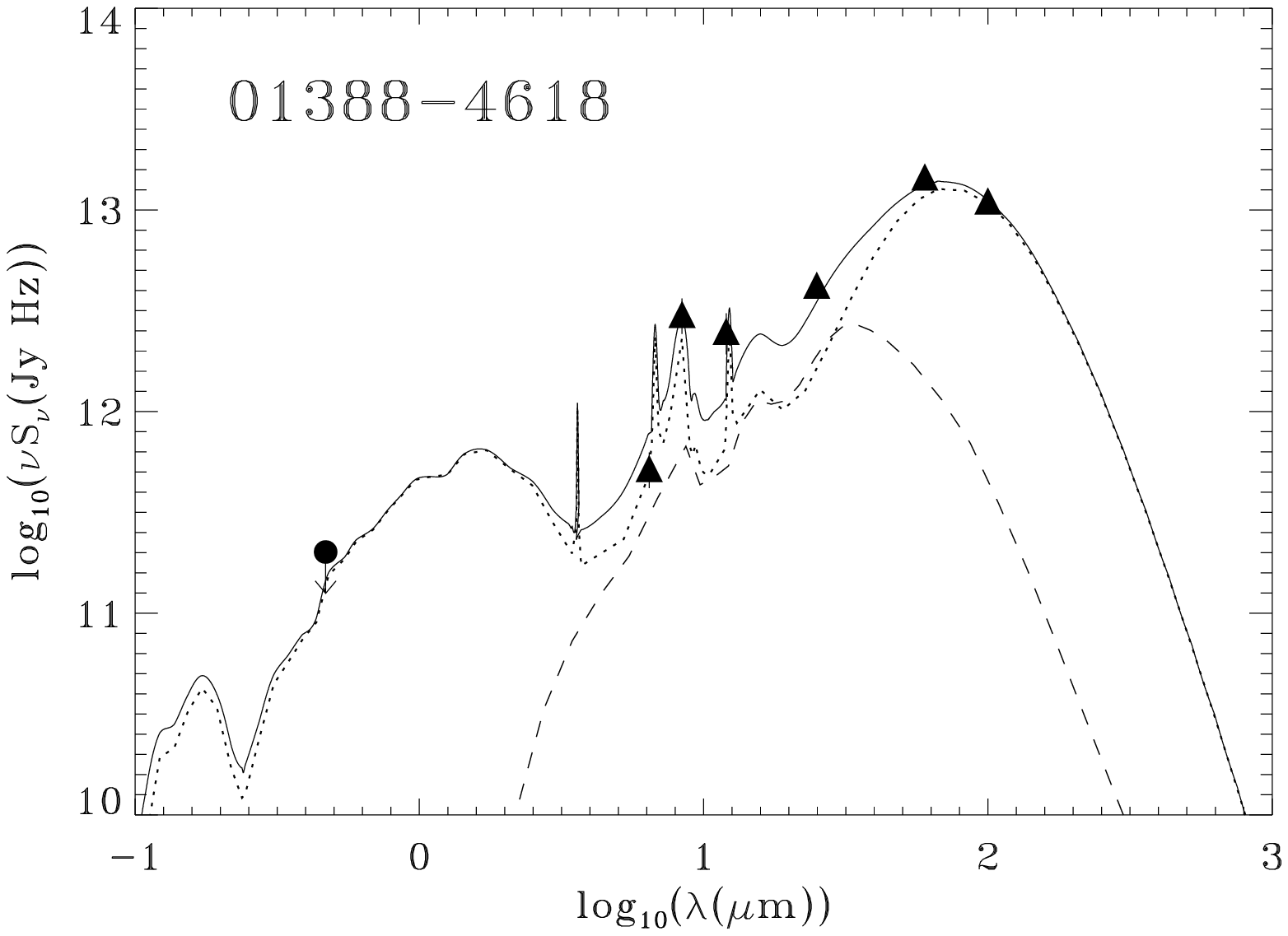,width=89mm}
\epsfig{figure=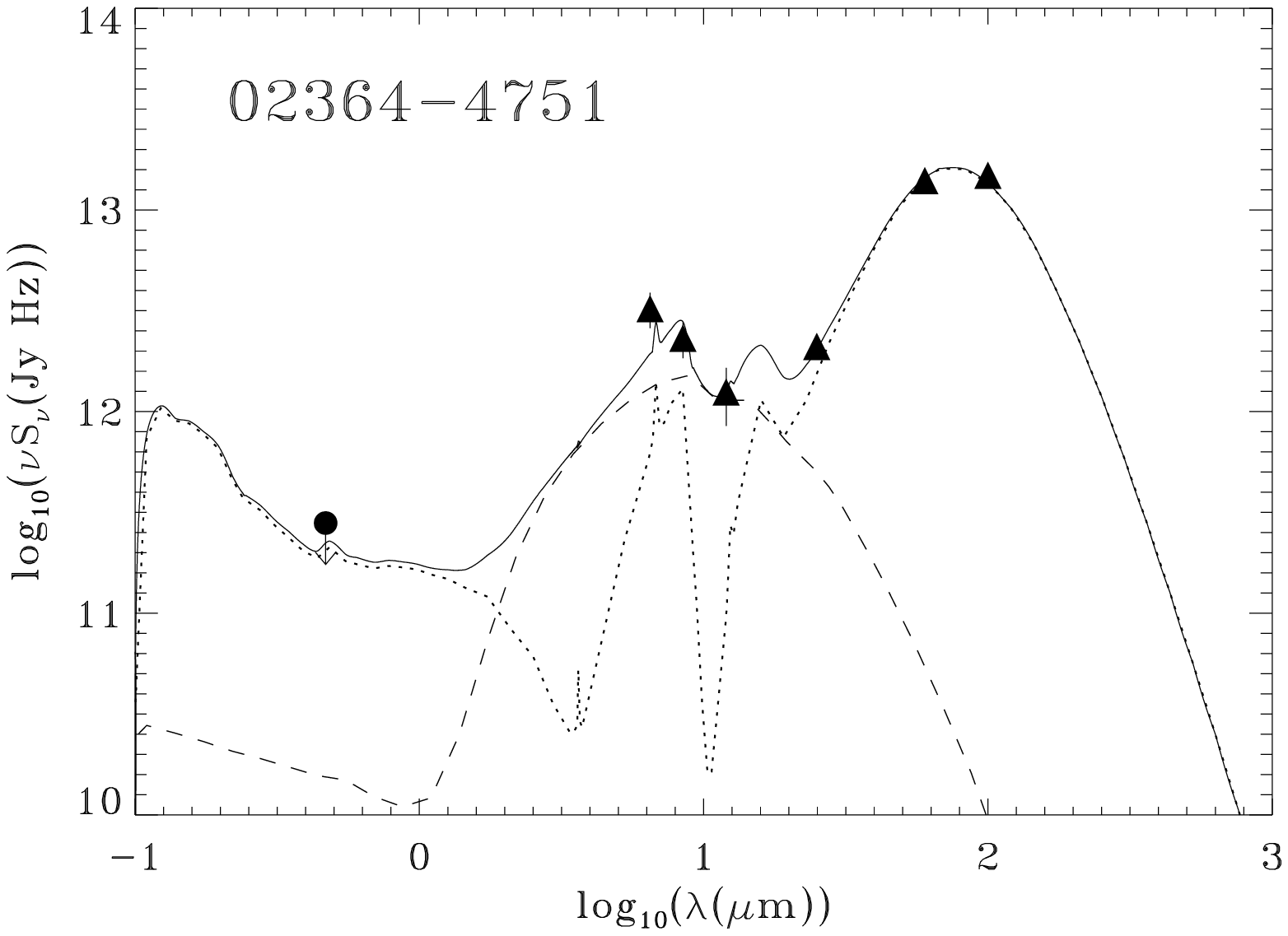,width=89mm}
\end{minipage}
\caption{ Best fit Spectral Energy Distributions for the 41 ULIRGs in our sample. In each 
case the solid line is the combined best-fit model, the dotted line is the Starburst 
component and the long dashed line is the AGN component.  
 \label{ulirg_seds}}
\end{figure*}

\begin{figure*}
\begin{minipage}{180mm}
\epsfig{figure=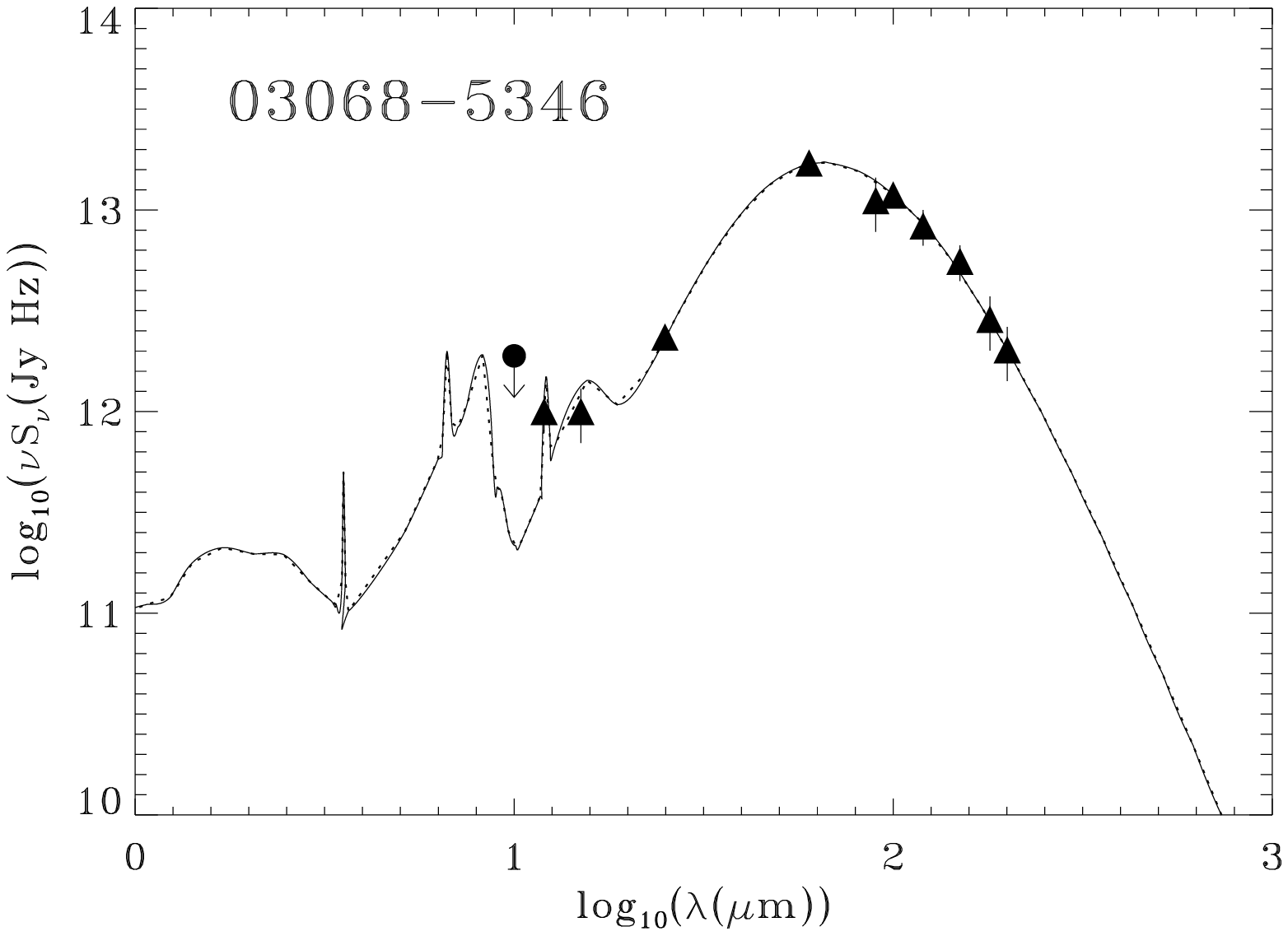,width=89mm}
\epsfig{figure=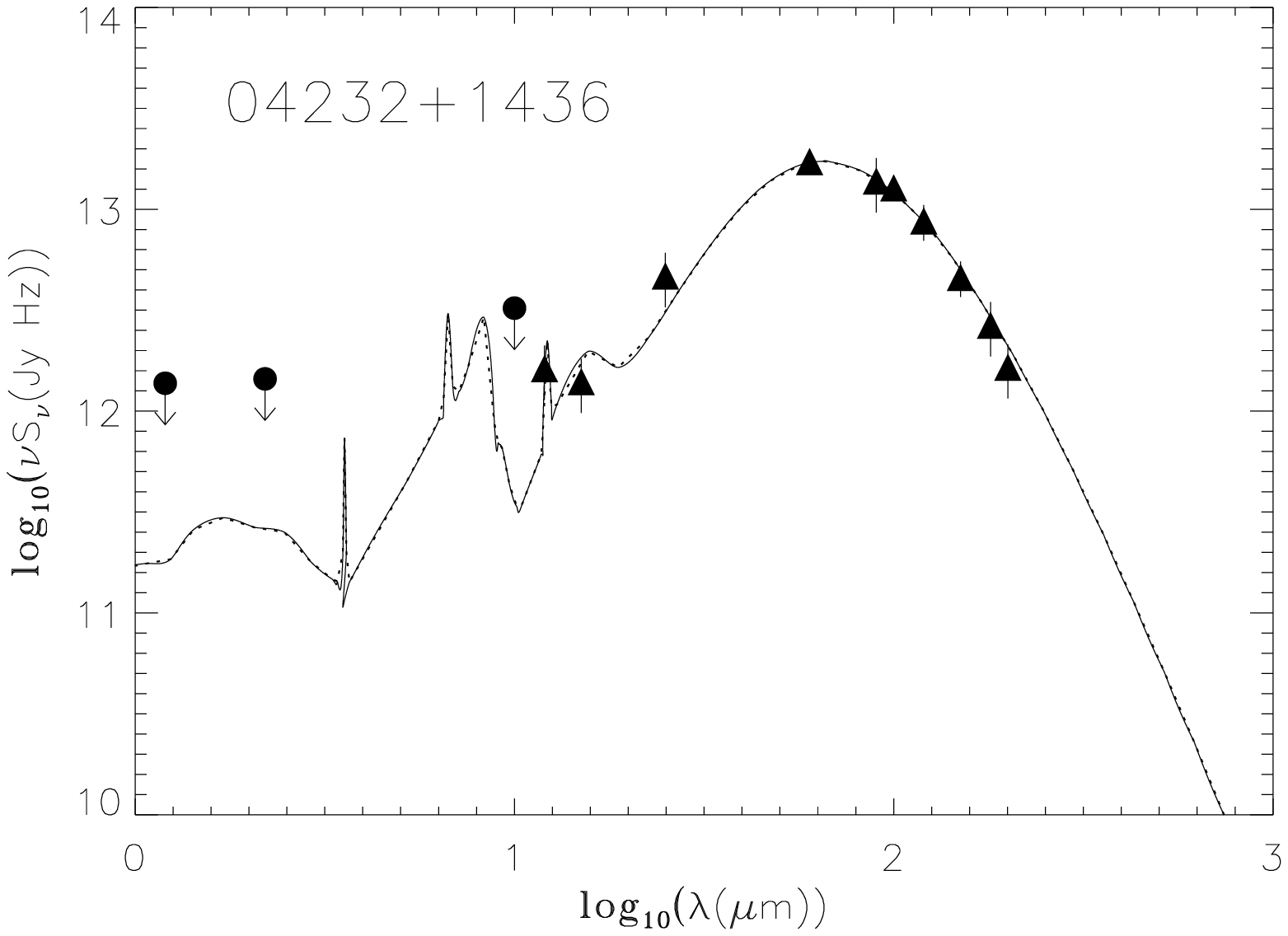,width=89mm}
\end{minipage}
\begin{minipage}{180mm}
\epsfig{figure=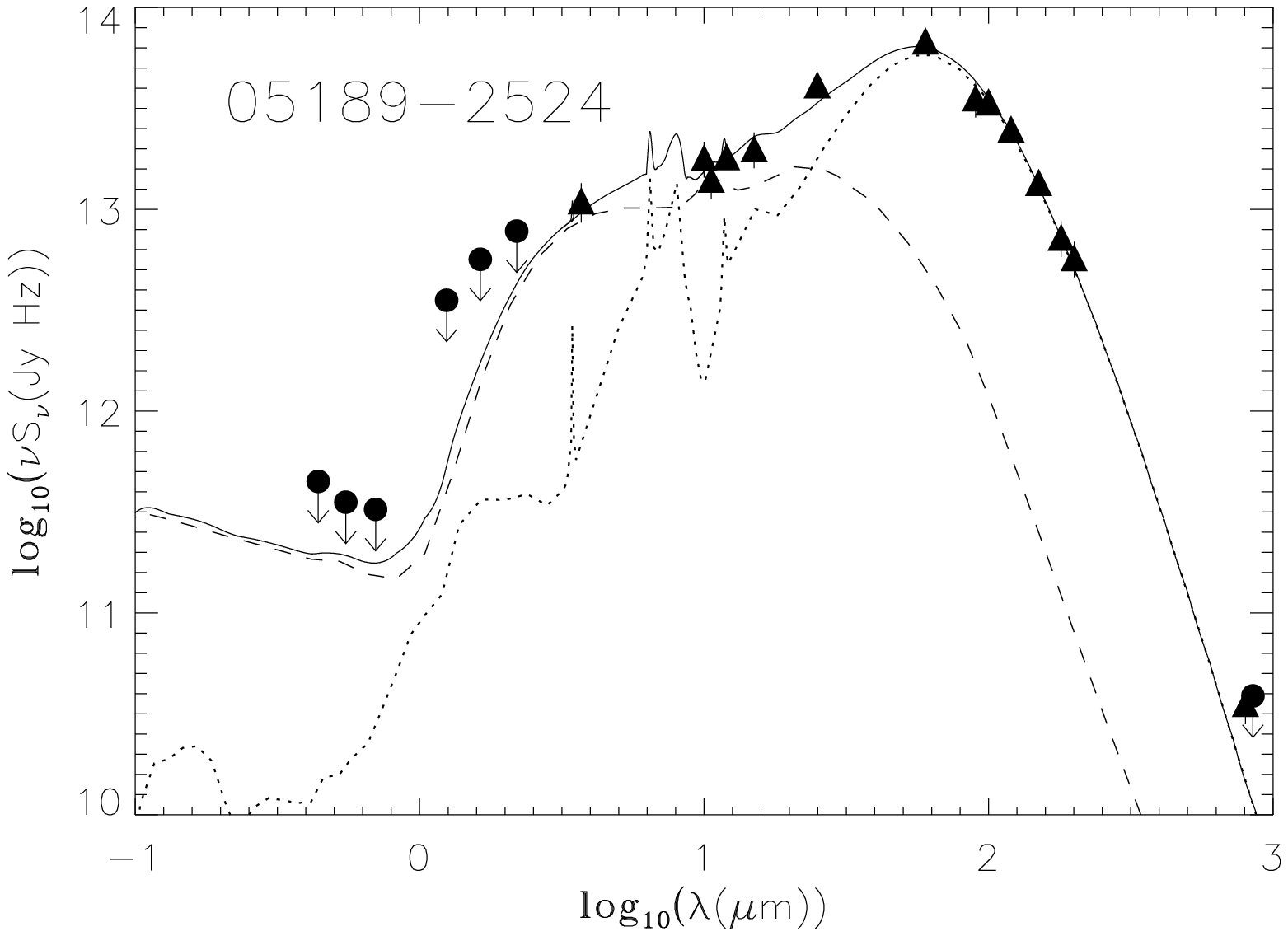,width=89mm}
\epsfig{figure=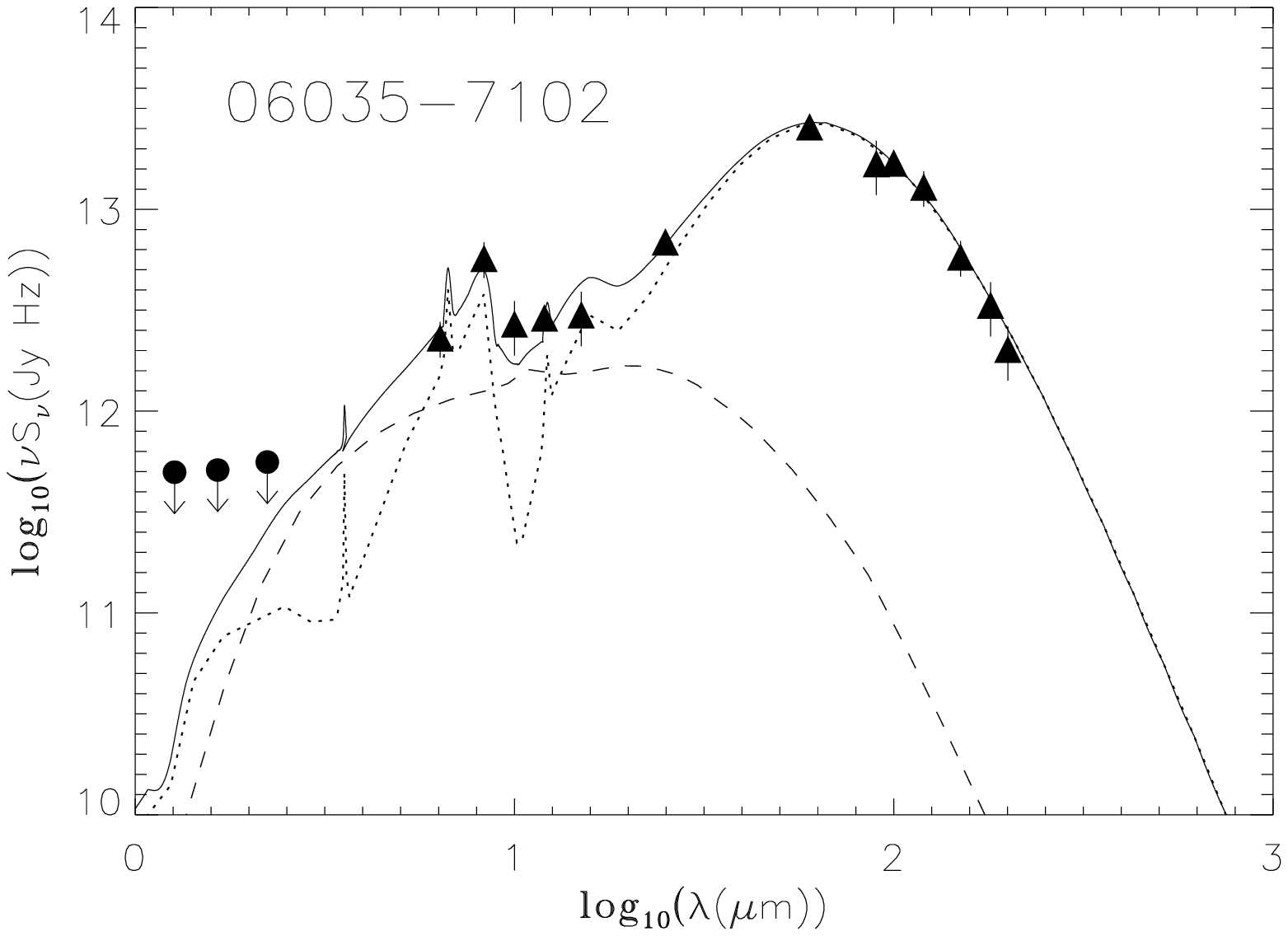,width=89mm}
\end{minipage}
\begin{minipage}{180mm}
\epsfig{figure=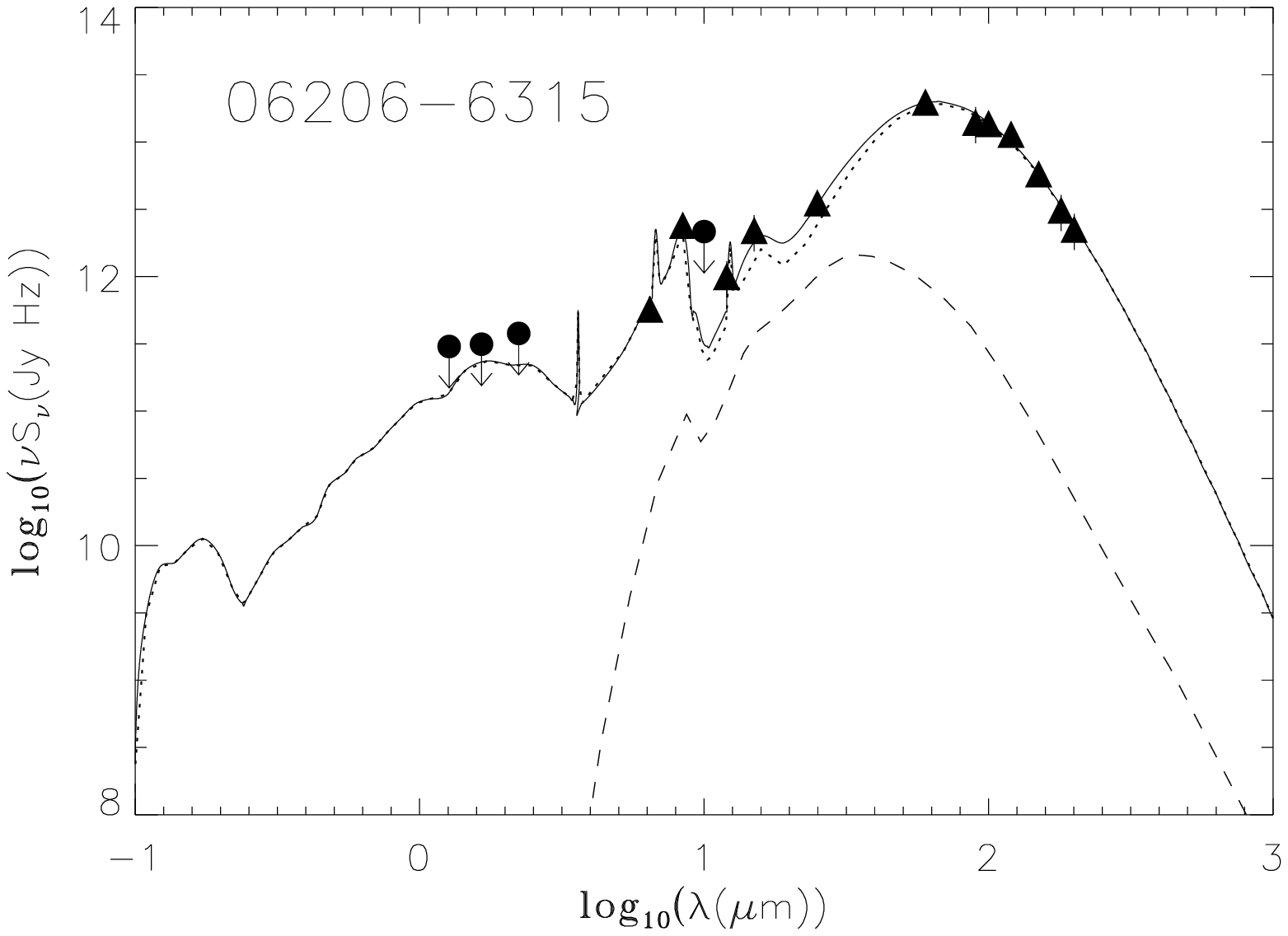,width=89mm}
\epsfig{figure=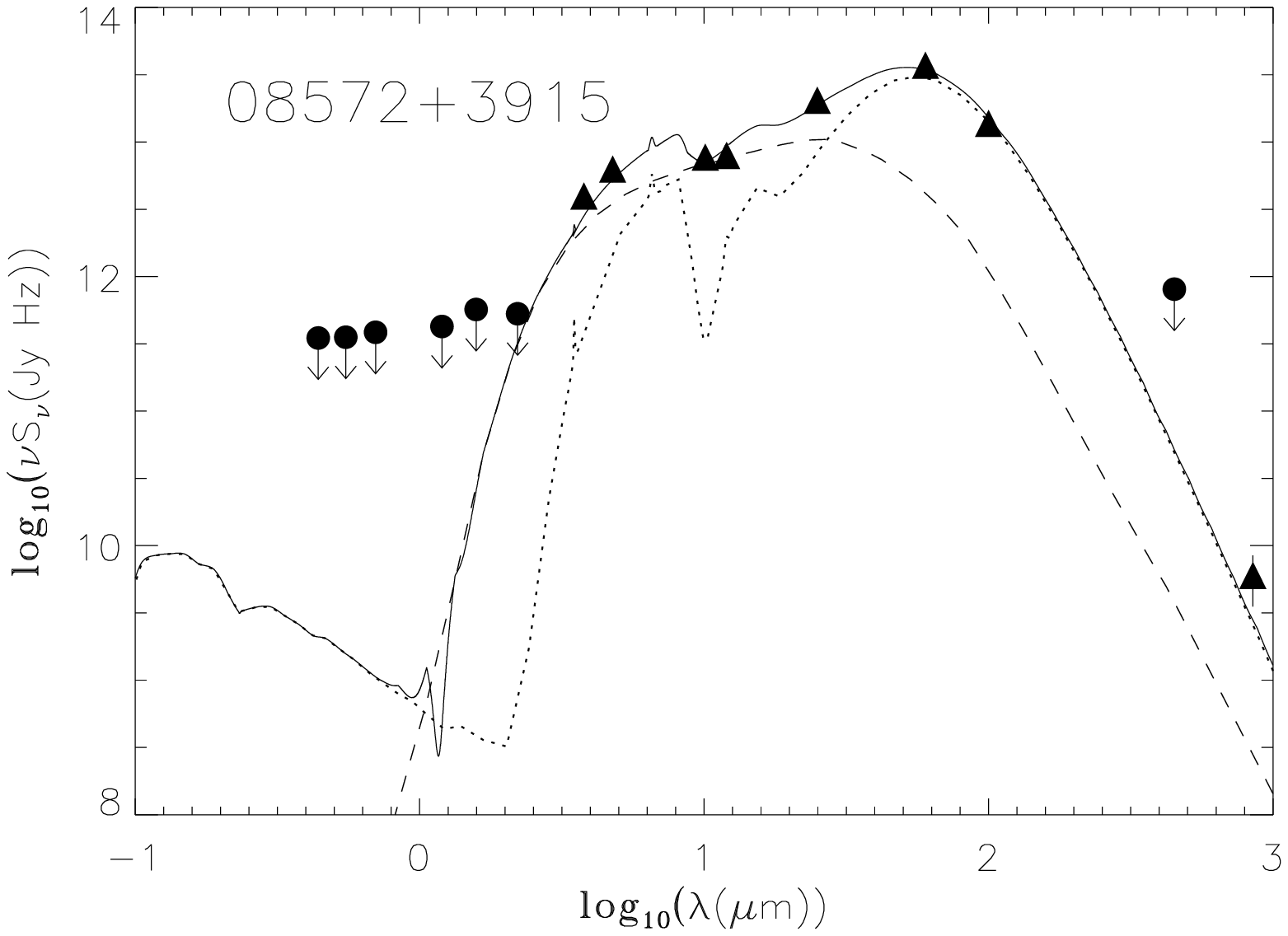,width=89mm}
\end{minipage}
\contcaption{}
\end{figure*}

\begin{figure*}
\begin{minipage}{180mm}
\epsfig{figure=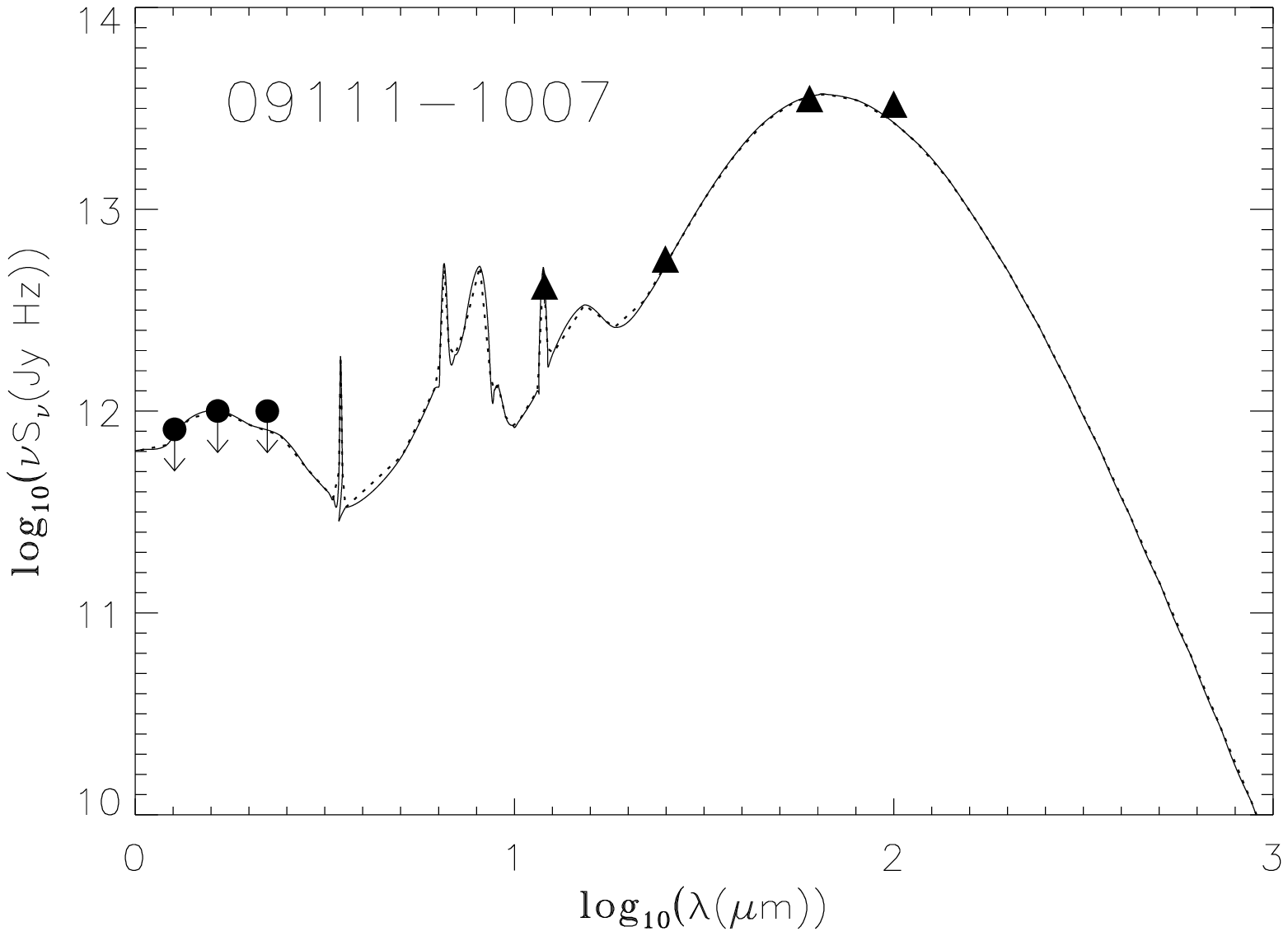,width=89mm}
\epsfig{figure=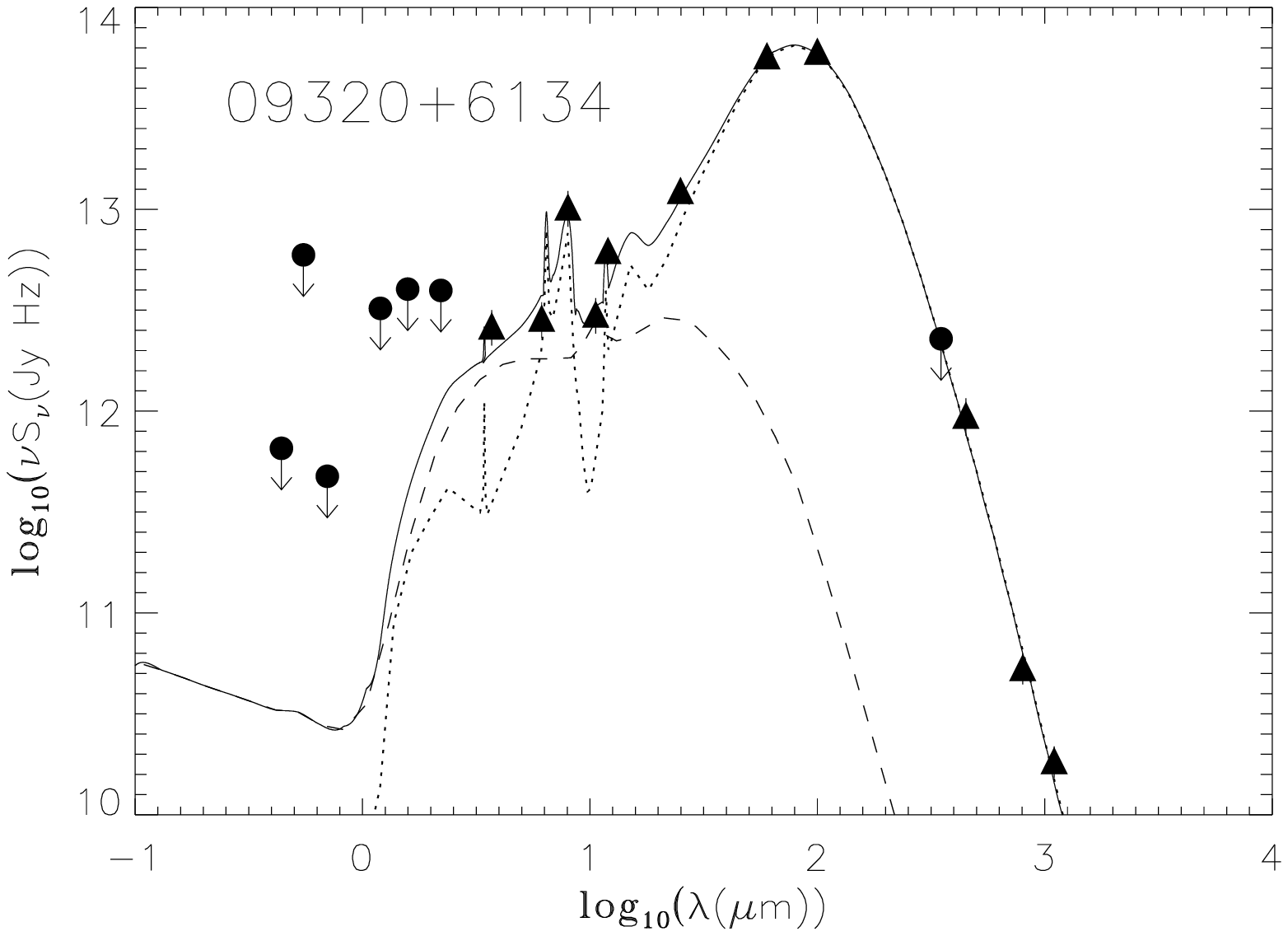,width=89mm}
\end{minipage}
\begin{minipage}{180mm}
\epsfig{figure=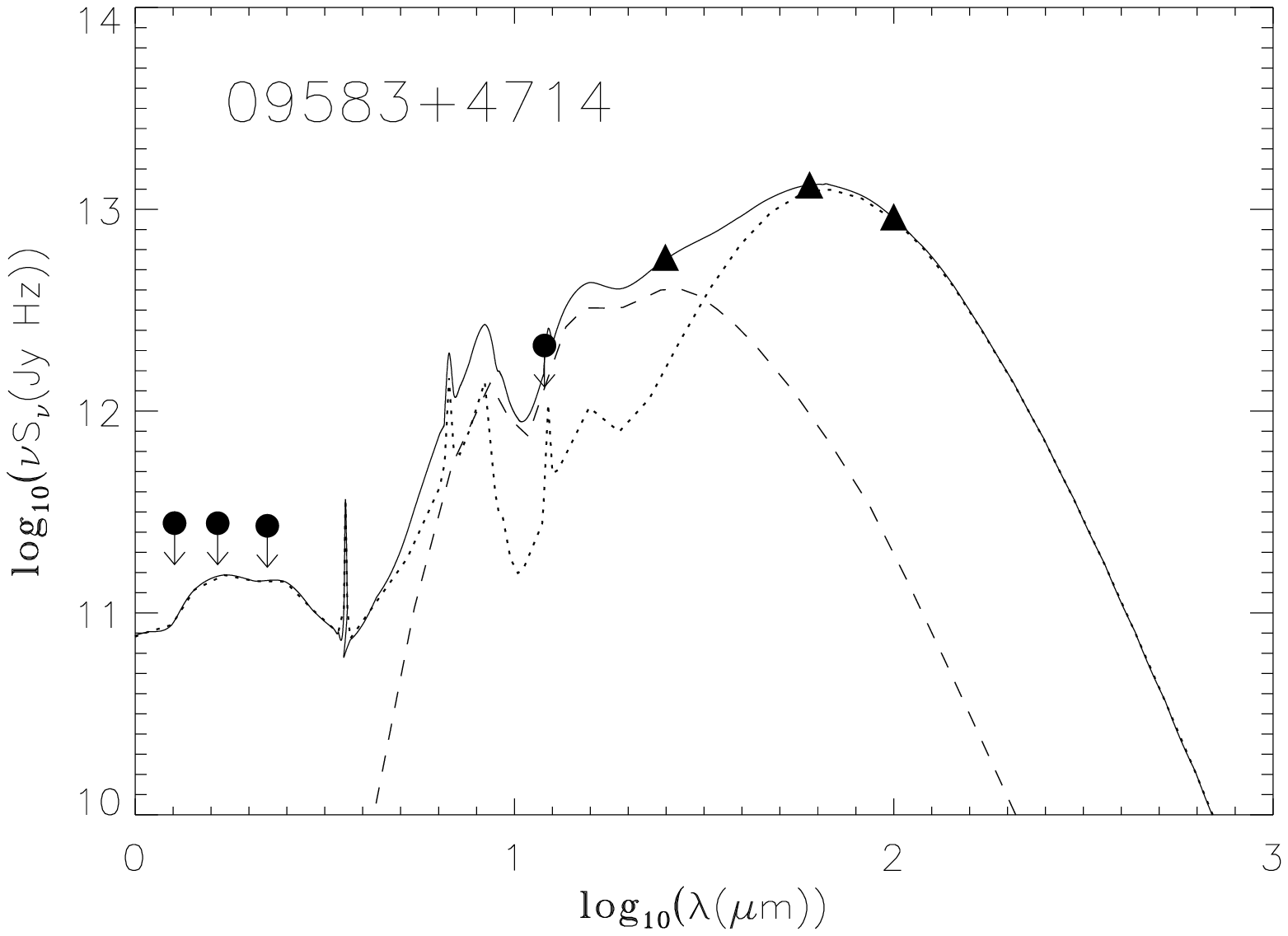,width=89mm}
\epsfig{figure=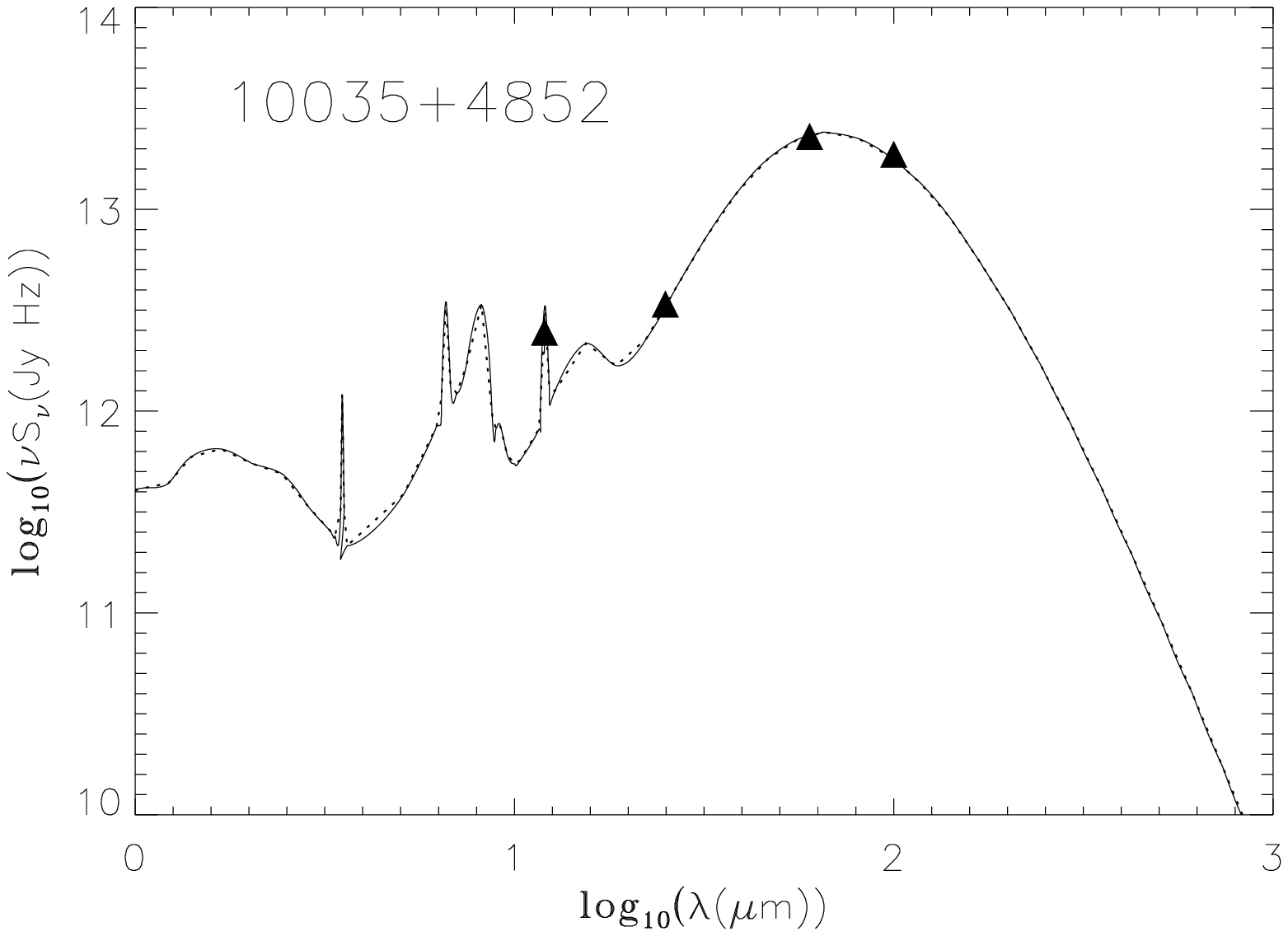,width=89mm}
\end{minipage}
\begin{minipage}{180mm}
\epsfig{figure=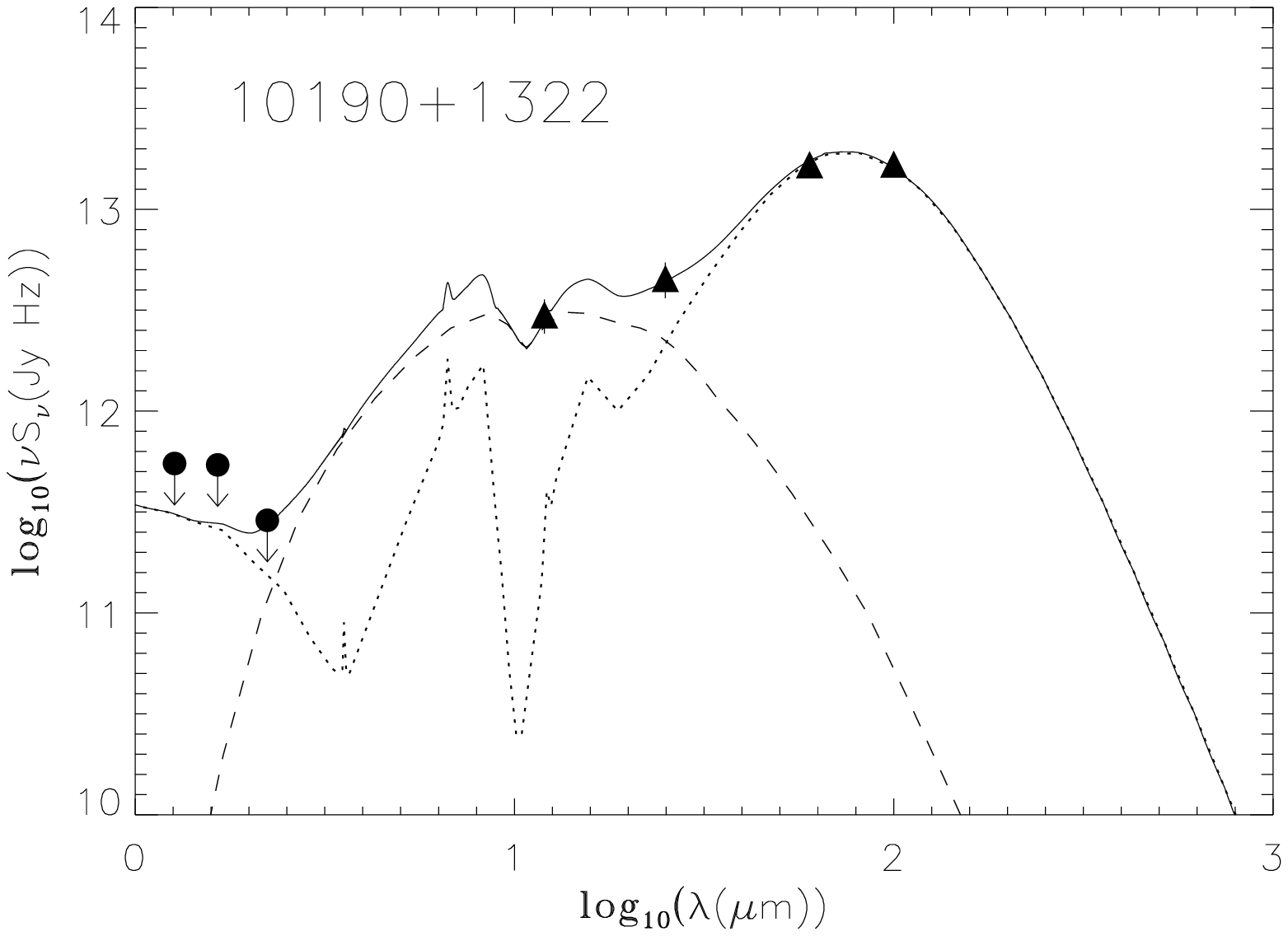,width=89mm}
\epsfig{figure=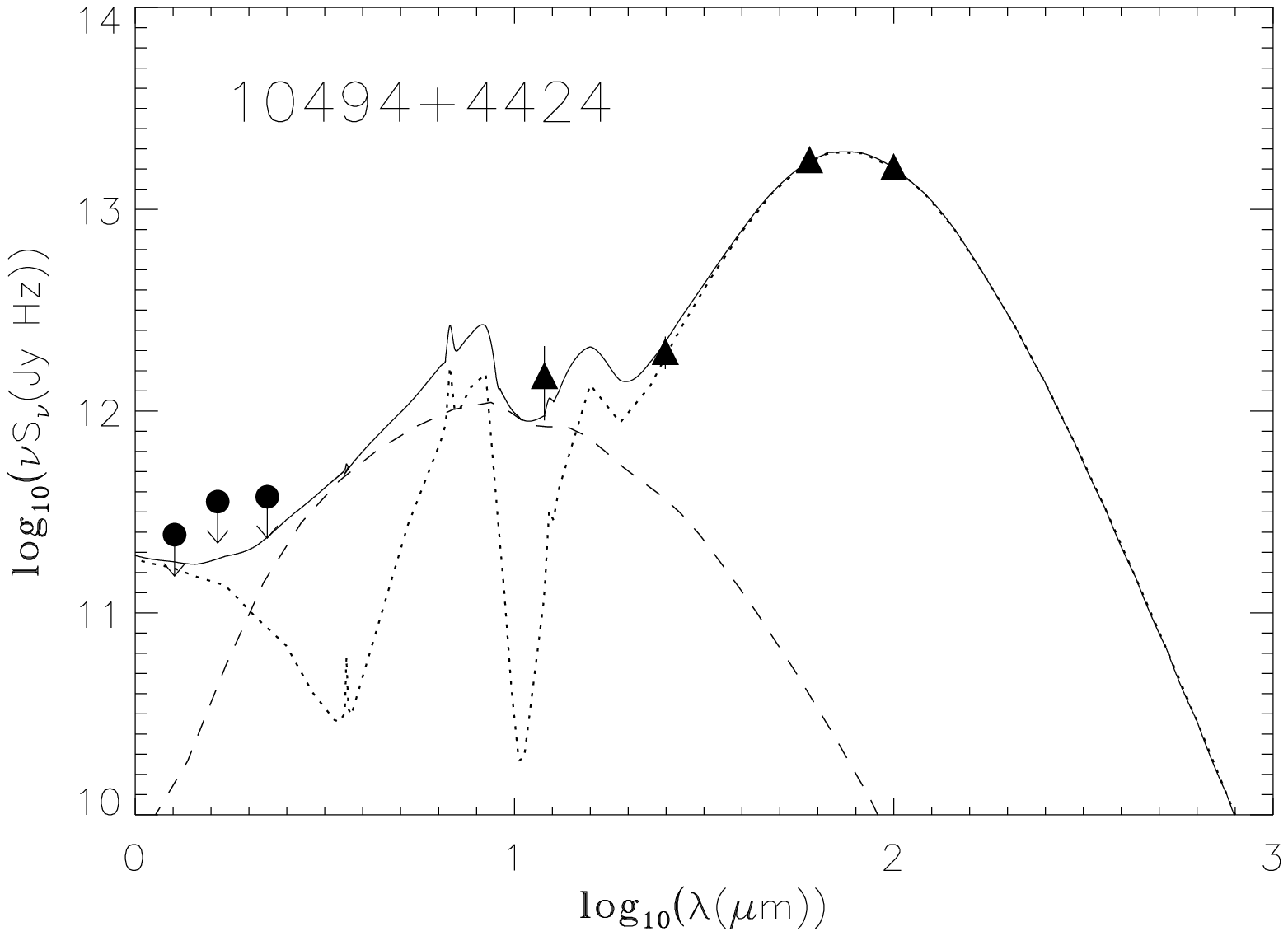,width=89mm}
\end{minipage}
\contcaption{}
\end{figure*}

\begin{figure*}
\begin{minipage}{180mm}
\epsfig{figure=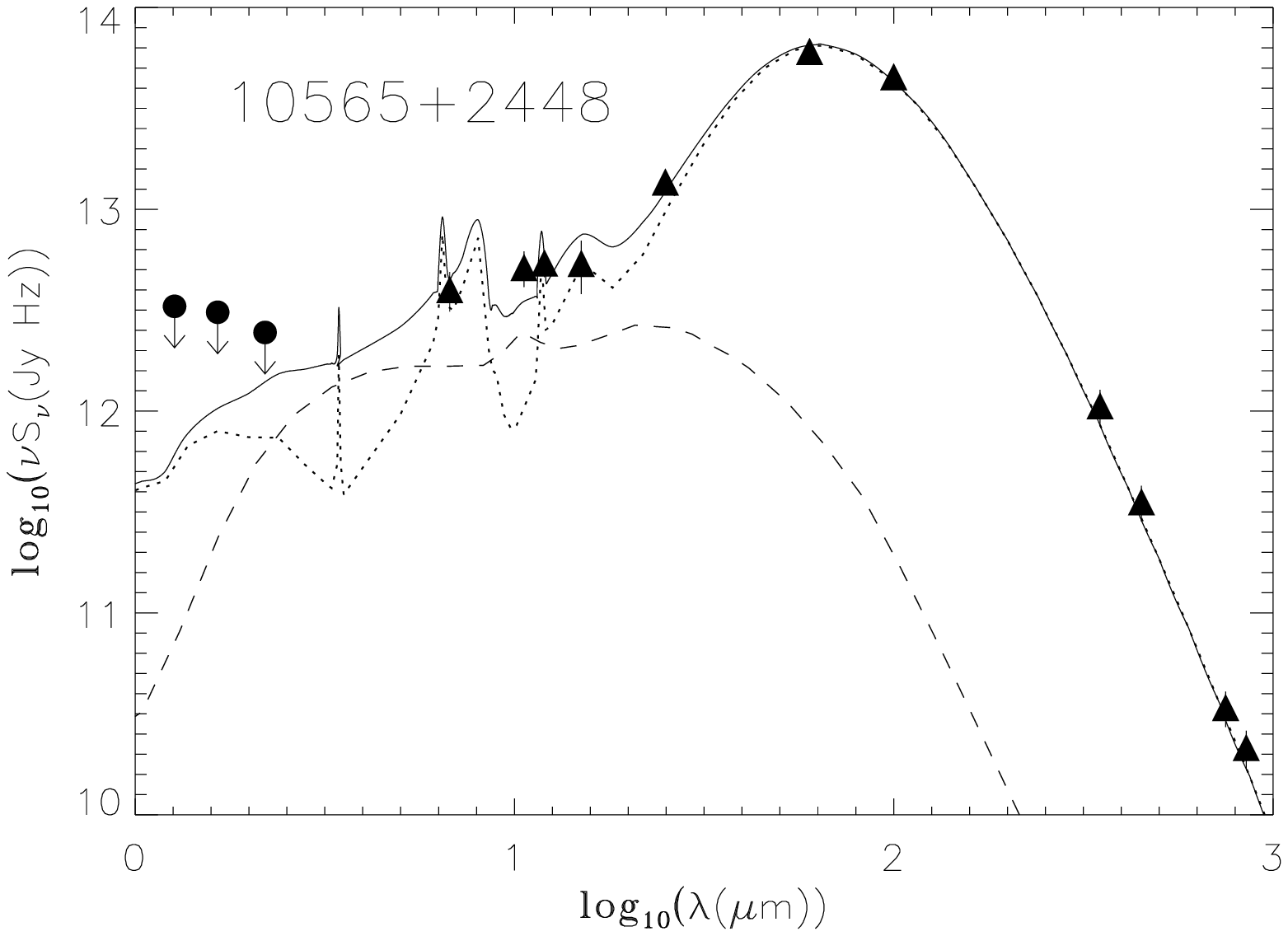,width=89mm}
\epsfig{figure=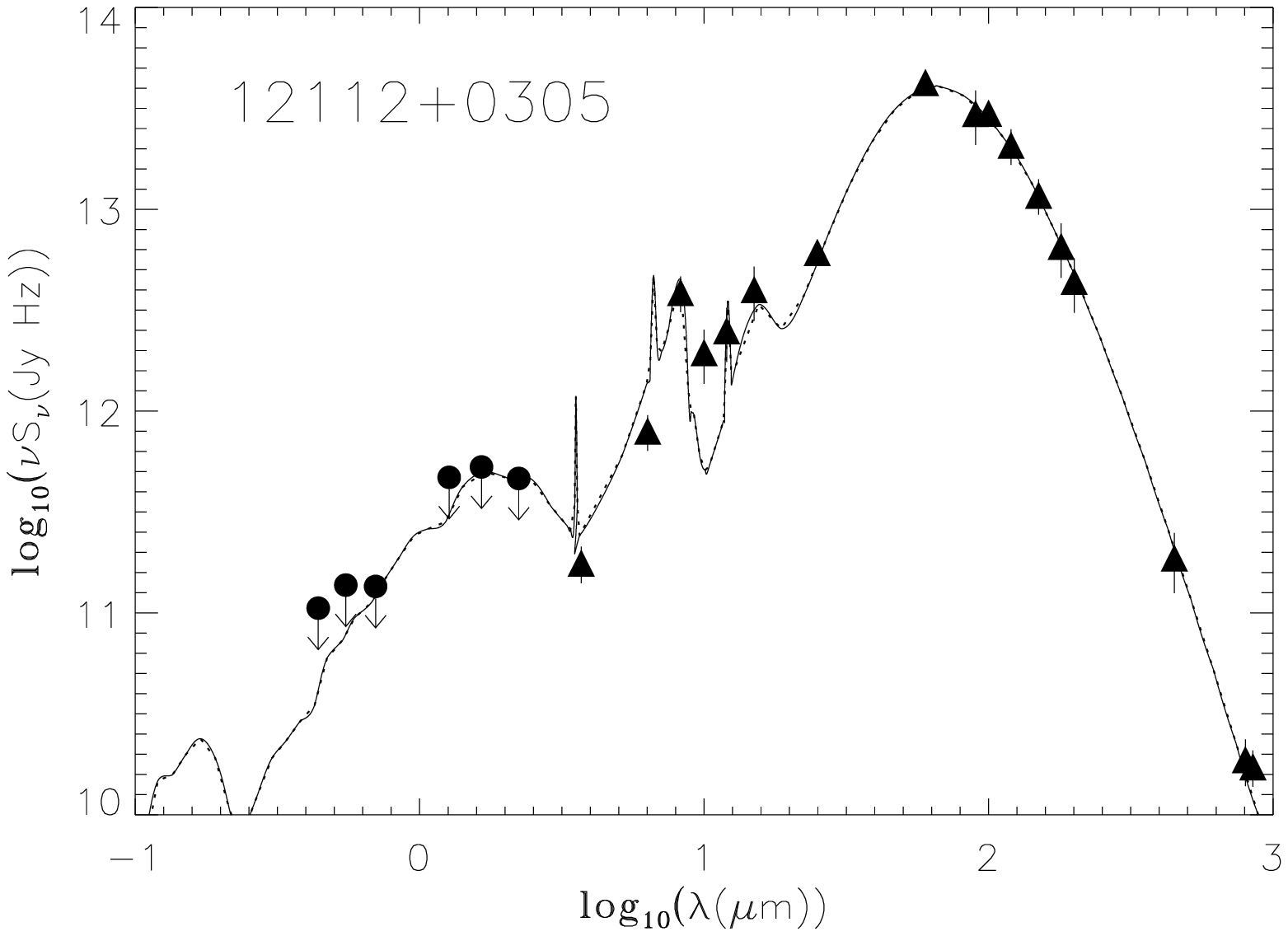,width=89mm}
\end{minipage}
\begin{minipage}{180mm}
\epsfig{figure=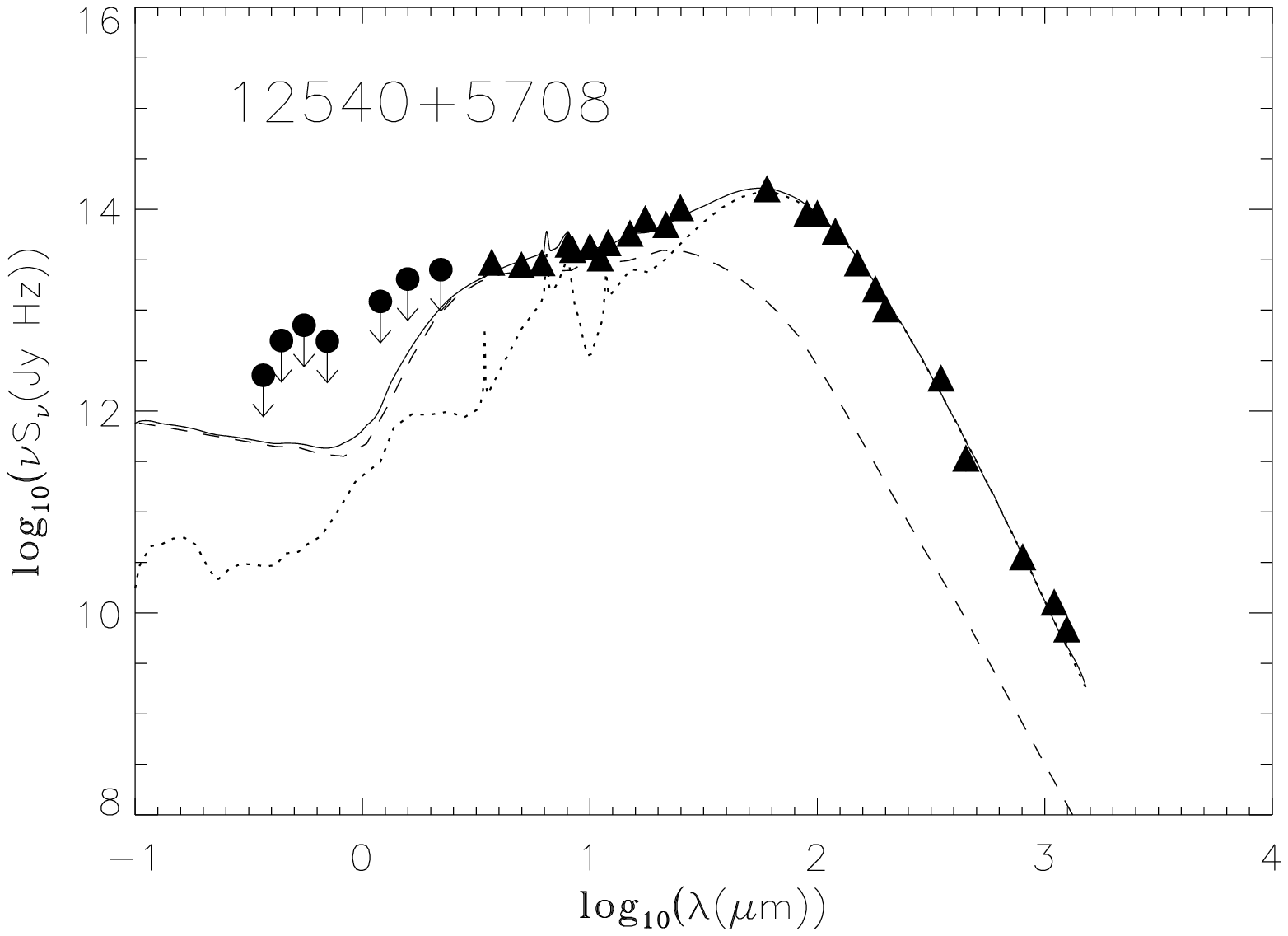,width=89mm}
\epsfig{figure=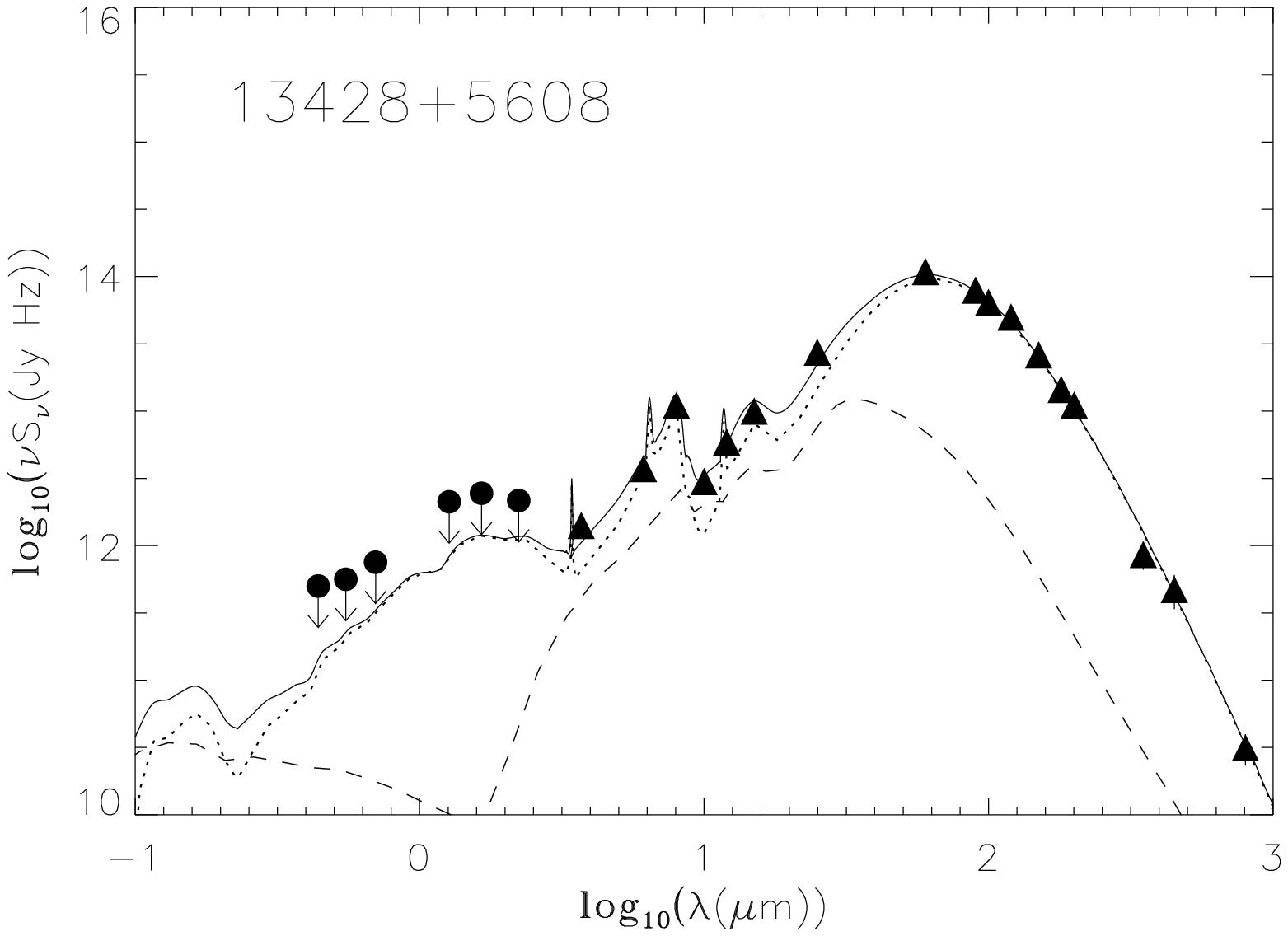,width=89mm}
\end{minipage}
\begin{minipage}{180mm}
\epsfig{figure=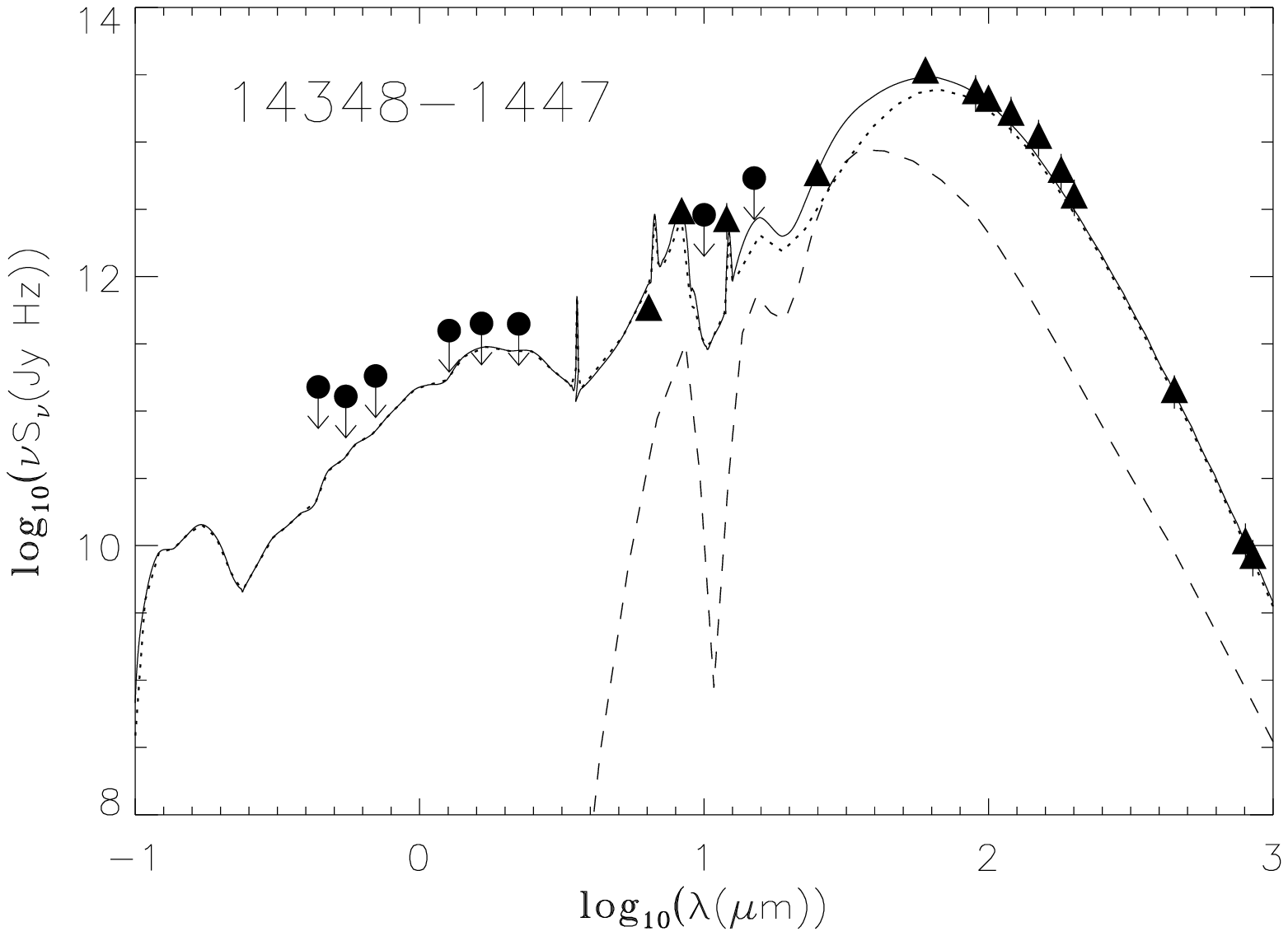,width=89mm}
\epsfig{figure=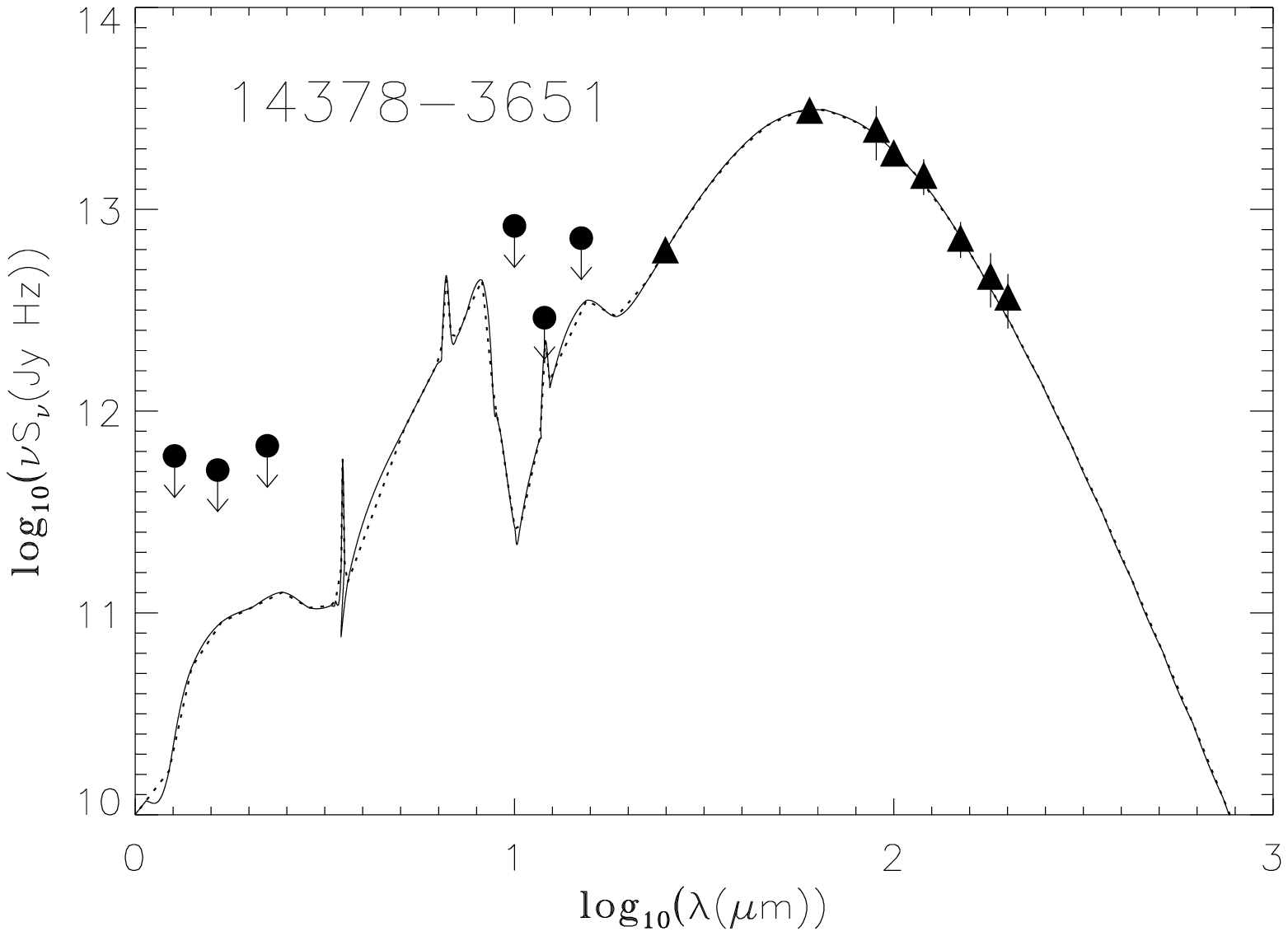,width=89mm}
\end{minipage}
\contcaption{}
\end{figure*}

\begin{figure*}
\begin{minipage}{180mm}
\epsfig{figure=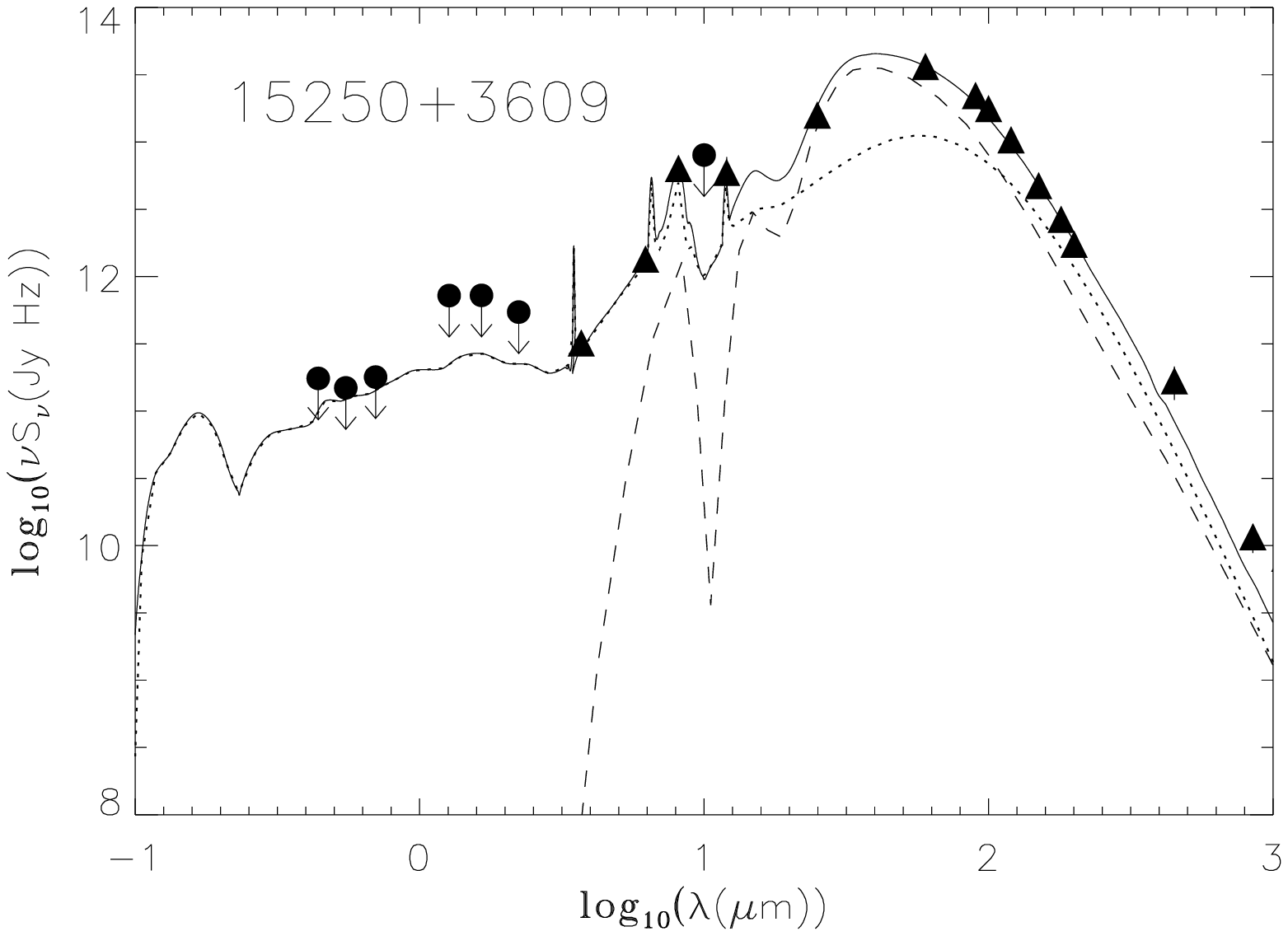,width=89mm}
\epsfig{figure=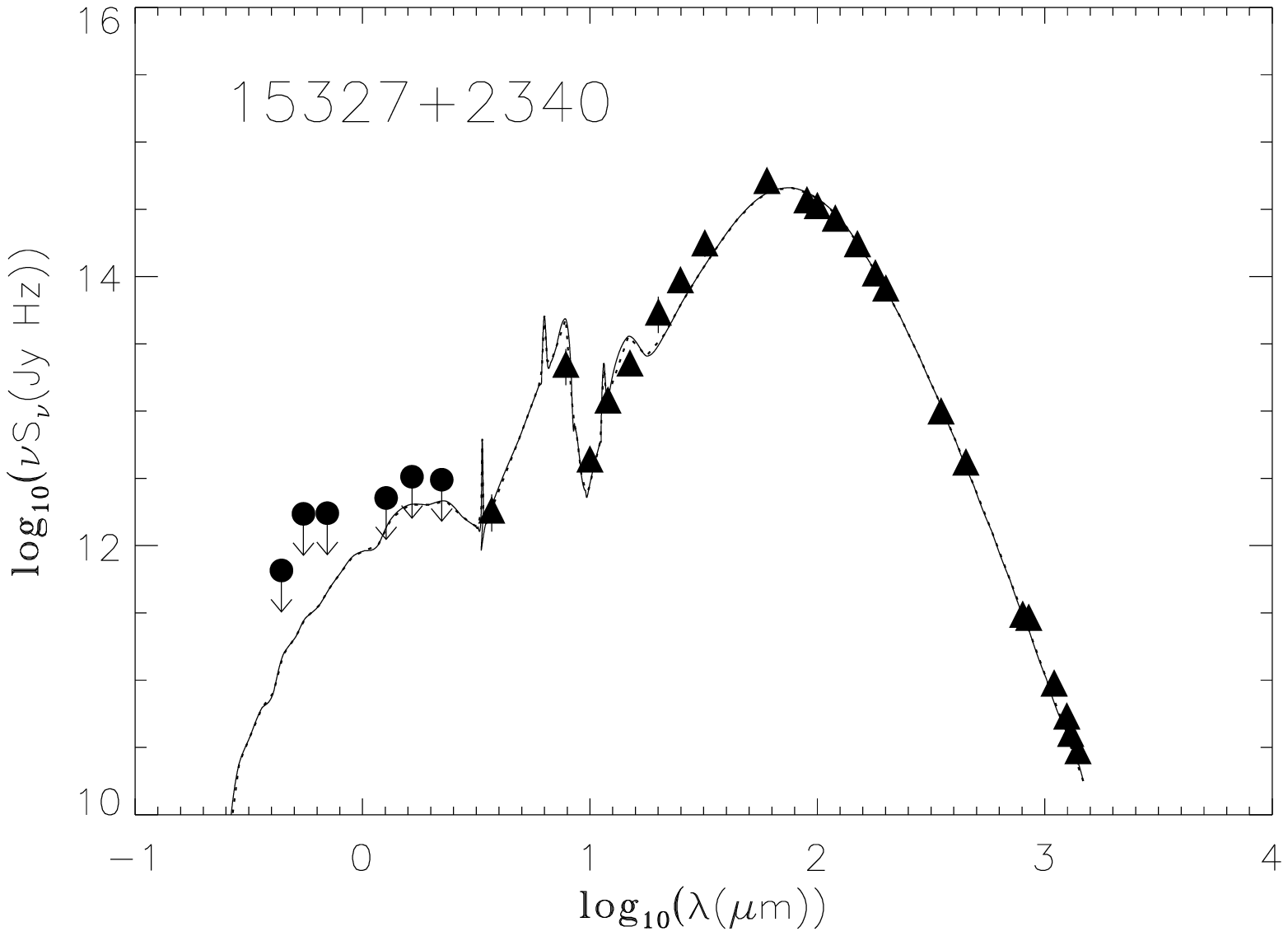,width=89mm}
\end{minipage}
\begin{minipage}{180mm}
\epsfig{figure=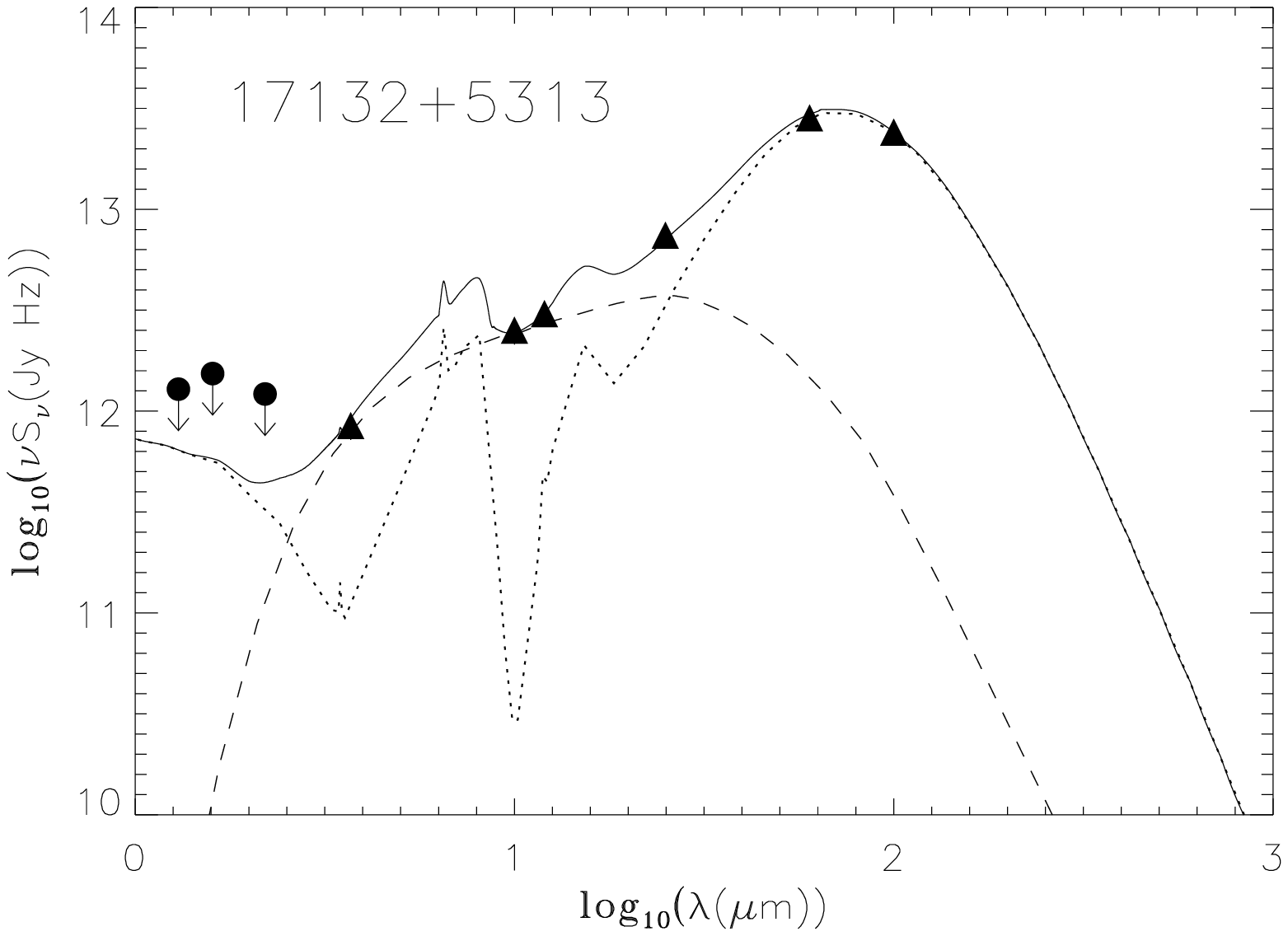,width=89mm}
\epsfig{figure=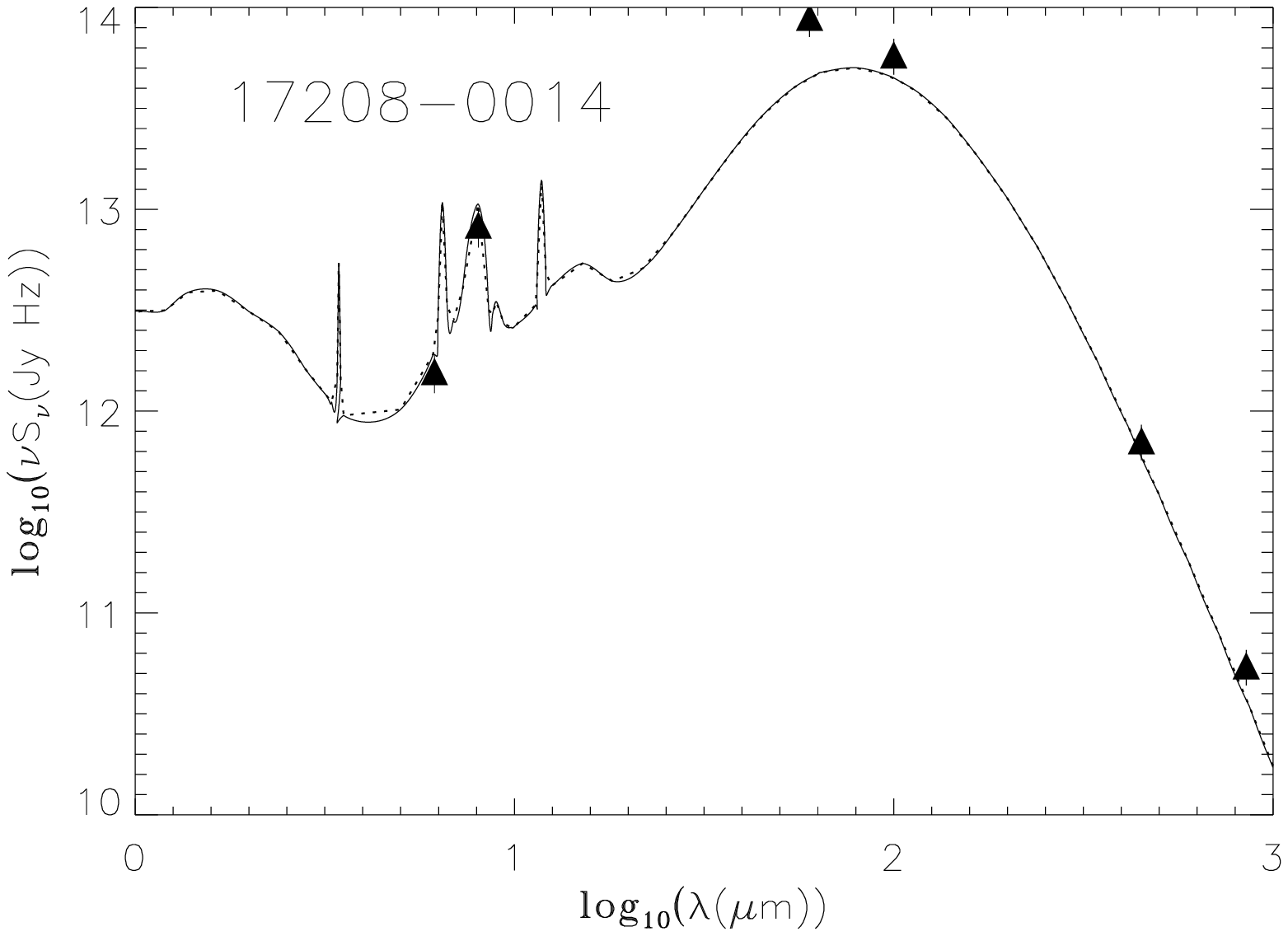,width=89mm}
\end{minipage}
\begin{minipage}{180mm}
\epsfig{figure=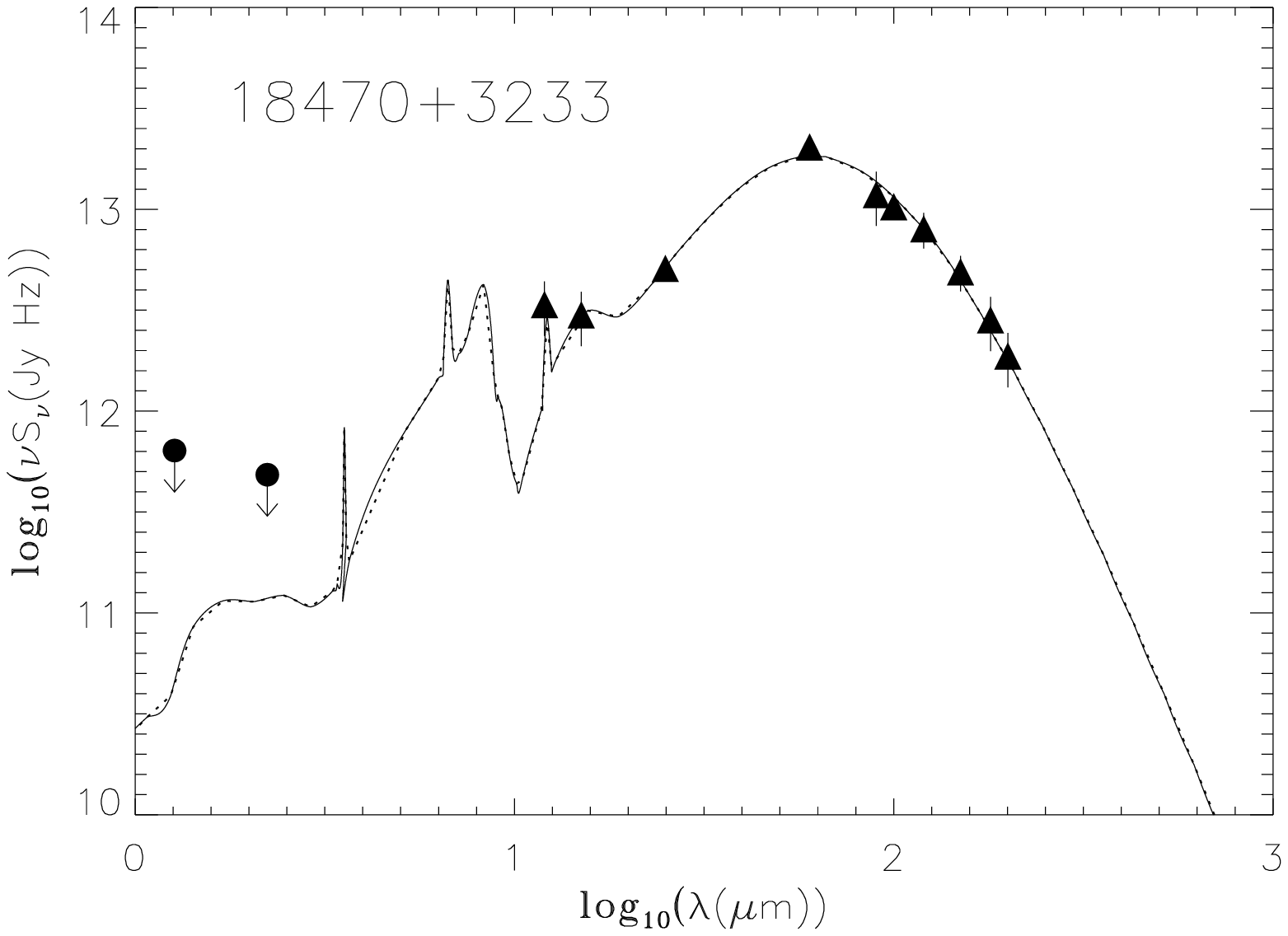,width=89mm}
\epsfig{figure=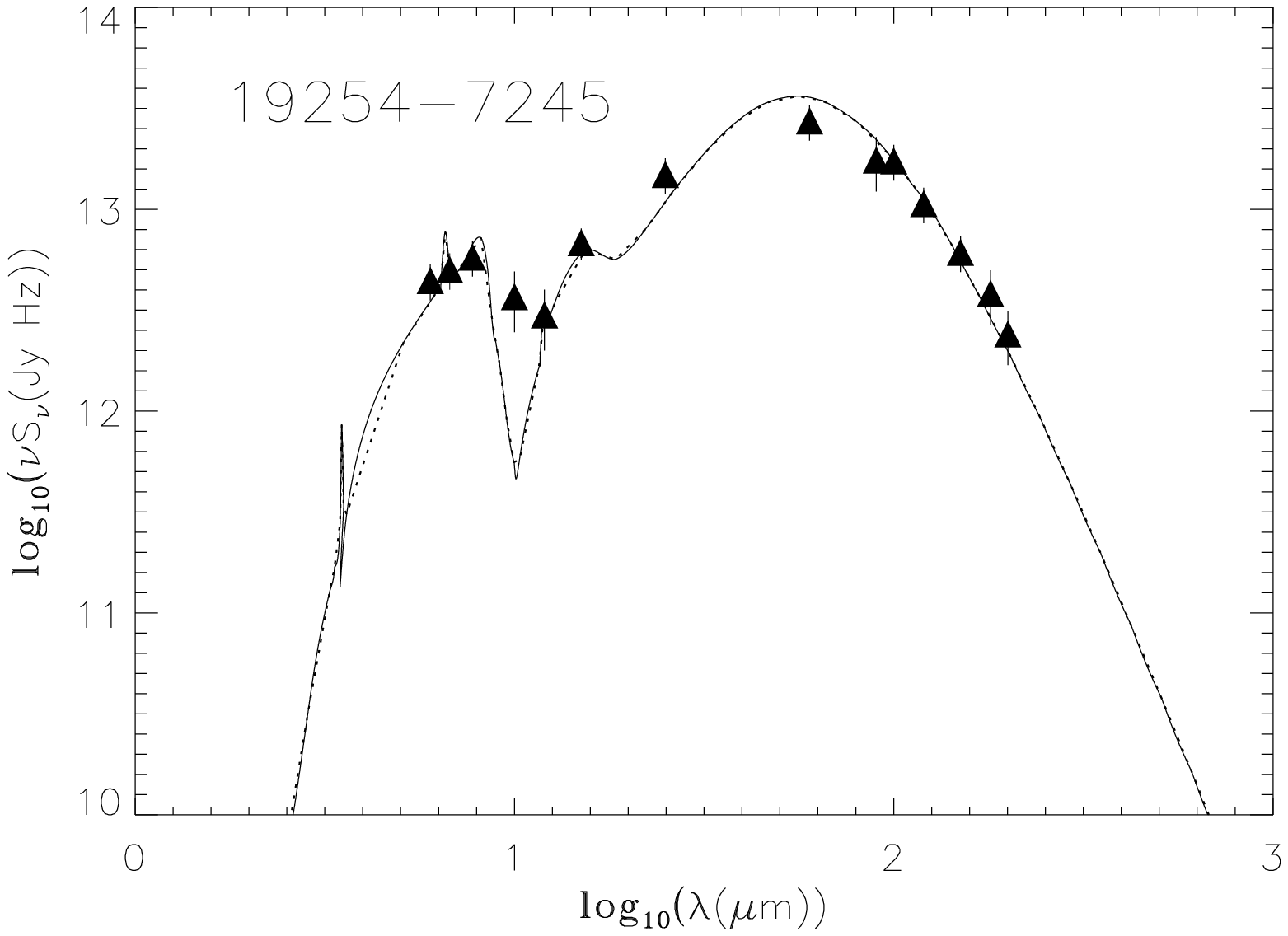,width=89mm}
\end{minipage}
\contcaption{}
\end{figure*}

\begin{figure*}
\begin{minipage}{180mm}
\epsfig{figure=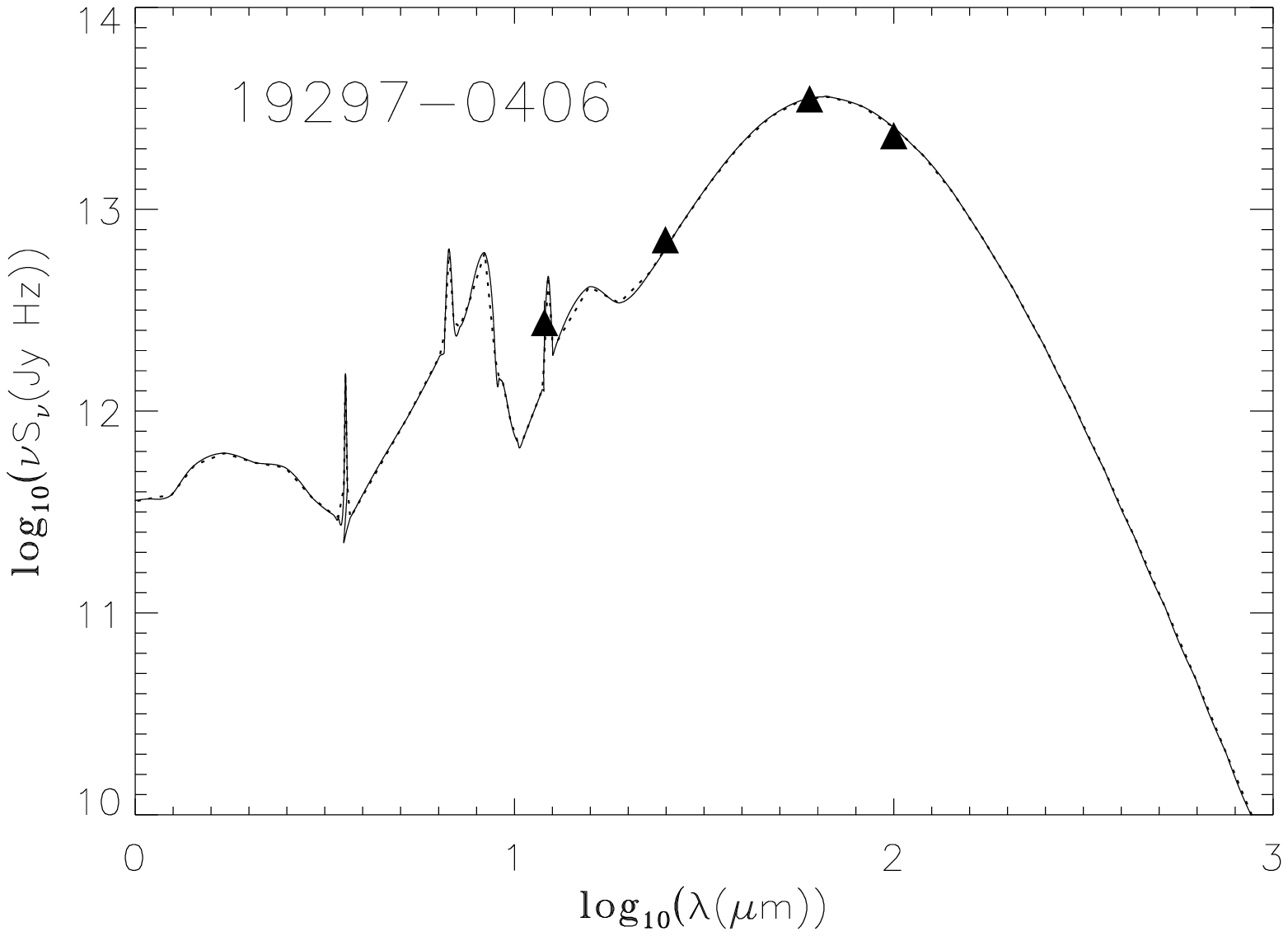,width=89mm}
\epsfig{figure=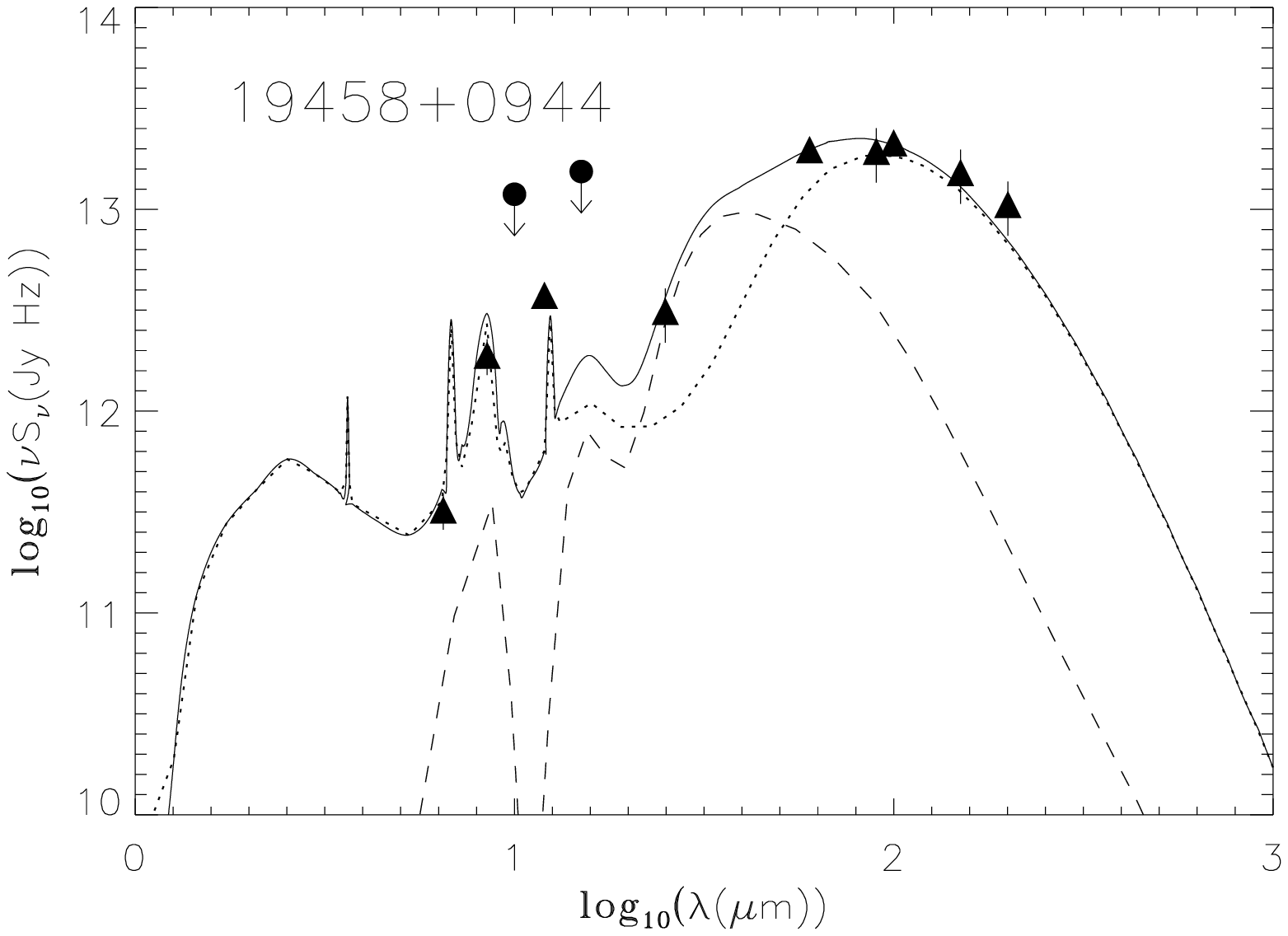,width=89mm}
\end{minipage}
\begin{minipage}{180mm}
\epsfig{figure=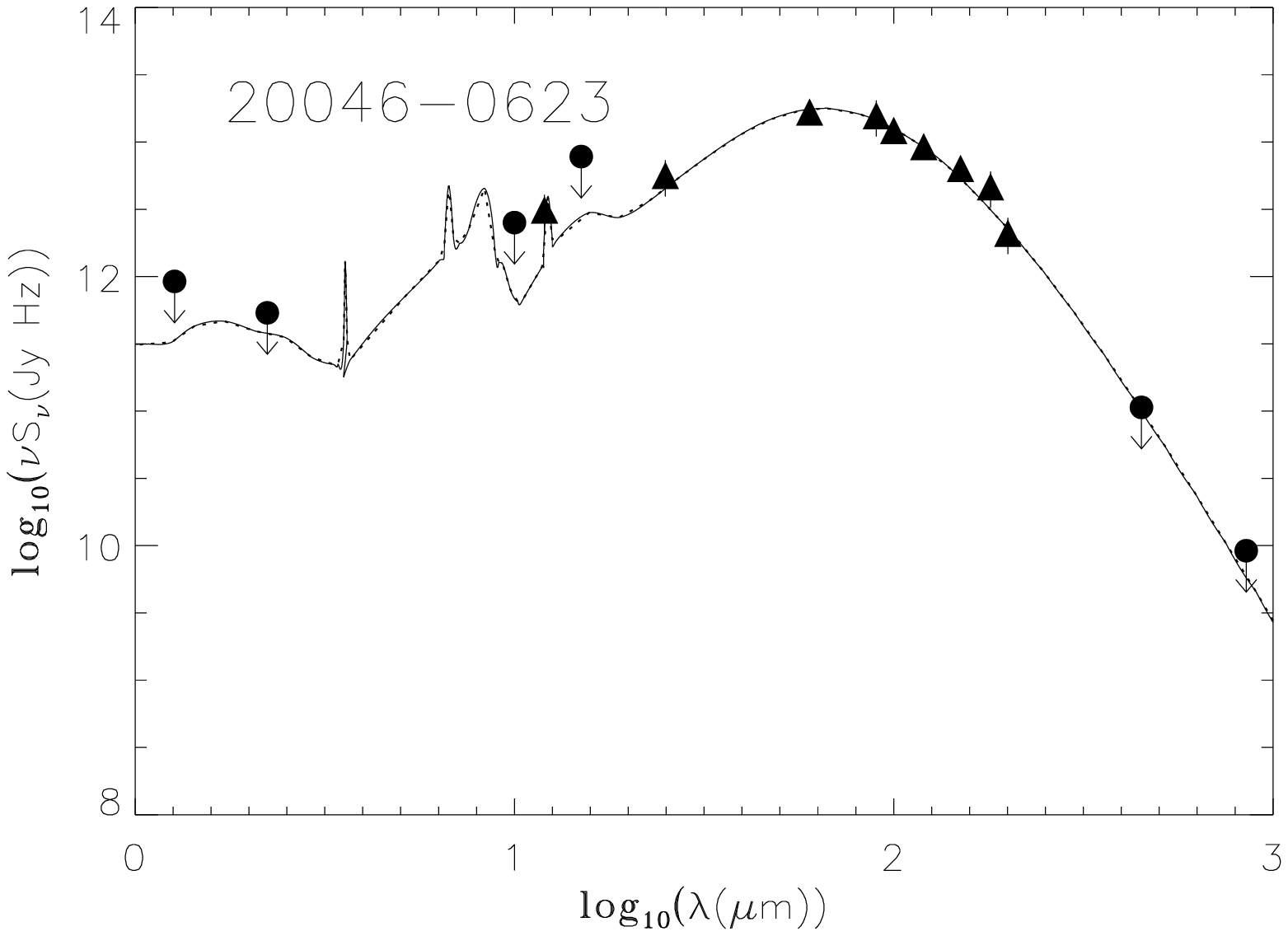,width=89mm}
\epsfig{figure=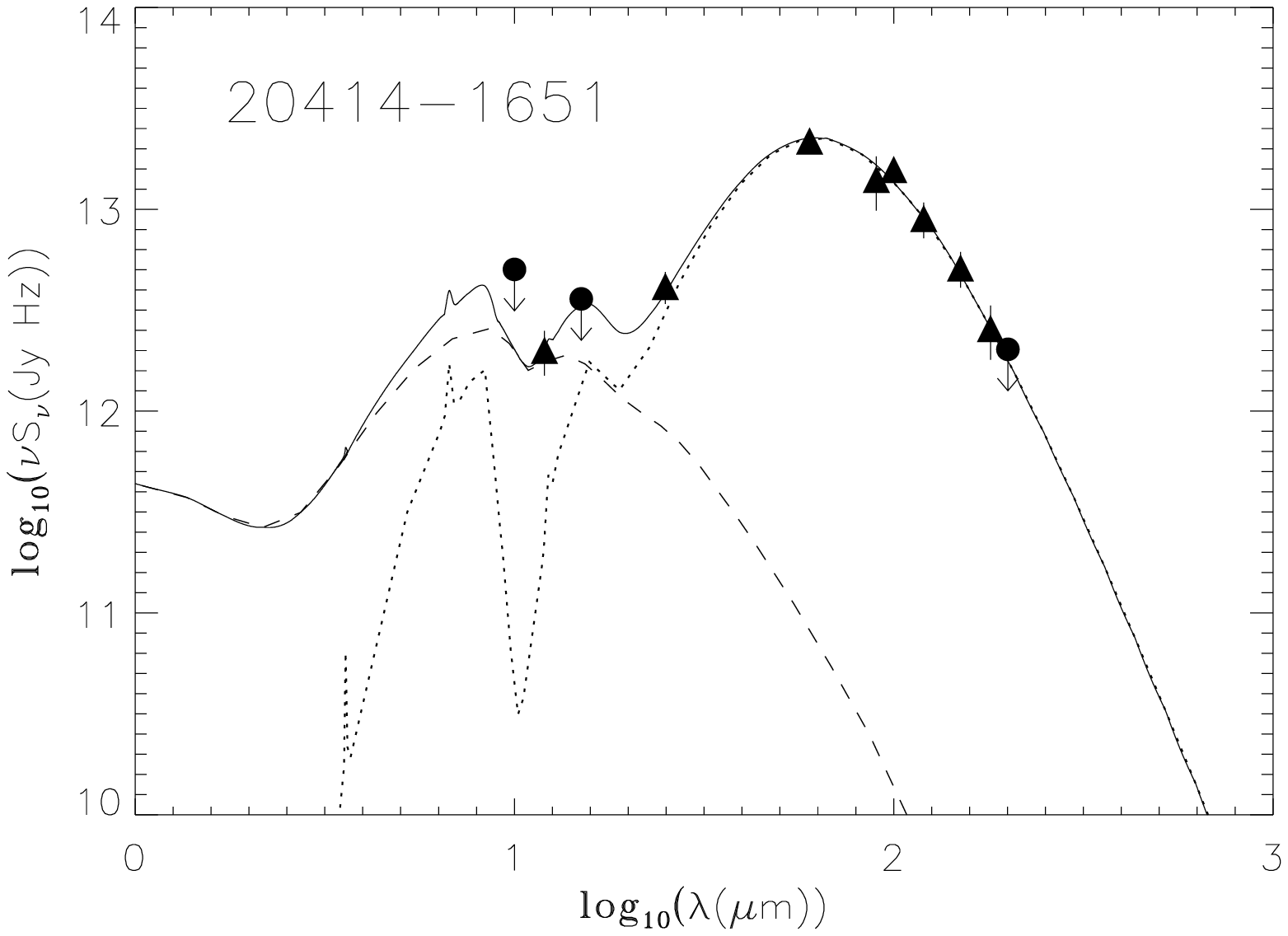,width=89mm}
\end{minipage}
\begin{minipage}{180mm}
\epsfig{figure=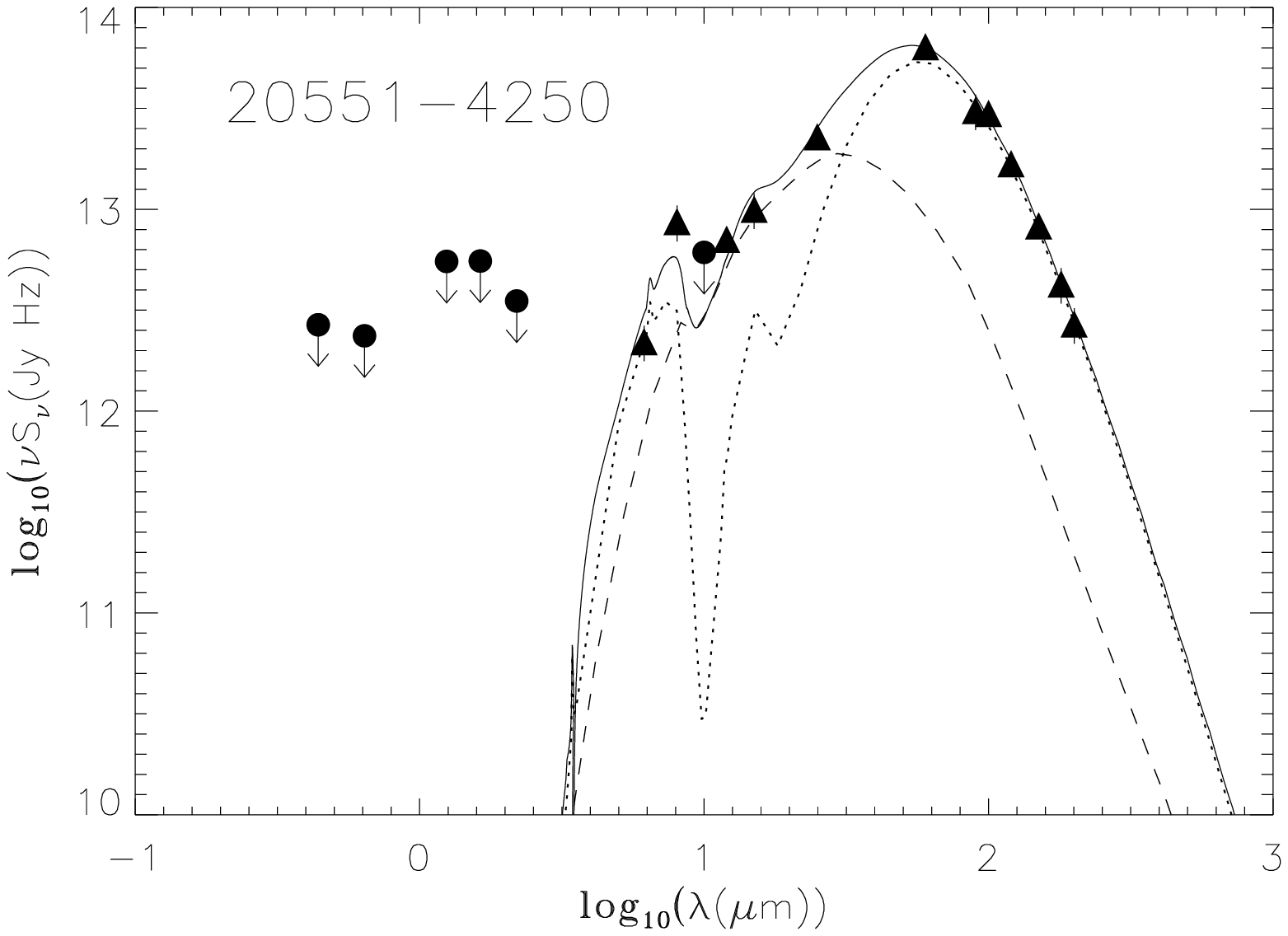,width=89mm}
\epsfig{figure=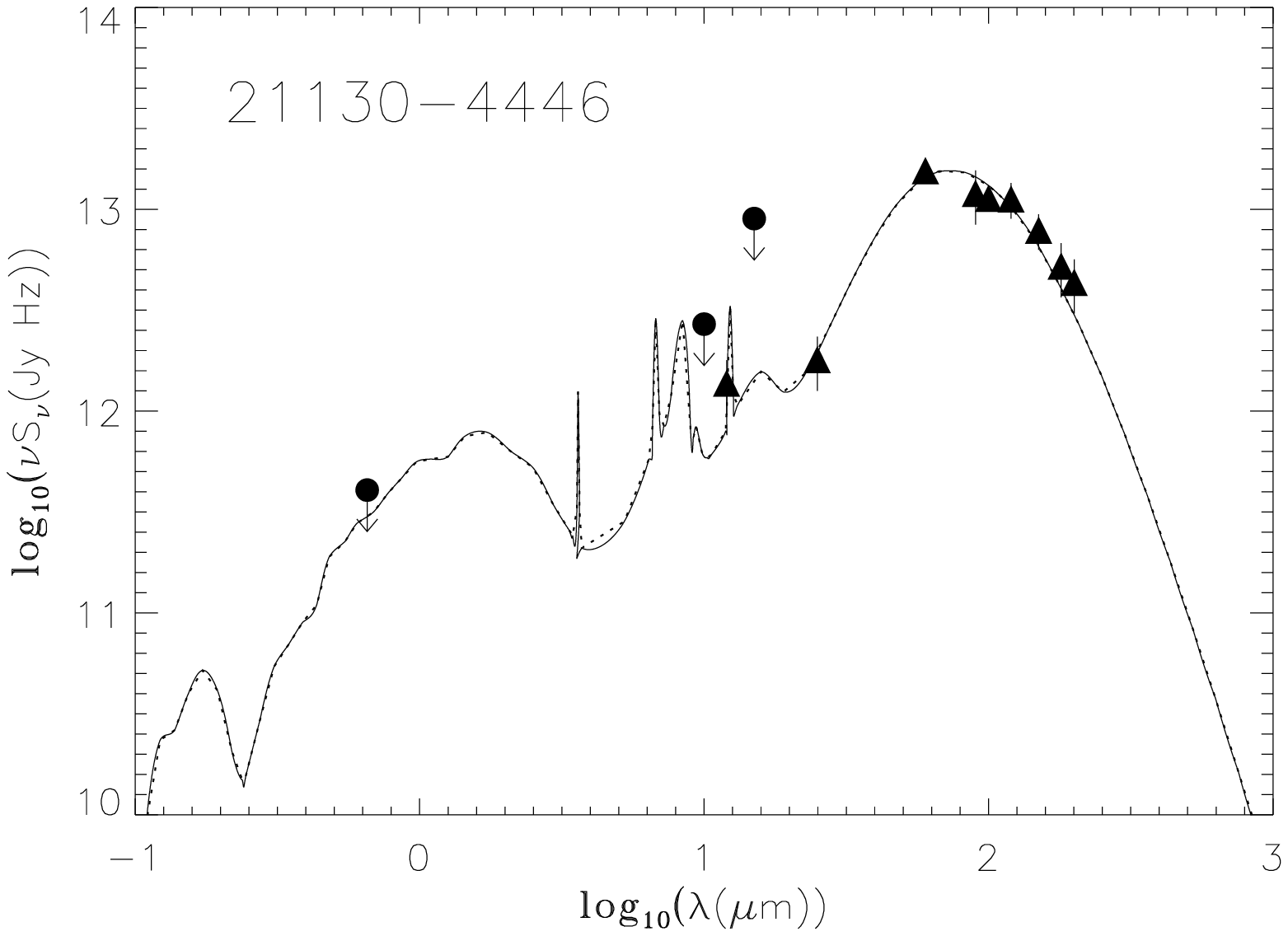,width=89mm}
\end{minipage}
\contcaption{}
\end{figure*}

\begin{figure*}
\begin{minipage}{180mm}
\epsfig{figure=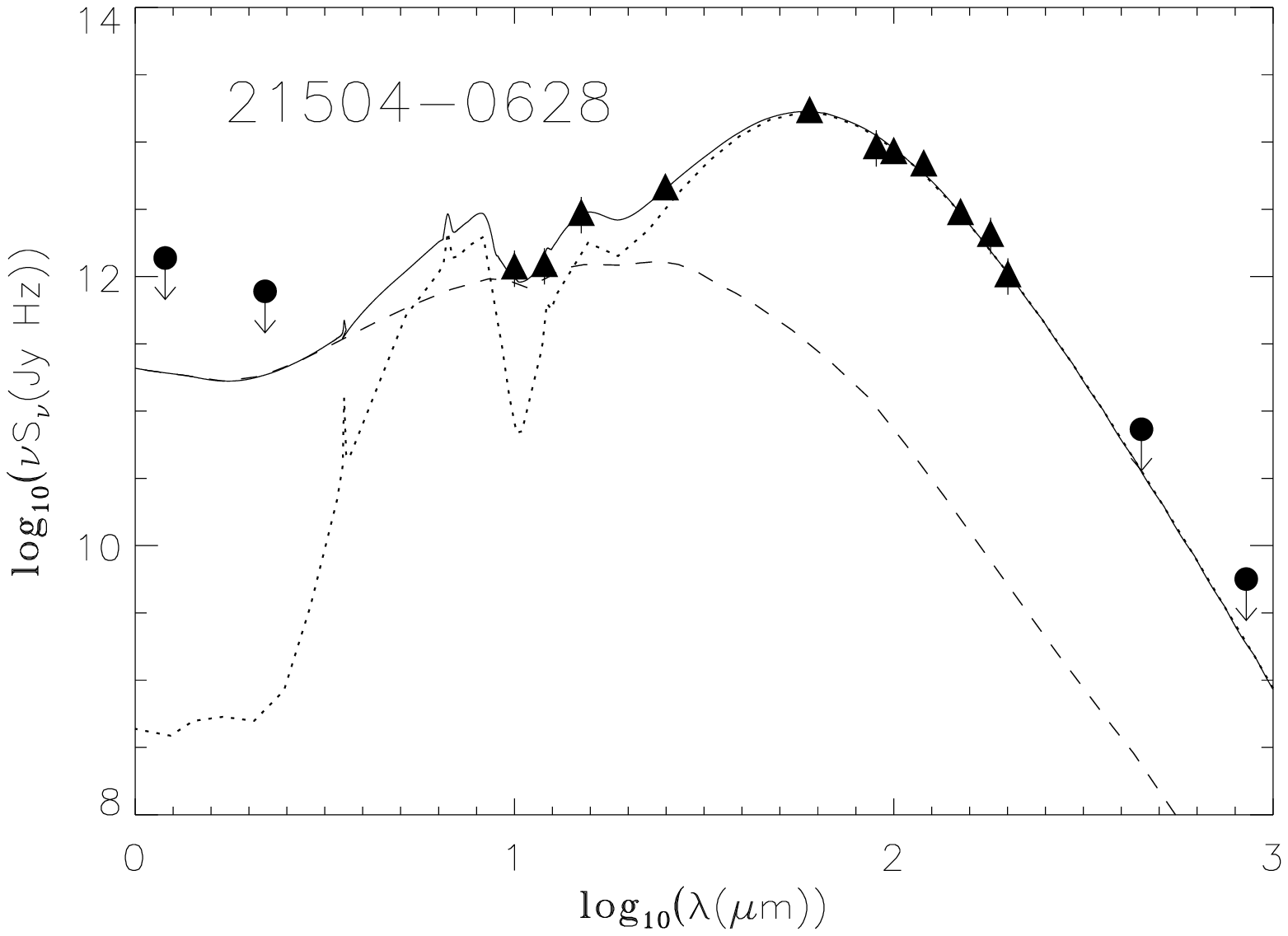,width=89mm}
\epsfig{figure=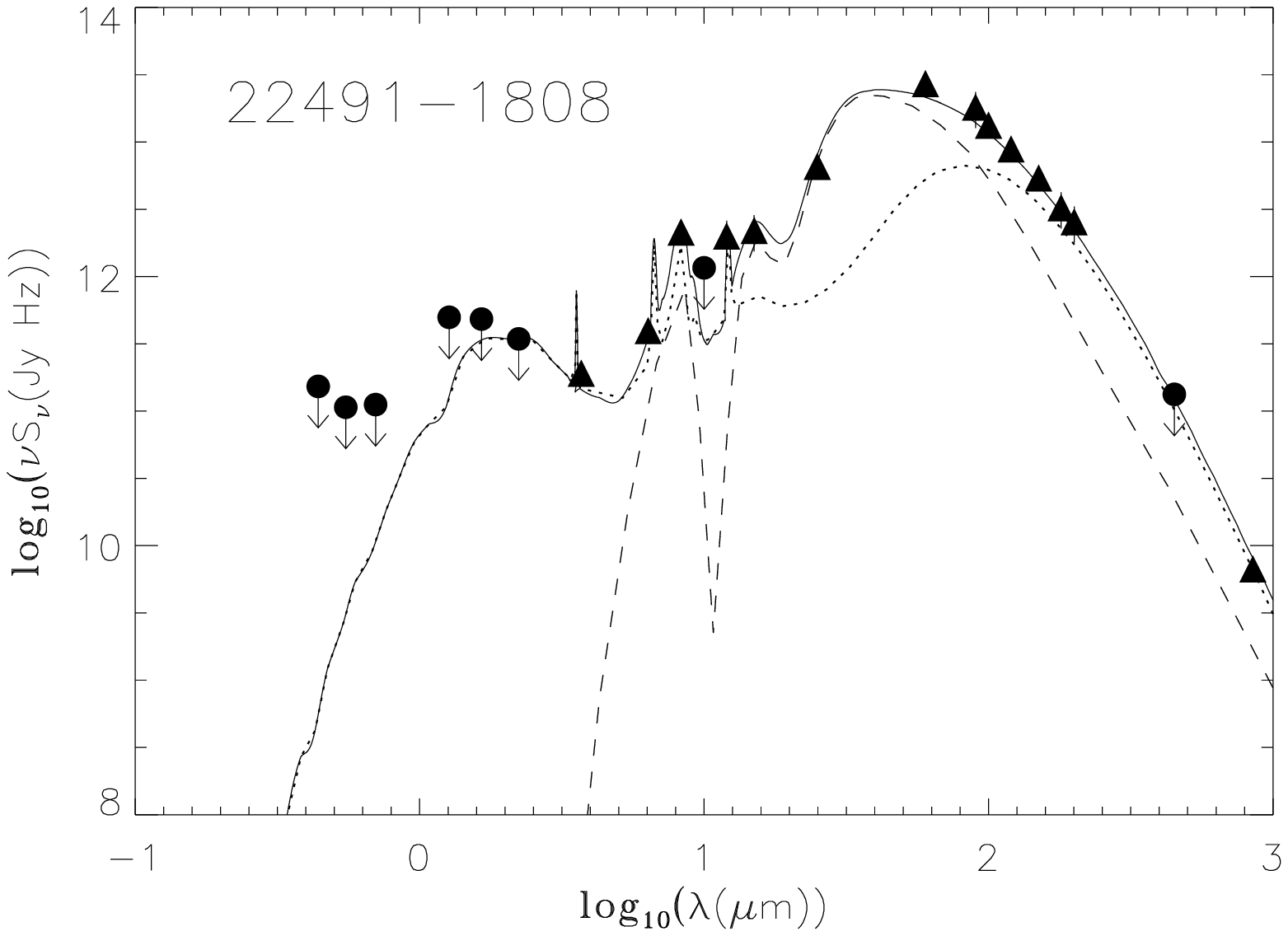,width=89mm}
\end{minipage}
\begin{minipage}{180mm}
\epsfig{figure=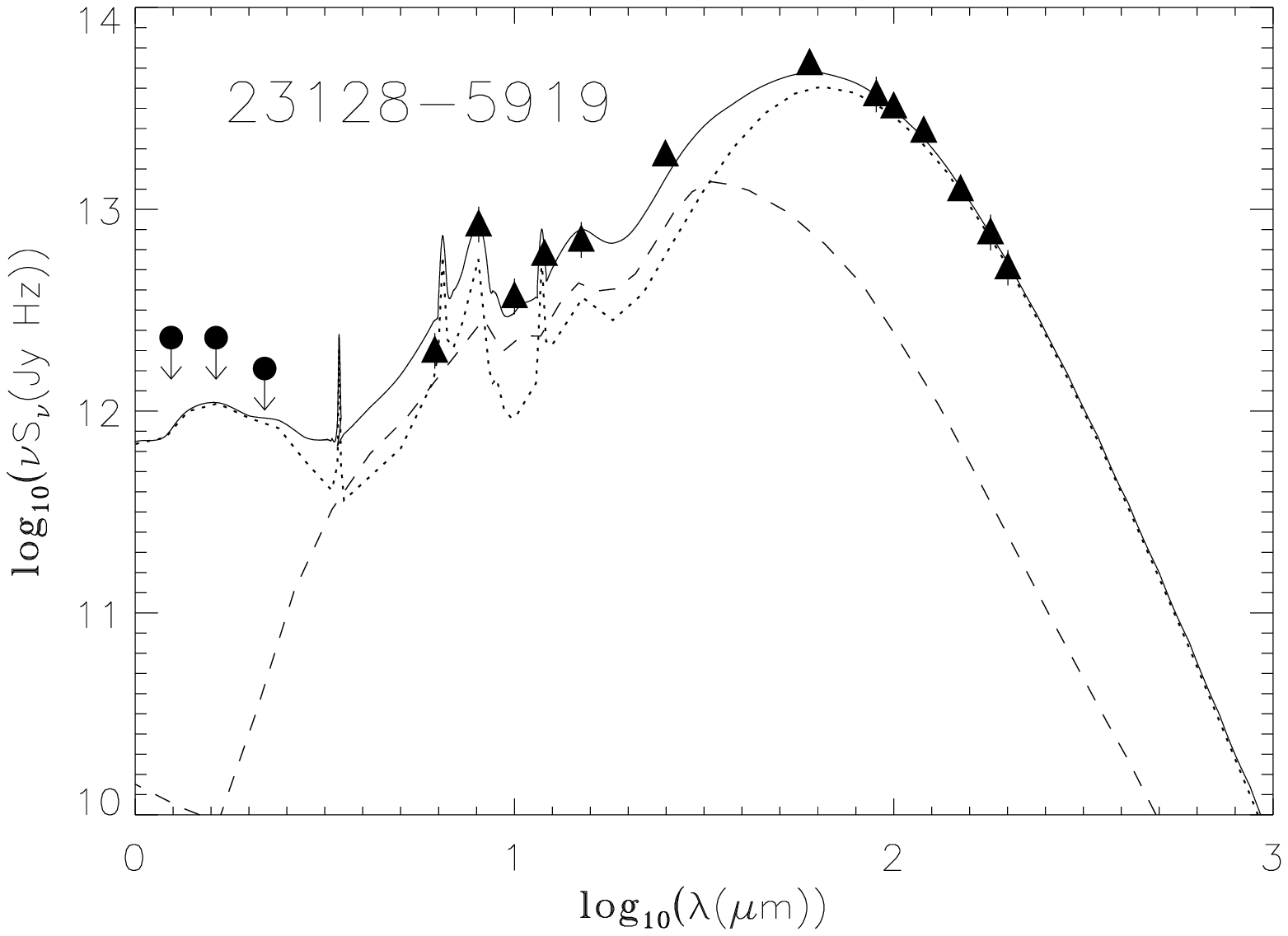,width=89mm}
\epsfig{figure=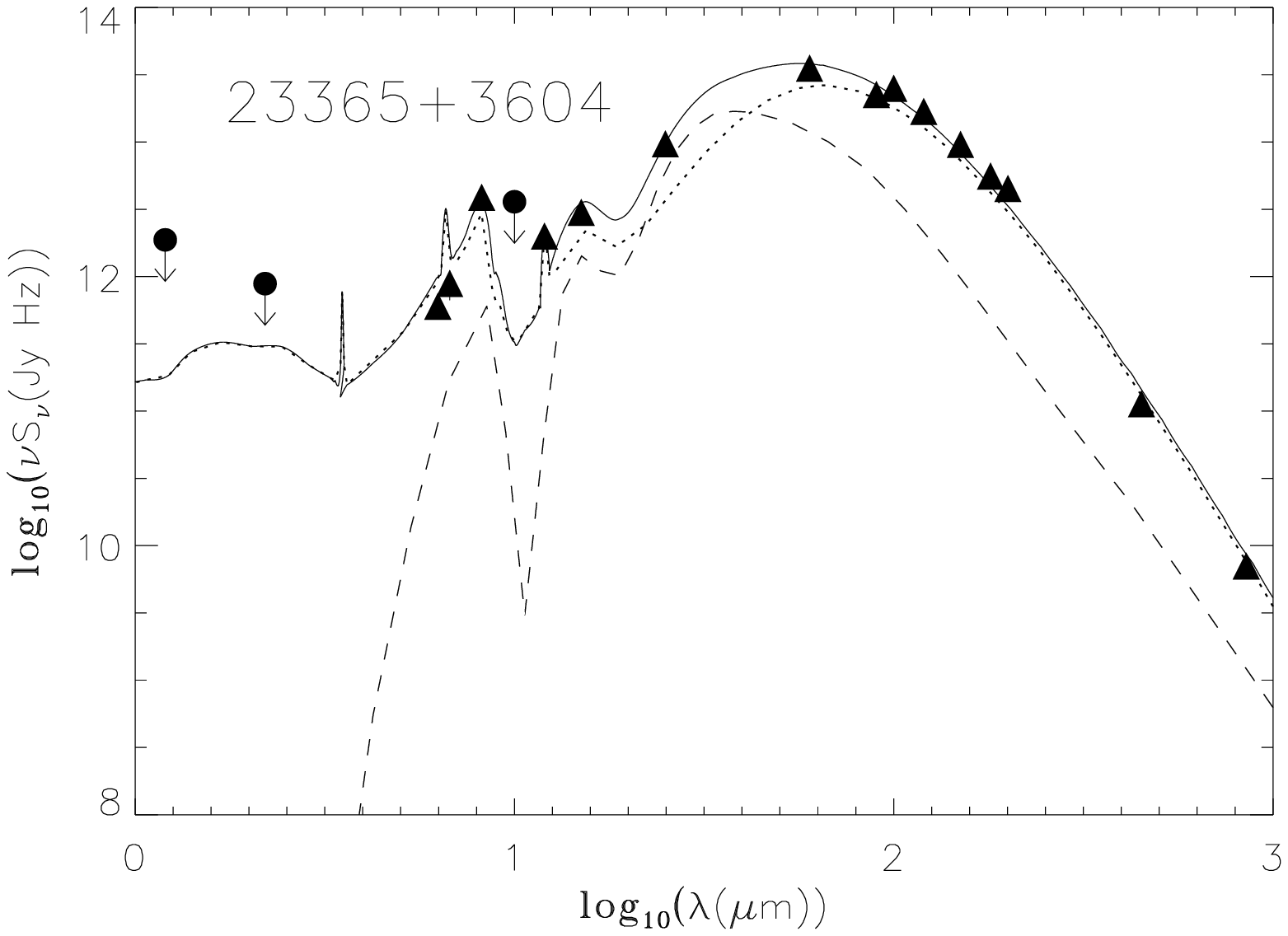,width=89mm}
\end{minipage}
\begin{minipage}{180mm}
\epsfig{figure=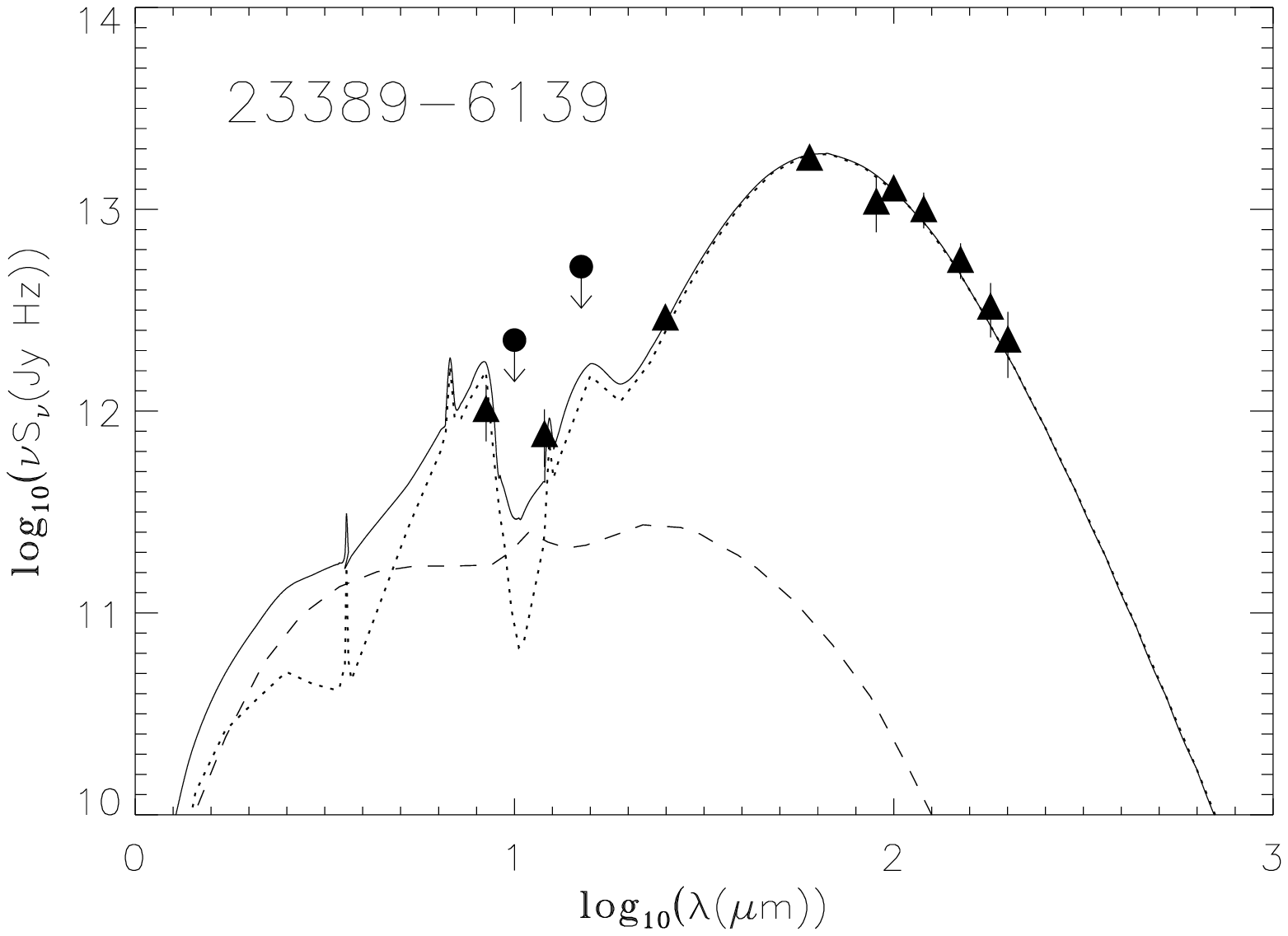,width=89mm}
\end{minipage}
\contcaption{}
\end{figure*}

\begin{table*}
\begin{minipage}{177mm}
\caption{Ultraluminous Galaxies: Luminosities and Model Parameters \label{ulirgparams}}
\begin{tabular}{@{}lcccccccccc}
\hline
Name       &$\chi^{2,a}$& $L^{Tot}_{IR}$        & $L^{Sb}_{IR}$         & $L^{AGN}_{IR}$        & Age$^{b}$ & SFR$^{c}$            & $\theta^{d}$ & $M_{d}^{e}$          & T$^{f}$ & $\beta^{g}$\\ 
           &          & $L_{\odot}$           & $L_{\odot}$           & $L_{\odot}$           & Myr       & $M_{\odot}$yr$^{-1}$ & \degr        & $M_{\odot}$          & kelvin  &            \\
\hline
00198-7926 & 3.50    &12.72$^{+0.01}_{-0.02}$&12.65$^{+0.05}_{-0.03}$& $<11.91$              & $64-71$   & $130\pm50$           & --           &7.31$^{+0.15}_{-0.10}$& $36\pm4$&$1.95\pm0.22$ \\ 
00199-7426 & 2.35    &12.36$^{+0.02}_{-0.02}$&12.30$^{+0.04}_{-0.01}$&11.34$^{+0.11}_{-0.17}$& $26-37$   & $360\pm50$           & --           &8.48$^{+0.09}_{-0.12}$& $27\pm7$&$1.96\pm0.40$ \\ 
00262+4251 & 1.66    &12.10$^{+0.01}_{-0.01}$&11.64$^{+0.08}_{-0.01}$&11.92$^{+0.01}_{-0.06}$& $26-45$   & $50\pm30$            & $<15$        &8.42$^{+0.20}_{-0.50}$& $25\pm5$&$1.87\pm0.05$ \\ 
00335-2732 & 1.33    &12.01$^{+0.04}_{-0.04}$&11.99$^{+0.03}_{-0.04}$& $<10.95$              & $<57$     & $170\pm20$           & --           &7.07$^{+0.09}_{-0.02}$& $42\pm3$&$1.97\pm0.20$ \\ 
01388-4618 & 2.08    &12.11$^{+0.03}_{-0.04}$&12.03$^{+0.03}_{-0.04}$& $<11.35$              & $>26$     & $200\pm20$           & --           &8.10$^{+0.09}_{-0.17}$& $28\pm4$&$1.92\pm0.10$ \\ 
02364-4751 & 1.93    &12.20$^{+0.01}_{-0.01}$&12.15$^{+0.01}_{-0.02}$&11.23$^{+0.07}_{-0.10}$& $>10$     & $260\pm30$           & --           &7.97$^{+0.02}_{-0.03}$& $32\pm7$&$1.91\pm0.20$ \\ 
03068-5346 & 1.56    &11.96$^{+0.01}_{-0.02}$&11.93$^{+0.04}_{-0.03}$& $<10.94$              & $1.6-6.6$ & $180\pm40$           & --           &7.62$^{+0.09}_{-0.12}$& $35\pm3$&$1.97\pm0.20$ \\
04232+1436 & 1.41    &12.07$^{+0.03}_{-0.04}$&11.95$^{+0.10}_{-0.11}$& $<11.70$              & $<37$     & $210\pm30$           & --           &7.87$^{+0.11}_{-0.30}$& $31\pm4$&$1.90\pm0.25$ \\ 
05189-2524 & 2.20    &12.16$^{+0.04}_{-0.02}$&11.99$^{+0.06}_{-0.06}$&11.67$^{+0.07}_{-0.12}$& $6.6-26$  & $170\pm30$           & $25-54$      &7.61$^{+0.28}_{-0.20}$& $33\pm5$&$1.94\pm0.06$ \\ 
06035-7102 & 1.00    &12.24$^{+0.01}_{-0.02}$&12.19$^{+0.01}_{-0.05}$&11.21$^{+0.20}_{-0.05}$& $16-26$   & $290\pm30$           & $>25$        &7.74$^{+0.04}_{-0.05}$& $35\pm2$&$1.95\pm0.03$ \\ 
06206-6315 & 1.00    &12.21$^{+0.01}_{-0.01}$&12.18$^{+0.01}_{-0.06}$&11.05$^{+0.40}_{-0.23}$& $16-26$   & $300\pm30$           & $<45$        &7.93$^{+0.02}_{-0.04}$& $31\pm2$&$1.92\pm0.02$ \\ 
08572+3915 & 2.22    &12.17$^{+0.01}_{-0.01}$&11.99$^{+0.01}_{-0.06}$&11.70$^{+0.09}_{-0.02}$& $<57$     & $180\pm20$           & --           &7.04$^{+0.20}_{-0.06}$& $45\pm5$&$1.92\pm0.10$ \\ 
09111-1007 & 1.92    &12.01$^{+0.02}_{-0.02}$&12.01$^{+0.02}_{-0.03}$& $<10.64$              & $26-37$   & $192\pm25$           & --           &7.90$^{+0.11}_{-0.16}$&$29\pm4$ &$1.94\pm0.28$ \\ 
09320+6134 & 2.11    &12.01$^{+0.01}_{-0.01}$&11.98$^{+0.01}_{-0.01}$&10.86$^{+0.03}_{-0.10}$& $>16$     & $160\pm20$           & $>25$        &8.33$^{+0.04}_{-0.11}$&$25\pm3$ &$1.93\pm0.03$ \\   
09583+4714 & 1.00    &12.06$^{+0.05}_{-0.06}$&11.90$^{+0.12}_{-0.10}$& $<11.80$              & --        & $160\pm70$           & --           & --                   &$34\pm15$& $1.95\pm0.70$ \\  
10035+4852 & 1.00    &11.98$^{+0.04}_{-0.05}$&11.97$^{+0.05}_{-0.07}$& $<11.50$              & --        & $180\pm30$           & --           &7.87$^{+0.10}_{-0.15}$&$29\pm5$ & $1.95\pm0.20$ \\ 
10190+1322 & 1.00    &12.10$^{+0.03}_{-0.02}$&12.01$^{+0.03}_{-0.03}$&11.36$^{+0.08}_{-0.13}$& $<16$     & $195\pm15$           & --           &7.85$^{+0.05}_{-0.05}$&$32\pm3$ &$1.97\pm0.15$  \\ 
10494+4424 & 2.04    &12.20$^{+0.02}_{-0.02}$&12.16$^{+0.02}_{-0.02}$& $<11.06$              & $<26$     & $280\pm40$           & --           &7.97$^{+0.05}_{-0.05}$&$32\pm5$ & $1.94\pm0.26$ \\ 
10565+2448 & 1.23    &12.06$^{+0.02}_{-0.02}$&12.03$^{+0.03}_{-0.02}$&10.80$^{+0.14}_{-0.10}$& $26-37$   & $200\pm20$           & --           &7.85$^{+0.05}_{-0.07}$&$30\pm2$ & $1.95\pm0.20$ \\ 
12112+0305 & 2.30    &12.29$^{+0.02}_{-0.02}$&12.29$^{+0.02}_{-0.03}$& $<9.50$               & $16-26$   & $390\pm20$           & --           &7.94$^{+0.08}_{-0.13}$&$33\pm2$ & $1.95\pm0.10$ \\ 
12540+5708 & 1.56    &12.58$^{+0.04}_{-0.03}$&12.42$^{+0.06}_{-0.06}$&12.05$^{+0.10}_{-0.01}$& $10-16$   & $470\pm50$           & $27-54$      &7.95$^{+0.02}_{-0.17}$&$36\pm2$ & $1.94\pm0.01$ \\ 
13428+5608 & 1.99    &12.15$^{+0.04}_{-0.03}$&12.09$^{+0.06}_{-0.03}$&11.21$^{+0.05}_{-0.50}$& $16-26$   & $260\pm30$           & --           &7.70$^{+0.17}_{-0.02}$&$31\pm4$ & $1.95\pm0.08$ \\ 
14348-1447 & 1.98    &12.30$^{+0.01}_{-0.03}$&12.18$^{+0.04}_{-0.04}$&11.54$^{+0.12}_{-0.25}$& $26-37$   & $310\pm30$           & $<32$        &7.97$^{+0.02}_{-0.02}$&$32\pm1$ & $1.91\pm0.02$ \\
14378-3651 & 1.00    &12.12$^{+0.03}_{-0.02}$&12.03$^{+0.10}_{-0.01}$& $<11.36$              & $16-26$   & $250\pm40$           & --           &7.81$^{+0.05}_{-0.20}$&$34\pm5$ &$1.95\pm0.3$   \\ 
15250+3609 & 2.32    &12.08$^{+0.03}_{-0.04}$&11.68$^{+0.08}_{-0.08}$&11.86$^{+0.07}_{-0.10}$& $10-37$   & $70\pm30$            & $<15$        &7.43$^{+0.08}_{-0.12}$&$37\pm4$ &$1.69\pm0.1$   \\ 
15327+2340 & 2.01    &12.14$^{+0.02}_{-0.01}$&12.14$^{+0.02}_{-0.01}$& $<9.90$               & $26-37$   & $240\pm30$           & --           &8.32$^{+0.05}_{-0.06}$&$27\pm3$ &$1.93\pm0.02$  \\  
17132+5313 & 1.00    &11.92$^{+0.03}_{-0.01}$&11.84$^{+0.02}_{-0.01}$&11.14$^{+0.01}_{-0.11}$& $10-26$   & $130\pm20$           & --           &7.64$^{+0.18}_{-0.05}$&$31\pm3$ &$1.95\pm0.03$  \\ 
17208-0014 & 4.70    &11.99$^{+0.03}_{-0.04}$&11.99$^{+0.04}_{-0.05}$& $<11.00$              & $>16$     & $150\pm30$           & --           &8.25$^{+0.19}_{-0.25}$&$26\pm6$ &$1.95\pm0.3$   \\  
18470+3233 & 1.00    &12.08$^{+0.03}_{-0.04}$&12.03$^{+0.04}_{-0.04}$& $<11.42$              & $6.6-37$  & $200\pm40$           & --           &7.54$^{+0.31}_{-0.30}$&$35\pm6$ &$1.95\pm0.05$  \\ 
19254-7245 & 1.52    &12.24$^{+0.08}_{-0.10}$&12.17$^{+0.10}_{-0.05}$& $<11.60$              & --        & $170\pm50$           & --           &7.60$^{+0.11}_{-0.14}$&$34\pm4$ & $1.96\pm0.1$  \\ 
19297-0406 & 1.00    &12.41$^{+0.02}_{-0.05}$&12.39$^{+0.03}_{-0.06}$& $<11.50$              & $<57$     & $480\pm150$          & --           &8.15$^{+0.10}_{-0.15}$&$31\pm6$ & $1.95\pm0.25$ \\ 
19458+0944 & 1.73    &12.40$^{+0.01}_{-0.02}$&12.25$^{+0.03}_{-0.01}$&11.83$^{+0.04}_{-0.08}$& $\sim26$  & $280\pm30$           & --           &8.99$^{+0.03}_{-0.12}$&$22\pm2$ & $1.89\pm0.01$ \\ 
20046-0623 & 1.00    &12.17$^{+0.03}_{-0.04}$&12.11$^{+0.03}_{-0.11}$& $<11.60$              & $<26$     & $210\pm60$           & --           &7.86$^{+0.02}_{-0.18}$&$32\pm9$ & $1.96\pm0.55$ \\   
20414-1651 & 1.18    &12.22$^{+0.01}_{-0.02}$&12.16$^{+0.01}_{-0.01}$&11.31$^{+0.05}_{-0.10}$& $6.6-16$  & $300\pm15$           & --           &7.49$^{+0.16}_{-0.14}$&$38\pm2$ & $1.96\pm0.10$ \\
20551-4250 & 1.95    &12.05$^{+0.04}_{-0.05}$&11.90$^{+0.01}_{-0.10}$&11.53$^{+0.03}_{-0.17}$& $10-57$   & $160\pm20$           & --           &7.05$^{+0.13}_{-0.05}$&$44\pm3$ & $1.91\pm0.04$ \\  
21130-4446 & 1.65    &12.14$^{+0.02}_{-0.01}$&12.11$^{+0.03}_{-0.02}$& $<10.90$              & $37-64$   & $250\pm15$           & --           &7.99$^{+0.05}_{-0.03}$&$30\pm2$ & $1.94\pm0.10$ \\  
21504-0628 & 1.00    &12.00$^{+0.03}_{-0.04}$&11.94$^{+0.04}_{-0.07}$&11.06$^{+0.25}_{-0.15}$& $<57$     & $180\pm20$           & --           &7.15$^{+0.11}_{-0.15}$&$42\pm4$ & $1.97\pm0.10$ \\  
22491-1808 & 2.38    &12.16$^{+0.01}_{-0.02}$&11.61$^{+0.02}_{-0.07}$&12.02$^{+0.03}_{-0.03}$& $\sim26$  & $60\pm10$            & $<10$        &8.04$^{+0.03}_{-0.04}$&$27\pm3$ & $1.80\pm0.02$ \\    
23128-5919 & 2.15    &12.02$^{+0.02}_{-0.02}$&11.93$^{+0.04}_{-0.07}$&11.30$^{+0.10}_{-0.07}$& $26-37$   & $150\pm30$           & --           &7.75$^{+0.07}_{-0.08}$&$30\pm3$ & $1.93\pm0.03$ \\    
23365+3604 & 2.90    &12.19$^{+0.01}_{-0.01}$&12.00$^{+0.01}_{-0.02}$&11.73$^{+0.02}_{-0.09}$& $\sim26$  & $200\pm20$           & $<10$        &7.81$^{+0.02}_{-0.03}$&$33\pm2$ & $1.86\pm0.02$ \\  
23389-6139 & 1.22    &12.16$^{+0.01}_{-0.01}$&12.15$^{+0.01}_{-0.02}$&10.48$^{+0.12}_{-0.18}$& $16-26$   & $290\pm30$           & --           &7.70$^{+0.20}_{-0.09}$&$33\pm4$ & $1.94\pm0.05$ \\ 
\hline
\end{tabular}

\medskip

Luminosities are the logarithm of the 1-1000$\mu$m luminosities obtained from the best fit combined starburst/AGN model, 
in units of bolometric solar luminosities. $^{a}$Reduced $\chi^{2}$ of the combined SED fit $^{b}$Starburst age $^{c}$Star 
formation rate $^{d}$Viewing angle of the AGN dust torus $^{e}$Dust mass $^{f}$Dust temperature $^{g}$Emissivity index
\end{minipage}
\end{table*}

\section{Individual objects}\label{sect:ind}

In this section we present further discussion on objects in our sample that are either 
well-studied by previous authors, or that show interesting or unusual features.

\noindent \underline{{\it 00198-7926:}} This source is a double nucleus system with a large tail, with a Sy2 
optical spectrum. The second nucleus is probably a large region of extranuclear unobscured star formation, and 
there are a number of HII regions embedded in the tail \citep{hev}. A deep Bepposax observation gave no 
detection, giving an upper limit to the X-ray flux of $F_{2-10} < 1\times10^{-13}$ erg cm$^{-2}$ s$^{-1}$ 
\citep{ris}. From our results, the properties of this object mark it as unusual. We 
find that the IR emission from this object is mostly starburst in origin. The starburst is comparatively old, 
and can thus account for both the sub-mm emission and all of the unobscured star formation observed by 
\citet{hev}. We interpret the `warm' infrared colour of this object as being due to the old starburst rather 
than an AGN. There is however comparatively little IR data available for this object and the resulting AGN 
upper limit is weak. The best fit model, shown in Figure \ref{ulirg_seds}, includes an AGN 
component. \\
\noindent \underline{{\it 05189-2524:}} This object appears to be a late-stage merger. It possesses a single, 
compact, very red nucleus with a Sy2 spectrum \citep{vei}, bisected by a dust lane \citep{sco}. \citet{you} 
observed broad lines in polarized flux in this object, suggesting the presence of an obscured AGN. Later 
observations \citep{dud} observed the $11.3\mu$m dust feature, suggesting that this object also contains a 
buried starburst. The X-ray spectrum \citep{ris} is well fitted by a two-component model, consisting of a power law with 
$\Gamma = 1.89^{+0.35}_{-0.34}$, absorbed by a column density of $N_{H} = 4.7^{+1.4}_{-1.1}$ cm$^{-2}$, 
and a thermal component with kT = $0.88^{+0.89}_{-0.35}$ keV. Overall the power law component can be 
interpreted as arising from a Compton thin AGN, with the thermal component either due to the AGN or 
to starburst activity. For our SED fitting we compiled an additional N band (10.6$\mu$m) flux of 498mJy 
$\pm$ 100mJy from \citet{mai}. From our results, we find that this object contains both a starburst and an 
AGN, where the AGN provides a significant fraction of the total IR luminosity. \\
\noindent \underline{{\it 09320+6134:}} This object possesses a single bright nucleus and a disturbed spiral 
structure \citep{sco}. There is one large tail extending $\sim38$kpc, and a second tail that forms almost a 
complete ring with a total length of $\sim65$kpc. The morphology has been interpreted as a late 
stage merger between two large spiral galaxies \citep{san2}, possibly involving a third gas-poor 
dwarf galaxy \citep{maj}. The optical spectrum is a combination of a LINER and a Sy2 \citep{gvv}. 
Near-IR spectroscopy by \citet{idm} shows that this object is likely to contain a heavily 
obscured AGN. Conversely, \citet{lvg} classify this object as starburst powered on the basis of 
mid-IR spectroscopy, and the sub-mm emission is also consistent with a starburst \citep{rlr}. 
We compiled an additional $350\mu$m upper limit of 2664mJy, and an additional $800\mu$m flux of 143mJy 
$\pm25$mJy from \citet{rlr}. 
We find that this system is a composite object containing both a starburst and an 
AGN, a result that is consistent with previous observations. From the SED fit presented in 
Figure \ref{ulirg_seds} it can be seen that the AGN has a higher near-IR flux than the starburst 
component, but that the AGN contribution in the mid-IR is small, and is negligible in the sub-mm. This 
object is a case in point that, in order to understand the power source in ULIRGs it is necessary 
to have data spanning a wide wavelength range. \\
\noindent \underline{{\it 10565+2448:}} We compiled an additional $350\mu$m flux of 1240mJy $\pm$ 248mJy, 
and a $750\mu$m flux of 85mJy $\pm17$mJy from \citet{rlr}. The AGN in this object, although weak compared 
to the starburst, is required for an acceptable SED fit. \\
\noindent \underline{{\it 12112+0305:}} This system contains two separate nuclei with a pair of tidal 
tails \citep{sco}. The optical spectrum is that of a LINER \citep{vei2}. This source is classified 
as a starburst on the basis of its mid-IR spectrum \citep{rig}. We compiled an additional $800\mu$m flux 
of 50mJy $\pm13$mJy from \citet{rlr}. The extensive IR data available for 
this object allow us to place strong constraints on the power source. We find that this object is 
powered by a starburst, with a severe upper limit on any AGN contribution. \\
\noindent \underline{{\it 12540+5708:}} This galaxy is commonly known as Mrk 231, and has a Sy1 optical 
spectrum. The IR luminosity is thought to be powered by both starburst and AGN activity \citep{crl,con1}. Mrk 231 possesses a 
single compact nucleus surrounded by irregular `rings' of recent star formation and a small 
tidal arm containing numerous blue starforming `knots', and is therefore thought to be an advanced 
merger. The nucleus is unresolved in the radio \citep{con1} and at $11.7\mu$m \citep{mil}, 
implying that the emission from the nucleus is powered by an AGN rather than a starburst. Mrk 231 
is classified as an AGN on the basis of both near-IR spectroscopy \citep{idm} and mid-IR 
spectroscopy \citep{rig}. Conversely, the X-ray emission from this object, although weak, cannot be 
fitted with a pure power law or thermal bremsstrahlung model \citep{iwa}. 
We compiled additional mid-IR data from \citet{rie0} and additional sub-mm data from \citet{rlr} and \citet{car0}. 
We find 
that Mrk 231 contains both a luminous starburst and an AGN, where the starburst contributes $\sim70\%$ 
of the $1-1000\mu$m IR luminosity. Interestingly, this result is in excellent agreement with independent 
estimates of the starburst and AGN luminosities in Mrk 231 derived from VLBI observations \citep{lon}. 
From Figure \ref{ulirg_seds} we can see that the AGN dominates the 
emission in the near- and mid-IR, in agreement with previous authors results, but that the starburst 
dominates at sub-mm wavelengths. \\
\noindent \underline{{\it 13428+5608:}} This object is commonly known as Mrk 273. It contains two 
nuclei with a single, long (41kpc) tidal tail and a `ring' of star formation almost 100kpc 
in diameter. The northern nucleus is resolved and is surrounded by unobscured star formation, whereas 
the southern nucleus is redder and unresolved \citep{sco}. The optical spectrum is that of a Sy2, and 
there are no broad lines in the near-IR spectrum, although there are strong narrow lines \citep{vei2}. 
Radio continuum and HI 21cm observations of this source classify it as a pure starburst \citep{con1,cat}. Conversely, 
mid-IR spectral observations \citep{rig,lvg} classify it as an AGN. This object is a strong X-ray 
source, with a complex spectrum. The best fit is a 4 component model, consisting of a 
direct and scattered AGN power law, a thermal component, and an iron line. Furthermore, the X-ray emission 
appears extended rather than unresolved \citep{lwh}. We compiled an additional $350\mu$m flux of 1004mJy 
$\pm$230mJy, and an aditional $800\mu$m flux of 84mJy $\pm$22mJy, both from \citet{rlr}. 
Our results are consistent with Mrk 273 containing 
both a starburst and an AGN, in agreement with X-ray observations. The SED fit in Figure \ref{ulirg_seds} 
is consistent with a starburst interpretation in the near-IR, and also gives a good fit in the mid-IR 
although the fit predicts the mid-IR emission is due to a starburst rather than an AGN. \\
\noindent \underline{{\it 14348-1447:}} We compiled an additional $800\mu$m flux of 29mJy $\pm$10mJy from 
\citet{rlr}. This object is very luminous, where the luminosity is dominated by a starburst. Whilst this 
object does contain a luminous AGN, the AGN only contributes $\sim17\%$ of the total infrared flux. \\
\noindent \underline{{\it 15250+3609:}} This object has one bright central nucleus, and a second, much dimmer 
nucleus $0.7\arcsec$ from the centre. There is also a large ringlike structure 25kpc 
in diameter \citep{sco}. The optical spectrum is a composite of HII and LINER features \citep{vei2}. 
\citet{lvg} classify this object as a starburst based on its mid-IR spectrum. Our results however 
show that this object contains both a starburst and an AGN. This is one of only 3 objects in the sample 
where the AGN provides more than half of the total IR emission. Interestingly, it can be seen from 
Figure \ref{ulirg_seds} that the sub-mm emission from this object is not dominated by a starburst, but 
instead arises in almost equal parts from a starburst and an AGN. We interpret this as arising from a 
heavily obscured AGN that is oriented nearly edge on to us, rather than an AGN with 
an extended torus. \\
\noindent \underline{{\it 15327+2340:}} This object, commonly known as Arp220, is both the closest ULIRG to 
us and the most intensely studied. The optical spectrum was first 
thought to be Sy2-like \citep{san2}, and this was interpreted as indicating an AGN power source. 
Conversely, the broad band optical-IR SEDs suggested a starburst as the main power source 
\citep{jow}. Observations and modelling at other wavelengths revealed a more complex 
picture. Ground-based mid-IR spectroscopy \citep{sar} suggested a hybrid nature for Arp220, including 
both starburst and AGN, whereas ISO mid-IR spectroscopy \citep{stu1} suggested a pure starburst. High 
resolution VLBA imaging \citep{smi1} revealed a number of unresolved sources in the nuclei, inferred to 
be radio supernovae, supporting the presence of a starburst. Optical integral field observations of the 
central regions \citep{acc} showed that the spectrum is LINER-like in most regions, but with 
a Sy2-like spectrum in the brightest emission line region. \citet{rie1} searched for hard X-rays in Arp220 
with HEAO-A1 data, and concluded that an AGN similar to those in quasars could not be the 
power source. Recent observations with Chandra \citep{cle2} detect weak hard X-ray emission from 
the nucleus, which has a possible origin in a weak AGN responsible for about $5\%$ of the bolometric 
luminosity. An alternative explanation, with a much larger AGN contribution to the bolometric luminosity, 
is possible if the AGN is obscured by a Compton-thick screen of $\sim10^{25}$ cm$^{-2}$ or more. There is 
some additional evidence for an obscured AGN in Arp220, including sub-mm 
observations \citep{haa} and LWS spectra that may indicate $\tau > 1$ at  $\sim100\mu$m \citep{fis}. 
We compiled additional IR photometry for Arp220 from \citet{kla0}, and additional sub-mm photometry 
from \citet{chi,rlr} and \citet{dun2}. 
Our results are consistent with a starburst interpretation for Arp220. The extensive IR data 
is well fitted by a pure starburst model, and the resulting upper limit on the IR luminosity of any 
AGN component is very strong. This upper limit on the AGN IR luminosity also allows us to set a lower 
limit in the UV optical depth of an IR luminous AGN in Arp220, of $\tau_{UV}=1500$. \\
\noindent \underline{{\it 17208-0014:}} This source has a single nucleus surrounded by a 
disturbed disk containing several compact star clusters, with a single tail. The optical spectrum is 
that of an HII region \citep{vei2} and near-IR imaging shows an extended nucleus, arguing against 
an AGN \citep{sco}. This object lies in a crowded field, and high spatial resolution 
$100\mu$m observations with the Kuiper Airborne Observatory \citep{zld} measure a $100\mu$m flux 
substantially below that from IRAS, implying that the IRAS fluxes for this object are confused. 
The $100\mu$m flux used in the fit is from \citet{zld}). This was the only object in our 
sample where the fit was relatively poor ($\chi^{2}_{red}=4.7$), however it can be seen from 
Figure \ref{ulirg_seds} that this poorness of fit is due to inconsistencies between the 
model and the IRAS $60\mu$m flux. As there are observed inconsistencies at $100\mu$m we infer that 
this inconsistency at $60\mu$m is also due to confusion. Overall, 
our results are consistent with a pure starburst interpretation for this object, although the AGN 
upper limit is weak.  \\
\noindent \underline{{\it 19254-7245:}} This object is more commonly known as the Super-Antena, and possesses a 
spectacular morphology. Two extremely long tail-like features extend on either side of the central 
regions, with a total length of 500kpc. These features can be explained as a result of a coplanar 
encounter between two massive disk galaxies \citep{mmi}. There are two optical nuclei in the central 
regions, one with a Sy2 spectrum and the other with a starburst spectrum \citep{mlm}. Further optical 
spectroscopy by \citet{clm} classify this source as an obscured AGN surrounded by starforming clouds.
The mid-IR spectrum of this object classifies it as an AGN \citep{lvg}.  X-ray observations \citep{pgs} 
were inconclusive, but suggested an AGN was at least present in this object. We compiled for this object 
an extra flux at $6.7\mu$m of 113mJy $\pm$23mJy from \citet{cha}. Of our sample 
the SED fit for this object is of reasonable quality, but deviates markedly from one point. We interpret 
this as being due to the very large angular size of this object, which is much larger than any other 
object in our sample. Whilst IRAS fluxes are likely to contain all the flux from the object, it is 
feasible that other observations have missed some flux. Our fit 
(Figure \ref{ulirg_seds}) is a pure starburst, but the AGN upper limit is weak. \\
\noindent \underline{{\it 22491-1808:}} This system possesses a complex morphology, with at least 3 possible nuclei 
and large numbers of starforming `knots'. There are also two small bright tails, in which many of the 
starforming knots are embedded. Based on the morphology, \citet{cui1} propose a multiple merger origin 
for this ULIRG. The mid-IR spectrum is consistent with a starburst \citep{rig,lvg}. We 
however find that this object contains both a starburst and an AGN. This is one of 
only three objects where the AGN contributes more than half the total IR luminosity. In 
the mid-IR the starburst dominates, but with a significant AGN contribution. The contribution 
from the AGN in the sub-mm is comparable to the sub-mm emission from the starburst, implying that 
the AGN torus is being viewed almost edge on. \\

\begin{figure*}
\rotatebox{0}{
\centering{
\scalebox{0.76}{
\includegraphics*[18,144][592,718]{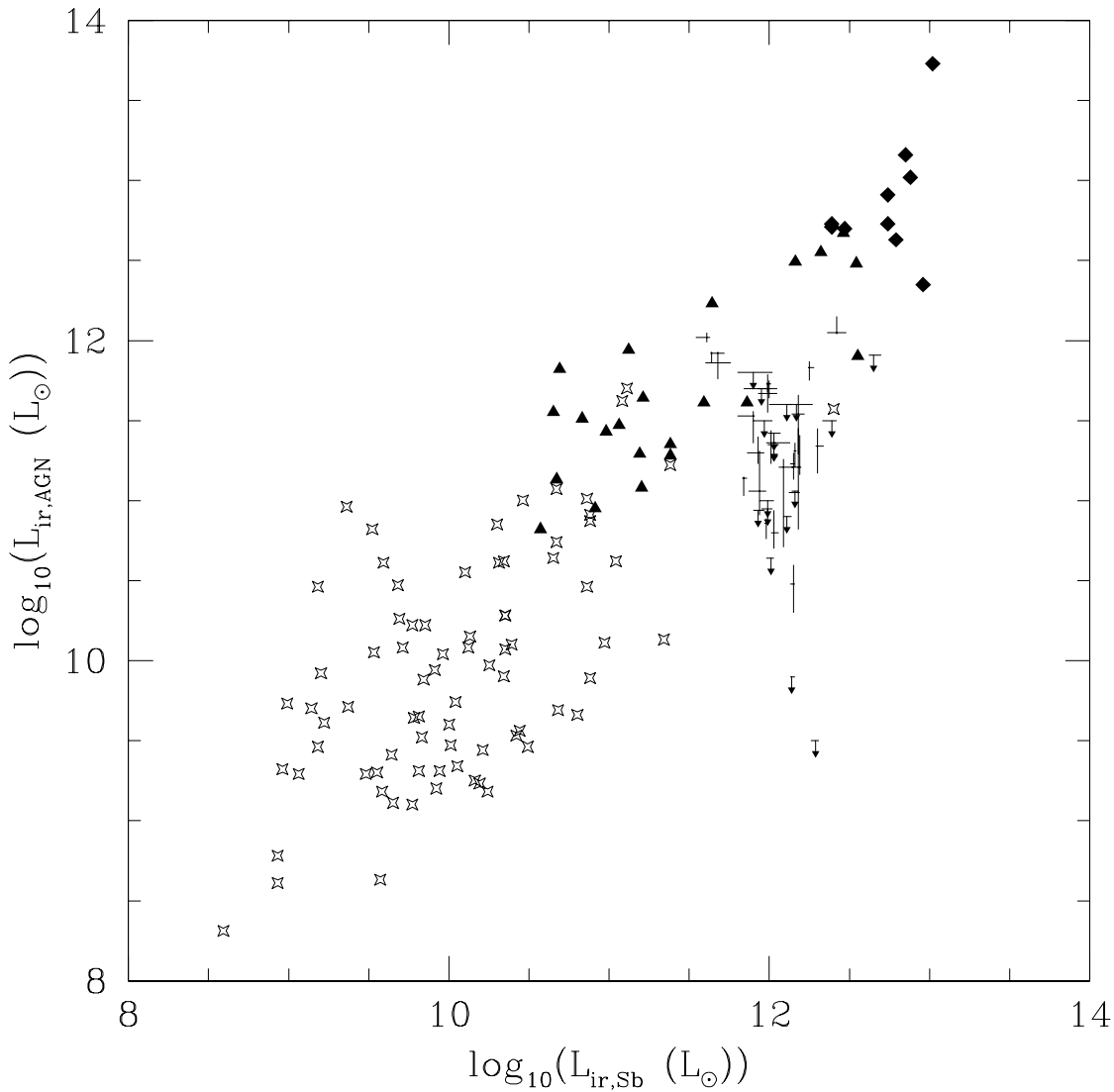}
}}}
\caption
{
AGN luminosity in the IR plotted against starburst luminosity in the IR. The ULIRGs from our sample are plotted 
as points, with a downward arrow indicating an upper limit on the AGN luminosity. Triangles are PG QSOs, and 
crosses are those galaxies with detections in all four IRAS bands, both these samples are taken from \citet{row2}. 
Diamonds are the Hyperluminous Infrared Galaxies taken from \citet{far3}.   
\label{sbvsagn}
}
\end{figure*}

\section{Discussion}\label{sect:discuss}

\subsection{Limitations of the models}

Before discussing our results, we examine the suitability of the starburst and AGN models used 
in this paper for understanding the nature of the power source in ULIRGs. Both sets of models are physical 
in nature, in that they do not assume an observational SED template for dusty star formation or AGN activity, 
and instead produce a predicted spectrum based on radiative transfer calculations of optical/UV light from 
a starburst or AGN propagating through a dusty medium. They are therefore well suited to investigating the IR 
emission from a population whose properties may vary markedly between individual objects. A complete 
description of the starburst and AGN models can be found in \citet{ef1} and \citet{ef0} respectively. 

It is however important to note that these models have drawbacks. In particular, the models cannot account 
for sub-mm emission from an extended (hundreds of parsecs instead of tens of parsecs) dust torus 
surrounding an AGN. If such a torus did exist in our sample the sub-mm emission from the 
torus could be mistakenly ascribed to star formation, leading to a substantially overestimated star formation 
rate. There do exist radiative transfer models for cool, extended dust surrounding an AGN. To test whether all or part 
of the sub-mm emission from our sample could arise from such a torus, we fitted the extended torus models of 
\citet{rr95} to the objects in our sample, both on their own and in combination with the starburst models described 
previously. Although in all cases the extended torus models were clearly rejected, we cannot completely discount 
this possibility, as the extended torus models do not cover a wide range of torus geometries and lines of sight.

Despite their limitations, the models appear to work very well. In a sample of 41 objects the models produce a good 
fit in all but one case. As described in Section 
\ref{sect:ind}, in this case there is reason to believe that the relative poorness of fit may be partially due 
to the quality of the data. The models successfully reproduce both a wide range of ULIRG SED shapes over the wavelength 
range $0.4 - 1000\mu$m (if the optical data are used as upper limits), and also  
reproduce spectral features such as the $7.7\mu$m PAH line, and the $9.7\mu$m silicate 
absorption feature. We therefore find that, under the caveats described previously, these models are well suited to 
studying the dust-shrouded power source in ULIRGs.

\begin{figure*}
\begin{minipage}{180mm}
\epsfig{figure=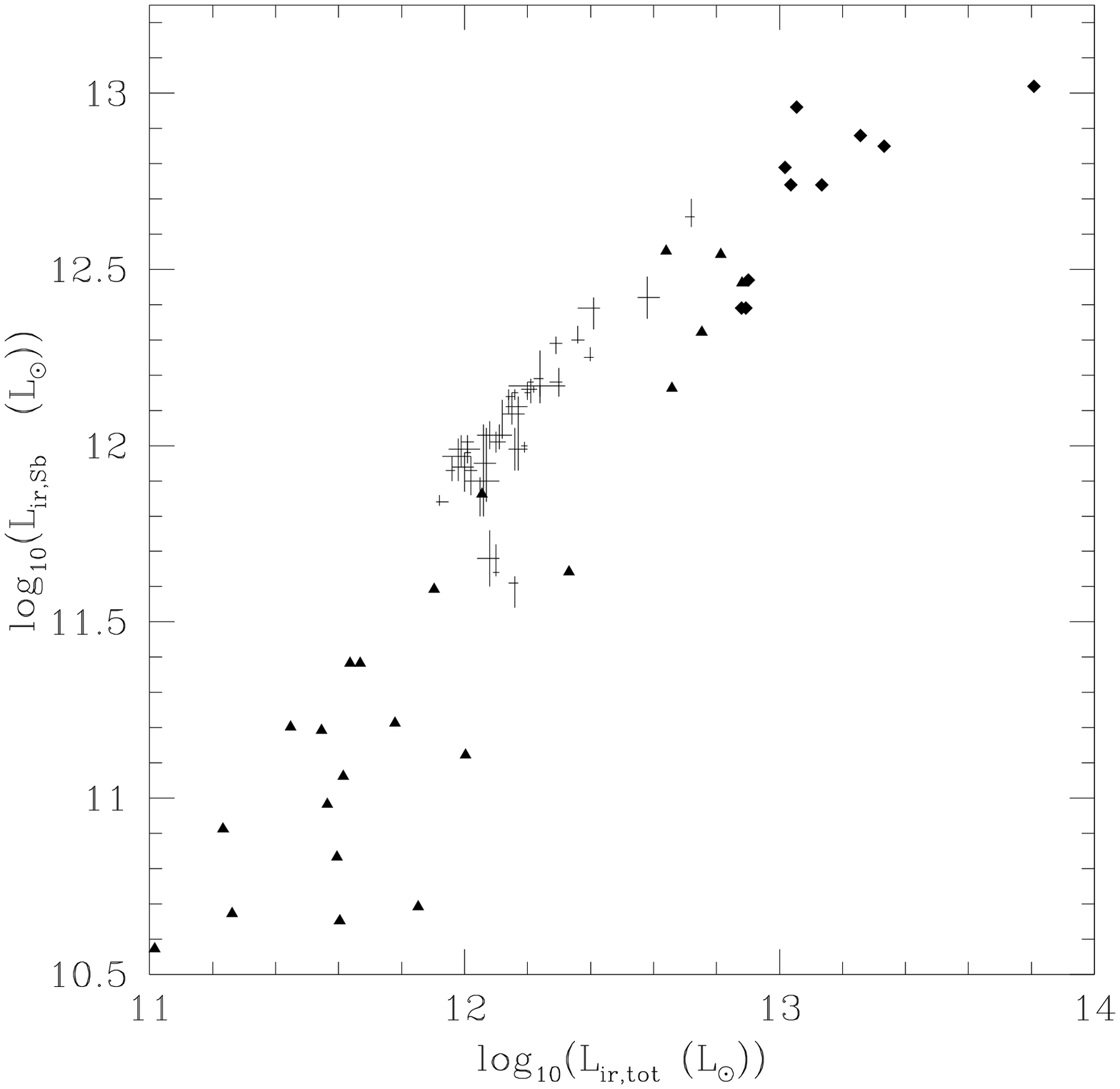,width=89mm}
\epsfig{figure=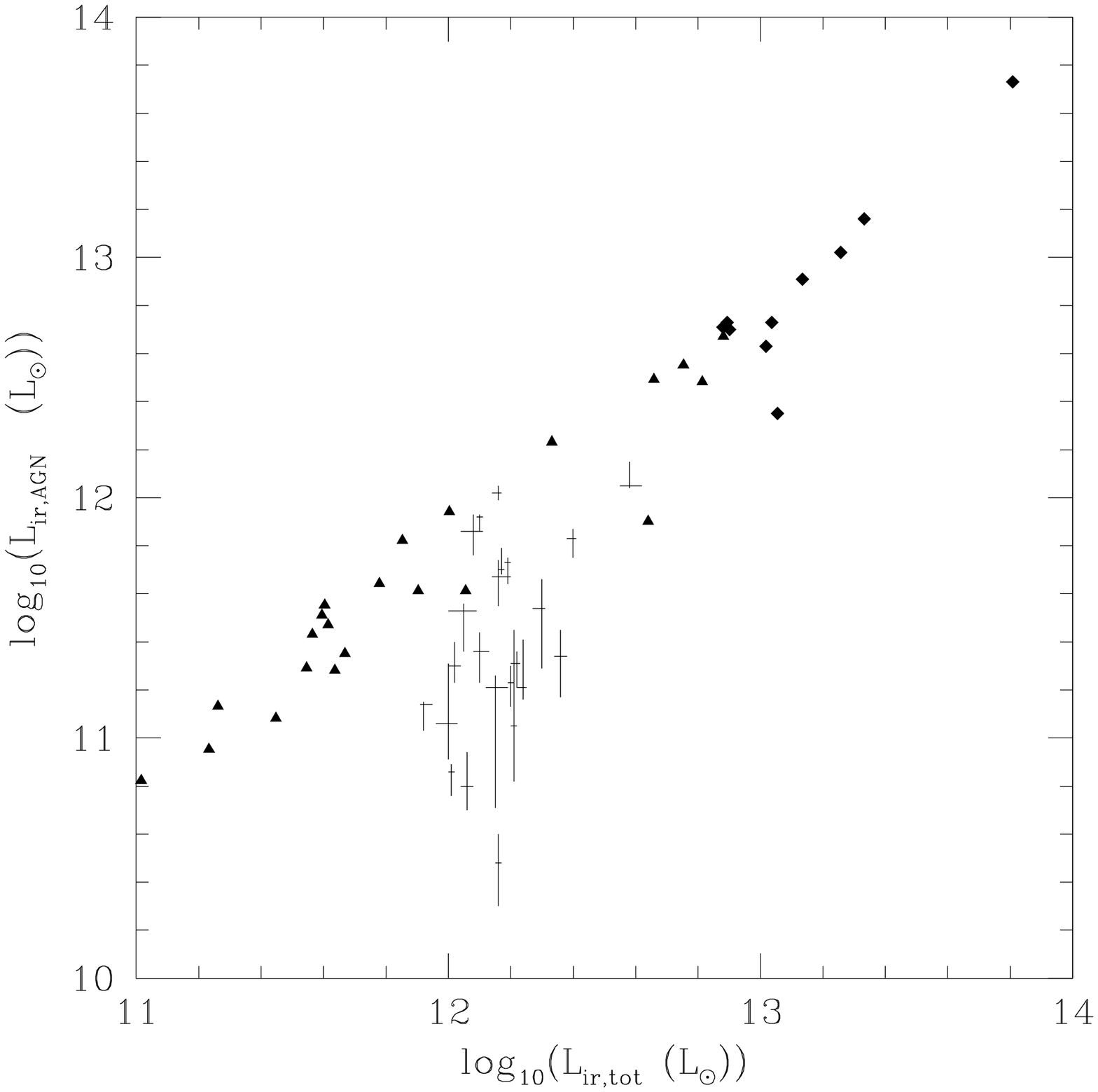,width=89mm}
\end{minipage}
\caption
{
Total IR luminosity plotted against (left) starburst luminosity and (right) AGN luminosity. Our ULIRGs are plotted as 
points, triangles are PG QSOs taken from \citet{row2}, and diamonds are the HLIRGs from \citet{far3}
\label{totvssbagn}
}
\end{figure*}

\subsection{Starburst and AGN activity in ULIRGs}

\subsubsection{What powers ultraluminous infrared galaxies?} \label{subsect:power}

Although ULIRGs have been studied for nearly two decades, the nature of the power source behind the IR emission has 
remained controversial. The most recent results on the power source in ULIRGs have come from mid-IR spectroscopy,
which show that most ULIRGs ($\sim80\%$) are powered mainly by starbursts \citep{rig,gen}, but that at least half of 
local ULIRGs show evidence for both starburst and AGN activity. The AGN fractional luminosity as a function of total IR 
luminosity has been studied since {\it IRAS} galaxies were discovered. Optical spectroscopy \citep{vei} found that the 
fraction of IRAS sources with AGN spectra, and the fraction of Seyfert galaxies amongst the AGN increases with increasing 
IR luminosity, reaching values of $62\%$ and $54\%$ respectively at IR luminosities $>10^{12}L_{\odot}$. Near-IR spectroscopy 
of ULIRGs \citep{vei2} shows that the fraction of ULIRGs with signs of AGN activity is at least $20\%-25\%$ but rises 
abruptly to $35\%-50\%$ for objects with $L_{ir}>10^{12.3}L_{\odot}$. Recent {\em ISO} spectroscopy of a small sample of ULIRGs 
\citep{tra2} found half of the sample to be starburst dominated and half to be AGN dominated. They also showed that, 
at IR luminosities below $10^{12.4}L_{\odot}$, most ULIRGs were starburst dominated, with the starburst 
contributing around $85\%$ to the IR emission. At IR luminosities above $10^{12.4}L_{\odot}$ the AGN contribution was 
much higher, with the starburst contributing about $50\%$ of the IR emission on average. Starburst 
dominated systems were found up to luminosities of around $10^{12.65}L_{\odot}$.

From our results, we find that both starburst and AGN activity are central in understanding the properties of local ULIRGs. 
All 41 objects in our sample contain a very luminous starburst that contributes significantly (at least $28\%$) to the total 
IR luminosity. There are no purely AGN powered systems in the sample. 23/41 objects have measured AGN luminosities where the 
AGN contributes significantly to the total IR emission, the remaining 18 objects have only upper limits on the AGN contribution. 
In three cases the AGN supplies more than half of the total IR emission, in the remaining cases the starburst is the dominant 
contributor. We therefore find that previous estimates of the fraction of local ULIRGs that are starburst dominated based on 
mid-IR spectroscopy have underestimated the true fraction, and that this fraction is $\sim90\%$ rather than $80\%$.
The mean starburst fractional luminosity for the sample is $82\%$, spanning the range $28\%$ to $\sim100\%$.  
Overall, the IR emission from ULIRGs as a class is either starburst in origin with a negligible AGN contribution, 
or arises from a combination of starburst and AGN activity with the starburst usually contributing the largest fraction. Given 
our sample size, the fraction of purely AGN-powered local ULIRGs must be less than $\sim2\%$. The derived star formation rates 
range from $50M_{\odot}$yr$^{-1}$ to $500M_{\odot}$yr$^{-1}$ and the derived dust masses span the range 
$10^{7} < M_{\odot} < 10^{9}$. These ranges are at least 1 order of magnitude higher than those observed in normal galaxies, 
indicating that all ULIRGs are going through a major star forming episode. It proved difficult to constrain the 
line of sight to the AGN torus, with meaningful constraints only achieved for a few objects. Although these values are discussed 
in section \ref{sect:ind} we cannot draw any general conclusions about relative AGN orientation in ULIRGs. 

\begin{figure}
\rotatebox{0}{
\centering{
\scalebox{0.38}{
\includegraphics*[18,144][592,718]{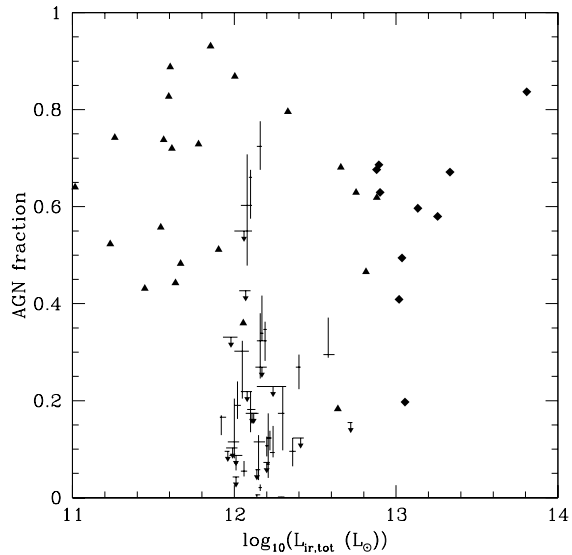}
}}}
\caption
{
Total luminosity in the IR plotted against AGN fraction. Key to the symbols is the same as for Figure \ref{totvssbagn}
\label{ltotvsagnfrac}
}
\end{figure}

Figure \ref{sbvsagn} shows a plot of starburst luminosity 
vs. AGN luminosity. Also plotted are the starburst and AGN luminosities of 10 Hyperluminous Infrared Galaxies (HLIRGs, diamonds) taken 
from \citet{far3}, as well as 22 PG QSOs (triangles) and 78 galaxies with detections in all four IRAS bands (crosses). 
The two latter samples are taken from \citet{row2} and rescaled to $H_{0}=65$ km s$^{-1}$ Mpc$^{-1}$. The figure shows 
that starburst and AGN luminosities are correlated, albeit with a large scatter, over 6 orders of magnitude in IR luminosity 
and over a wide range of galaxy types. Therefore, there may be a common physical factor governing the IR luminosity of both 
starbursts and AGN in these galaxies. We postulate that this factor is the available quantities of gas 
and dust in the nuclear regions of these systems, as this affects both the rate and duration of black hole accretion, 
and the number of stars that can form. 

We next examine trends in starburst and AGN luminosity against total IR luminosity. Figure \ref{totvssbagn} shows total IR 
luminosity plotted against starburst luminosity and AGN luminosity. Also plotted 
are the PG QSO and HLIRG samples, however in this section we only discuss the ULIRGs plotted 
in these figures; comparisons between the ULIRGs, PG QSOs and HLIRGs can be found in \S \ref{ulirglowhighz}. The left 
hand panel shows that the ULIRG starburst luminosity is strongly correlated with the total luminosity. From the right-hand plot 
in Figure \ref{totvssbagn} we can see that, although there is a trend for more luminous ULIRGs to contain a more luminous AGN, 
the trend is less clear than for the starburst luminosity. 
Figure \ref{ltotvsagnfrac} shows total IR luminosity plotted against {\it fractional} AGN luminosity. It can be seen that there 
is no trend for increasing AGN dominance with increasing total luminosity in ULIRGs, contrary to previous claims. 
We find that these claims were based on finding generally more luminous AGN in more luminous ULIRGs, but that there is no evolution 
of fractional AGN luminosity with total luminosity in local ULIRGs.

\subsubsection{Multiple starbursts and multiple mergers}\label{subsubsect:mult}

The number of individual starburst events that occur over the lifetime of a ULIRG is an important parameter in understanding 
starburst triggering and evolution in all classes of active galaxy. Similarly, the number of progenitors in a 
typical ULIRG merger can be used to estimate which fraction of ULIRGs occur in the field galaxy population between 2 or 3 
progenitors, and which fraction occur in compact groups of galaxies between multiple progenitors. 

Both these parameters have proved difficult to estimate. Numerical simulations of mergers between two galaxies have 
shown that multiple, discrete starburst events can be triggered even in double mergers. The timing of the starbursts is 
thought to depend on the progenitor galaxy morphologies, with early starbursts (before the galaxies start to merge) occurring 
in bulgeless disk galaxies, and late starbursts (after the galaxies have started to merge) occurring in disk galaxies with 
a bulge component \citep{mih,mih2}. Simulations of multiple galaxy mergers have however shown 
that more than two starburst or AGN nuclei is evidence that there were more than two merger progenitors \citep{ta0}, 
and that repetitive starbursts are characteristic of multiple mergers \citep{bek}. Observationally, multiple 
mergers have been linked to Arp 220 \citep{dia} and to larger samples of ULIRGs \citep{bor,far1}. Other 
authors have estimated the lifetime of a single starburst event in ULIRGs. \citet{thor} derive a starburst lifetime range 
of $10^{6} - 10^{7}$ years, and infer that galactic 
superwinds produced by supernovae may be responsible for the short duration. \citet{gen} derive an upper limit on the starbust 
lifetime in ULIRGs of $\sim10^{8}$ years. The lifetime of a ULIRG has been previously estimated to lie in the range 
$10^{8} - 10^{9}$ years \citep{mur2,far1} based on the observed range of morphologies in large samples of ULIRGs, the 
apparent discrepancy between the ULIRG and starburst lifetimes has led to the suggestion that multiple starbursts and multiple 
mergers may be common in ULIRGs \citep{far1}. 

The starburst ages derived for our sample are listed in Table \ref{ulirgparams}. As our sample is large and 
the selection is robust we are confident that this range in ages is representative of the ULIRG population 
as a whole. Excluding those objects with upper or lower limits on the starburst age, the derived ages 
span the range $1.6\times10^{6} \leq$ years $\leq 7.1\times10^{7}$, with most lying in the range 
$1.0\times10^{7} \leq$ years $\leq 3.7\times10^{7}$. These lifetime ranges are clearly inconsistent with a single starburst 
event powering the IR emission for the lifetime of a ULIRG. From section \ref{subsect:power} however, starbursts are 
present in at least $98\%$ of ULIRGs and in most cases dominate the IR emission; AGN activity cannot 
therefore explain this discrepancy in lifetimes. We therefore find that multiple starburst events must be common in the ULIRG 
population, and by inference that multiple mergers must be common as well. 

From this analysis, coeval starbursts with different ages are feasible in ULIRGs, and it is important to note 
the effects this may have on our SED fitting approach. Whilst the available photometry is sufficient to separate starburst 
and AGN emission, it is not sufficient to resolve the starburst emission into multiple components. The derived age ranges 
in Table \ref{ulirgparams} should however encompass the age ranges of any coeval starbursts, due to the effect of age on the shape of the 
starburst SED. As the age of the starburst increases then (with reference to figure 3 of \citet{ef1}), the peak of the 
SED shifts to longer wavelengths, PAH features become stronger, the 9.8$\mu$m silicate absorption feature becomes shallower, 
and the optical/UV flux increases. Hence, the mid- and far-IR photometry, in constraining the strength of the mid-IR spectral 
features and the peak of the SED respectively, provide a lower limit to ages of any coeval starbursts, whereas the optical 
photometry provides an upper limit. Therefore, even if some or all of the ULIRGs in our sample contained two or more starbursts 
with different ages, we are confident that the starburst age ranges given in Table \ref{ulirgparams} encompass the ages of 
these starbursts. 

\begin{figure*}
\begin{minipage}{180mm}
\epsfig{figure=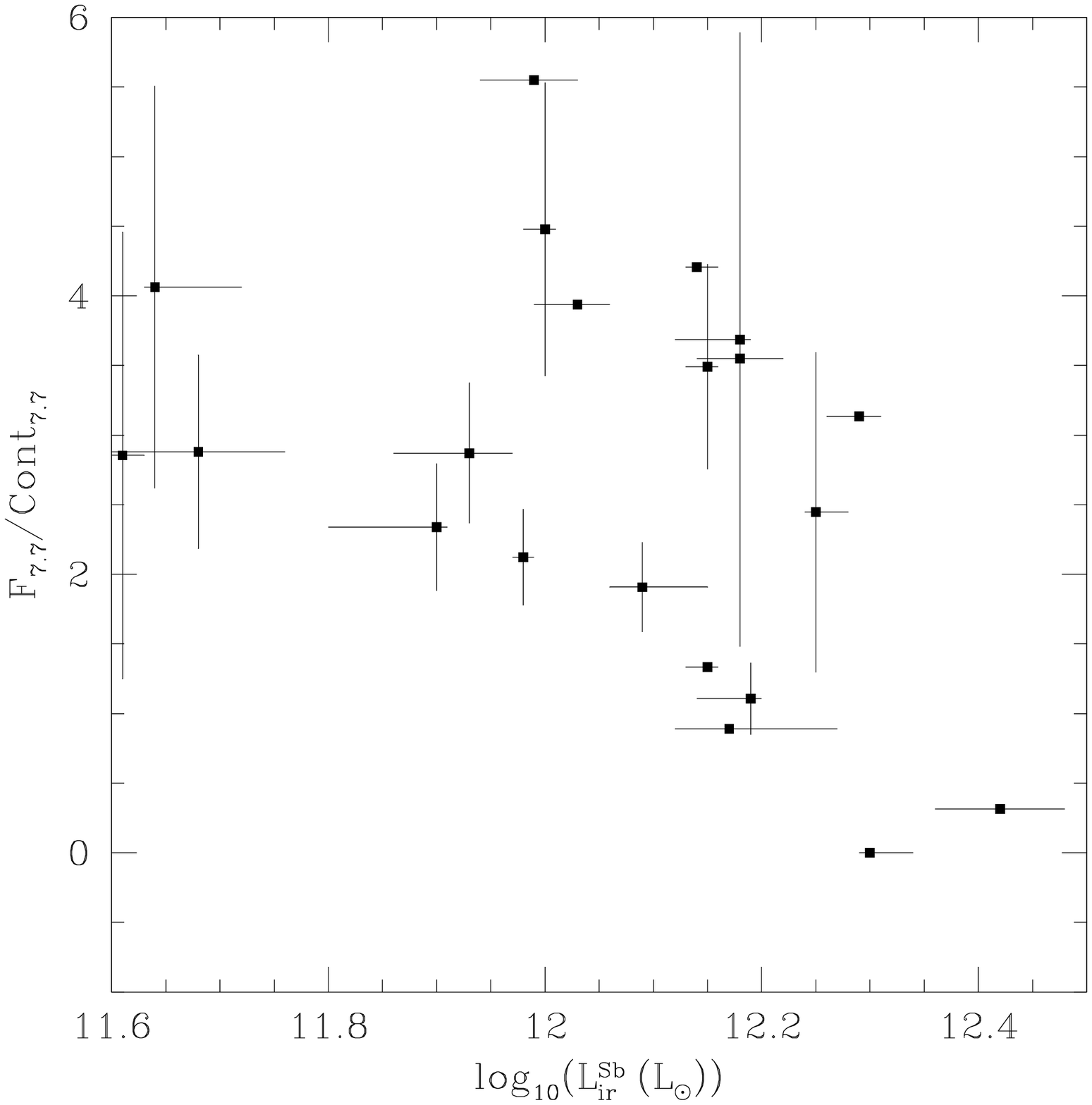,width=89mm}
\epsfig{figure=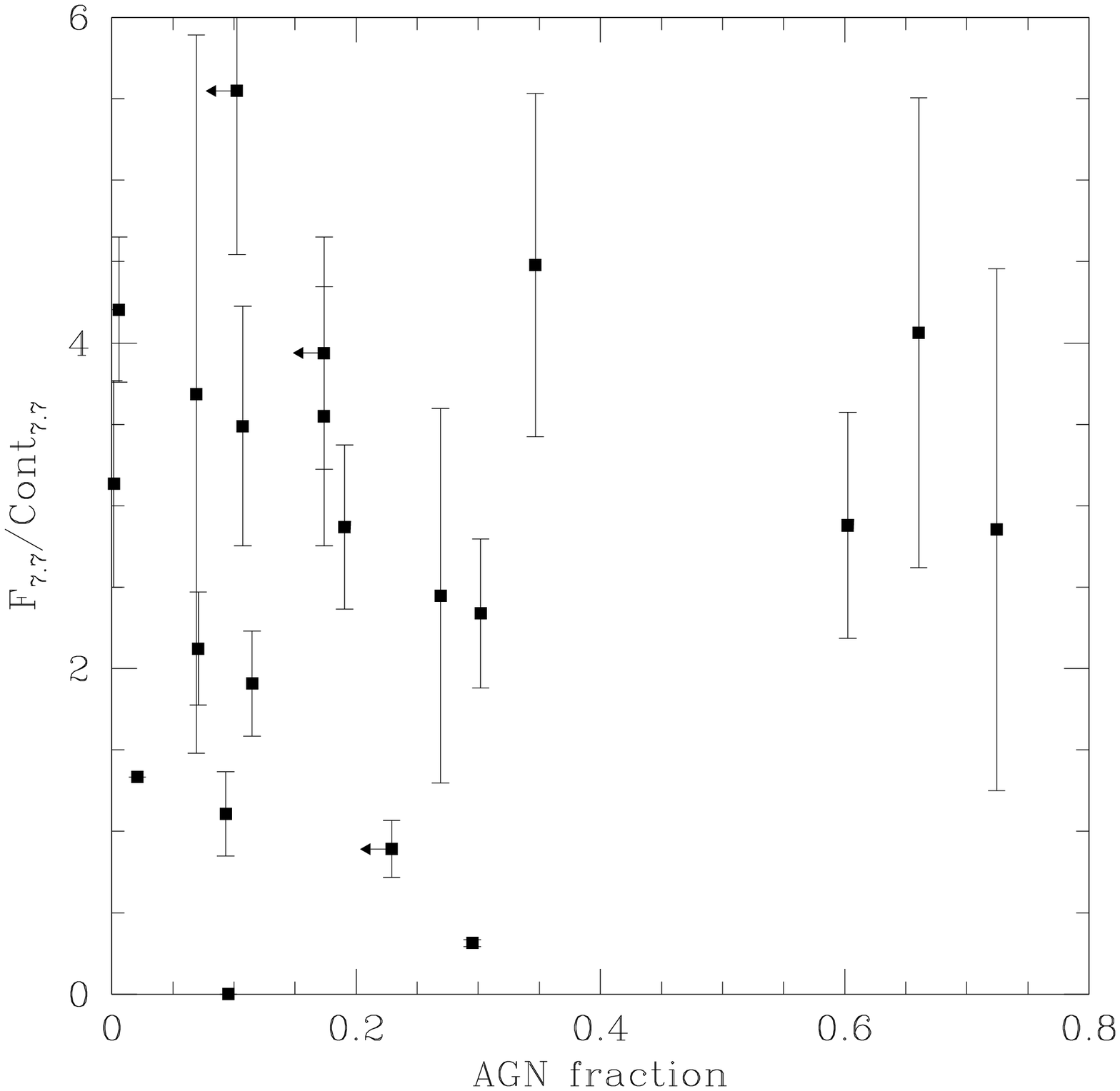,width=89mm}
\end{minipage}
\caption
{
(Left) Starburst luminosity plotted against $7.7\mu$m line/continuum ratio from \citet{rig}. (Right) Fractional AGN luminosity 
plotted against $7.7\mu$m line/continuum ratio
\label{lineconts}
}
\end{figure*}

\subsection{Observational diagnostics of ULIRGs}

\subsubsection{Mid-infrared spectroscopy}

Several groups have constructed diagnostics 
for the power source in ULIRGs based on mid-IR spectra from ISO observations, specifically using the relative 
strengths of the Unitentified 
Infrared Emission Band (UIB) features lying between $5\mu$m and $12\mu$m. Currently, 
the most prevalent of these diagnostics is that proposed by \citet{gen}. In this diagnostic the ratio of the 
$7.7\mu$m emission, thought to arise from PAHs, relative to the 
continuum emission at the same wavelength is used to differentiate between starburst and AGN emission. A high value of 
$F_{7.7}/C_{7.7}$ ($>1$) indicates starburst power whereas a lower value of $F_{7.7}/C_{7.7}$ indicates 
AGN power. This is generally borne out by mid-IR observations of lower luminosity starburst and AGN systems 
(e.g. \citet{rig}). There are however potential problems with this approach. Firstly, since the active regions in 
ULIRGs are small and dusty it is not clear that even mid-IR 
observations can penetrate to the central regions, although it is argued that this is unlikely to be significant 
\citep{gen,rig,tra2}. Secondly, some metal-poor dwarf starburst galaxies show no UIB features, 
but as ULIRGs are not metal poor this is also argued not to be important \citep{rig}. 

By comparing the results from diagnostics using the $F_{7.7}/C_{7.7}$ ratio to our results for 
the 22 ULIRGs common to our sample and to \citet{rig}, we can examine the reliability of the $F_{7.7}/C_{7.7}$ 
technique. In Figure \ref{lineconts} we plot $F_{7.7}/C_{7.7}$ against starburst luminosity and fractional AGN 
luminosity. From the left hand plot it can be seen there is, at best, a weak trend for 
more luminous starbursts to have {\it lower} values of $F_{7.7}/C_{7.7}$. The immediate possibility is that the mid-IR 
diagnostics are correct and our SED fitting approach is wrong, however our SED fits reproduce the 
observed $F_{7.7}/C_{7.7}$ ratios in all 22 objects, as well as the observed SED from the near-UV to the sub-mm. 
We therefore infer that, of the two approaches, ours is more robust. Therefore, it seems that the relative strength 
of the $7.7\mu$m PAH feature is not a reliable indicator of the luminosity of the starburst, but merely of its presence. We 
postulate that the discrepancy between the nature of the $7.7\mu$m PAH feature in ULIRGs and in less luminous 
starburst galaxies is due to two reasons. Firstly, in ULIRGs it is more likely than in moderate starbursts that the 
obscuration is so high that spectral features are not apparent in the mid-IR. Secondly, the intense UV radiation in 
ULIRG-like starbursts may be capable of dissociating the hydrocarbon bonds in PAHs. Both these reasons would explain 
why more luminous starbursts have smaller $F_{7.7}/C_{7.7}$ ratios.

More serious however is that the right hand panel of Figure \ref{lineconts} shows no correlation between 
$F_{7.7}/C_{7.7}$ and the fractional AGN luminosity, from which we infer that the $F_{7.7}/C_{7.7}$ ratio 
{\it on its own} is not an accurate indicator of the overall power source in ULIRGs. This discrepancy between 
our results and those of \citet{rig} can also be explained by the more extreme nature of ULIRGs. From 
our results, all ULIRGs contain a starburst, explaining the prevalence of PAH features in their mid-IR spectra, 
however the majority of ULIRGs are composite systems and may contain a heavily 
obscured starburst or AGN that does not emit significantly in the mid-IR but that does emit strongly in the 
far-IR/sub-mm. As an example we consider IRAS 00262+4251. This object, with $F_{7.7}/C_{7.7} = 4.063$ is 
classified as a pure starburst by \citet{rig}. From our SED fit however (Figure \ref{ulirg_seds}), 
the AGN does not contribute in the mid-IR as the torus is viewed almost edge on and the 
corresponding obscuration in the mid-IR is very high. At wavelengths between $50\mu$m and $\sim120\mu$m the 
AGN contribution is however dominant. Although this object does contain a starburst, overall the emission is 
mainly due to AGN activity. We conclude that the $F_{7.7}/C_{7.7}$ ratio on its own is only an indicator of 
whether a ULIRG contains a moderately obscured and moderately luminous starburst. It cannot be 
used to probe either heavily obscured or very luminous starburst activity, or AGN that are either isotropically 
obscured or oriented nearly edge on relative to us. As such, it cannot be used on its own to probe the overall 
power source in ULIRGs.

\begin{table}
\caption{Ultraluminous Infrared Galaxies detected in the radio \label{ulirgradio}}
\begin{tabular}{@{}lccccc}
\hline
Name        & $F_{1.4GHz}$ & $q_{total}$ & $q_{starburst}$ & $q_{AGN}$  \\
\hline
00262+4251  &  28.0    	   & 1.92        & 1.32   	   & 1.80       \\
00335-2732  &  11.25   	   & 2.60        & 2.58       	   & $<0.70$    \\
04232+1436  &  29.39       & 2.10        & 2.10    	   & $<-0.4$    \\
05189-2524  &  28.8        & 2.66        & 2.62     	   & 1.53       \\
08572+3915  &  4.3     	   & 3.21        & 3.16            & 2.27 	\\
09320+6134  &  170.1       & 1.85        & 1.85            & 0.00 	\\
09583+4714  &  36.5        & 1.90        & 1.87            & $<0.60$    \\
10035+4852  &  28.1        & 2.26        & 2.26            & $<-0.40$   \\
10494+4424  &  21.2        & 2.28        & 2.28    	   & $<-0.30$   \\
10190+1322  &  16.8        & 2.38        & 2.37    	   & 0.48       \\
10565+2448  &  56.9        & 2.38        & 2.37    	   & 0.45 	\\
12112+0305  &  23.3        & 2.57        & 2.57    	   & $<-0.37$   \\
12540+5708  &  308         & 2.03        & 2.00    	   & 0.88 	\\
13428+5608  &  144         & 2.22        & 2.22    	   & -0.15      \\
14348-1447  &  35.8        & 2.26        & 2.17    	   & 1.53 	\\
14378-3651  &  33.8        & 2.29        & 2.29    	   & $<-0.50$   \\
15250+3609  &  14.5        & 2.70        & 2.21    	   & 2.53 	\\
15327+2340  &  326.5       & 2.42        & 2.42    	   & $<-1.0$	\\
17132+5313  &  29.3        & 2.34        & 2.32    	   & 0.98 	\\
17208-0014  &  81.8        & 2.07        & 2.07    	   & $<0.76$	\\
18470+3233  &  12.4        & 2.50        & 2.50    	   & $<-0.08$   \\
19458+0944  &  14.0        & 2.52        & 2.37    	   & 1.99 	\\
20046-0623  &  14.6        & 2.42        & 2.42    	   & $<-0.16$   \\
20414-1651  &  23.4        & 2.32        & 2.32    	   & -0.23 	\\
21504-0628  &  13.8        & 2.41        & 2.40    	   & 0.62	\\
22491-1808  &  5.9         & 2.87        & 2.33    	   & 2.72 	\\
23365+3604  &  27.2        & 2.47        & 2.31    	   & 1.93 	\\
\hline
\end{tabular}

\medskip

1.4GHz radio fluxes were taken from the NVSS catalogues and are in mJy. The $q$ values 
were calculated using Equation \ref{equnfircorr}.

\end{table}

\begin{figure*}
\begin{minipage}{180mm}
\epsfig{figure=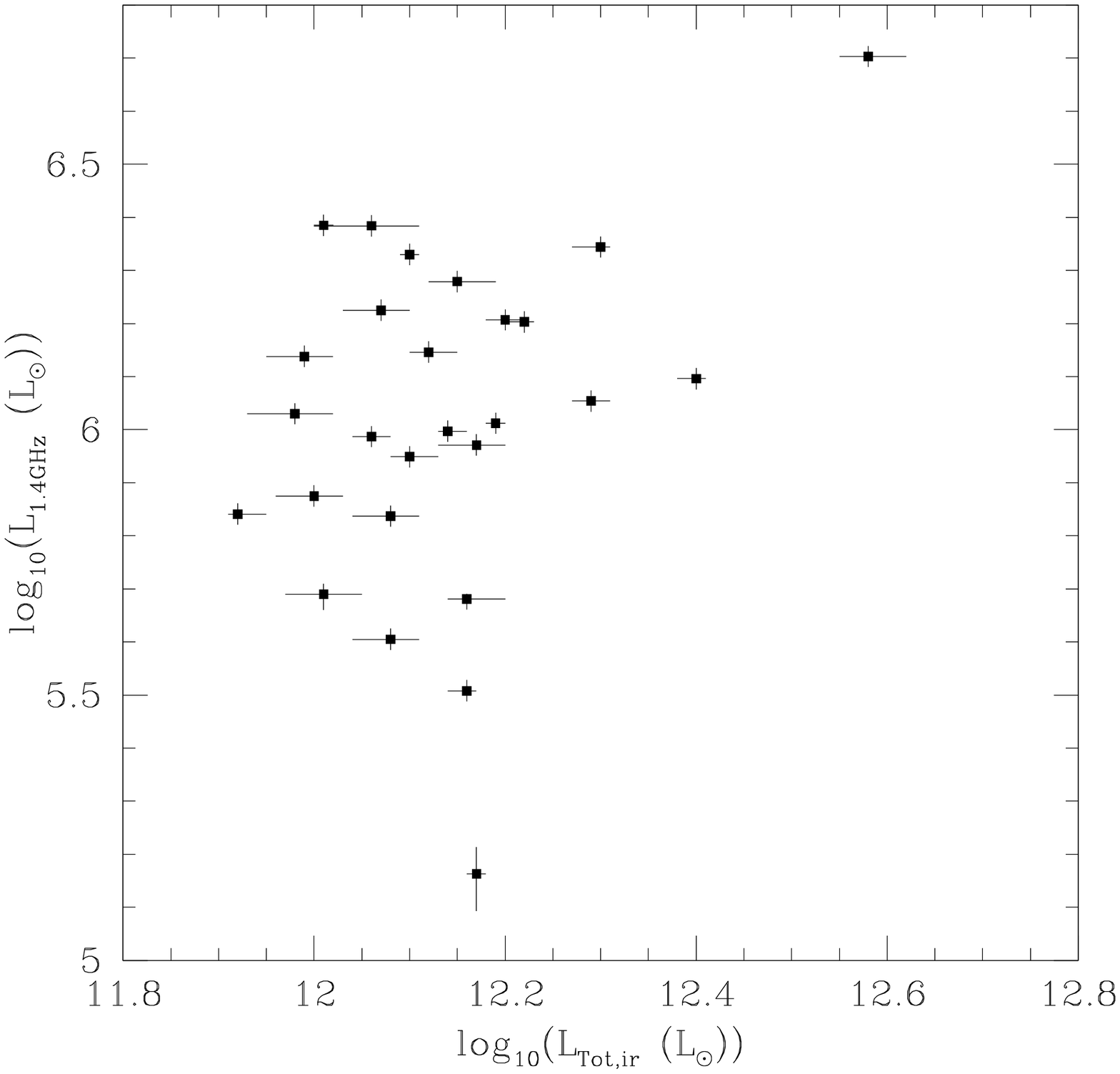,width=89mm}
\epsfig{figure=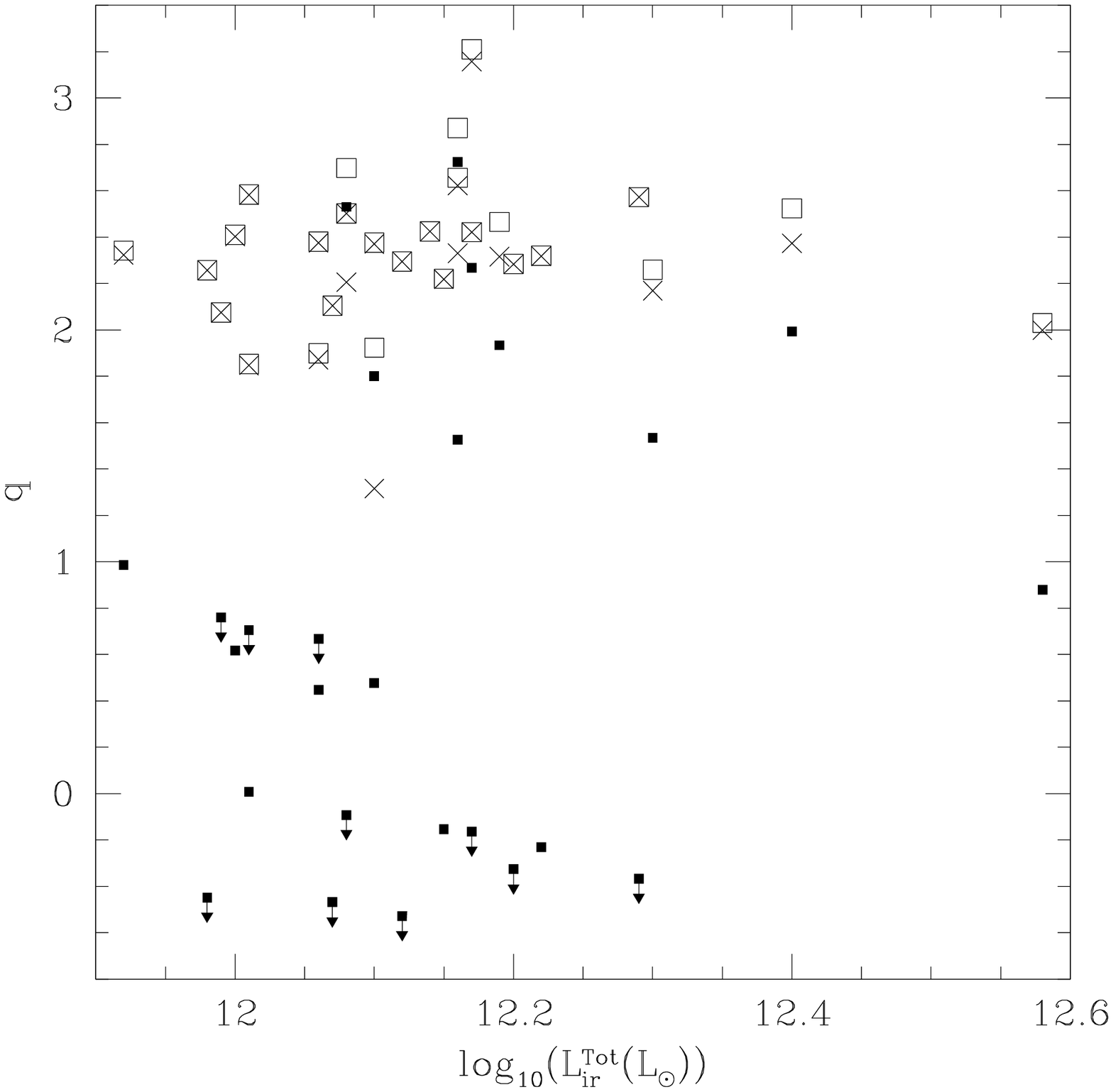,width=89mm}
\end{minipage}
\caption
{
(Left) Total IR luminosity plotted against total radio luminosity. (Right) Total IR 
luminosity vs. $q$ parameter. Open squares are $q_{total}$, crosses are $q_{starburst}$, 
and small filled squares are $q_{AGN}$. A downward arrow indicates 
an upper limit.
\label{totvslradio}
}
\end{figure*}

\subsubsection{The radio-infrared correlation}

There is observed to exist a tight correlation between radio and infrared flux for both normal and active 
galaxies, extending over several orders of magnitude \citep{hsr}. The physical origin of this relation is 
believed to be a population of relativistic electrons accelerated by supernova remnanats produced in regions 
of massive star formation \citep{hpa}. This relation is usually expressed using the parameter $q$, the 
logarithm of the ratio of infrared flux to radio flux. For starburst galaxies and ULIRGs, $q$ is observed to lie in the range 
$2.0 < q < 2.6$ \citep{con1}. Radio Loud Quasars on the other hand have typical $q$ values in the range 
$0 < q < 1$ (e.g. \citet{roy}), whereas Radio Quiet Quasars have $q$ values lying in a similar range to local 
starburst galaxies and ULIRGs \citep{spa}. It has therefore been suggested that the radio-IR correlation in 
both ULIRGs and Radio Quiet Quasars is due to star formation \citep{con1,cns}. 

\begin{figure*}
\begin{minipage}{180mm}
\epsfig{figure=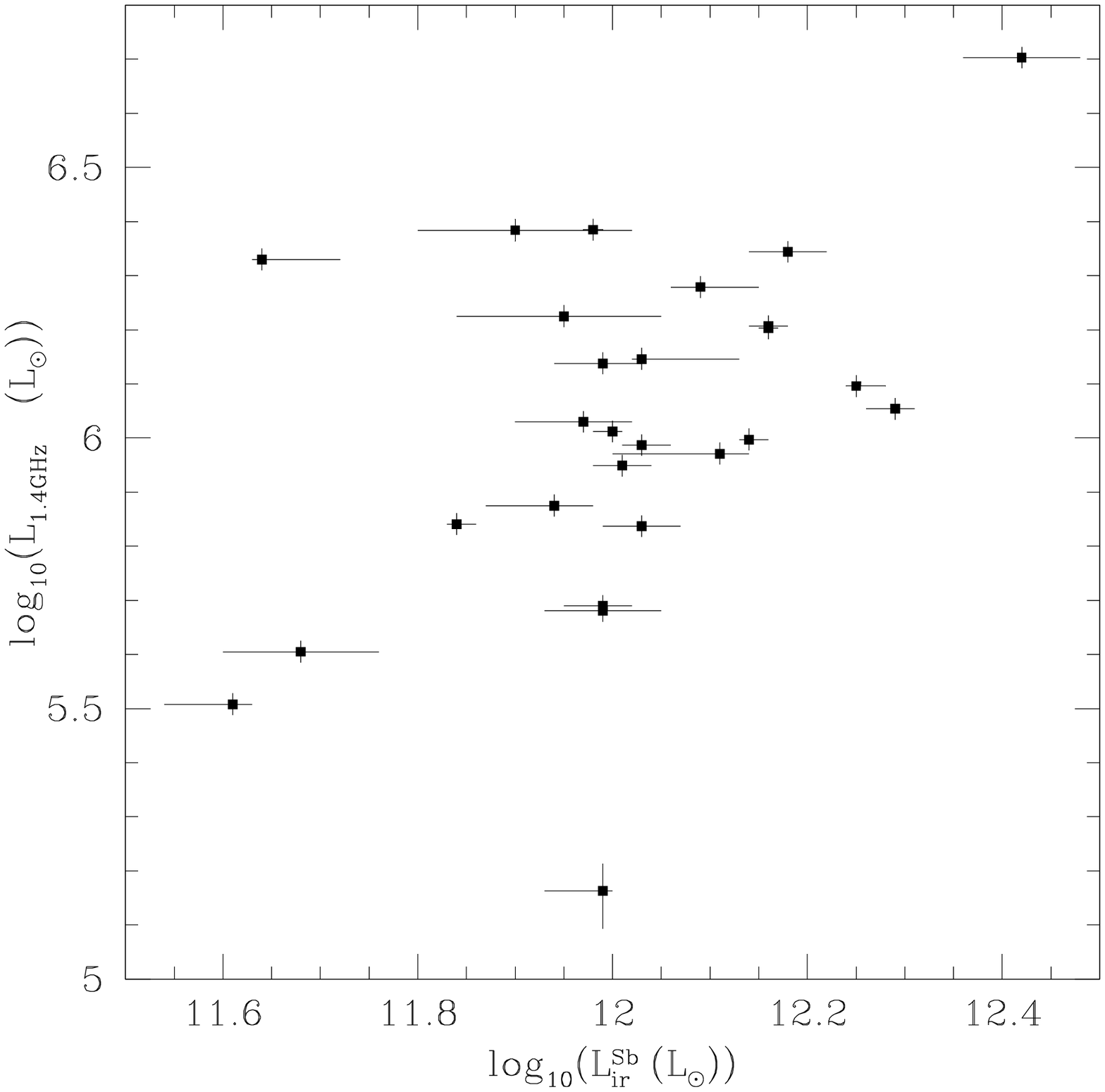,width=89mm}
\epsfig{figure=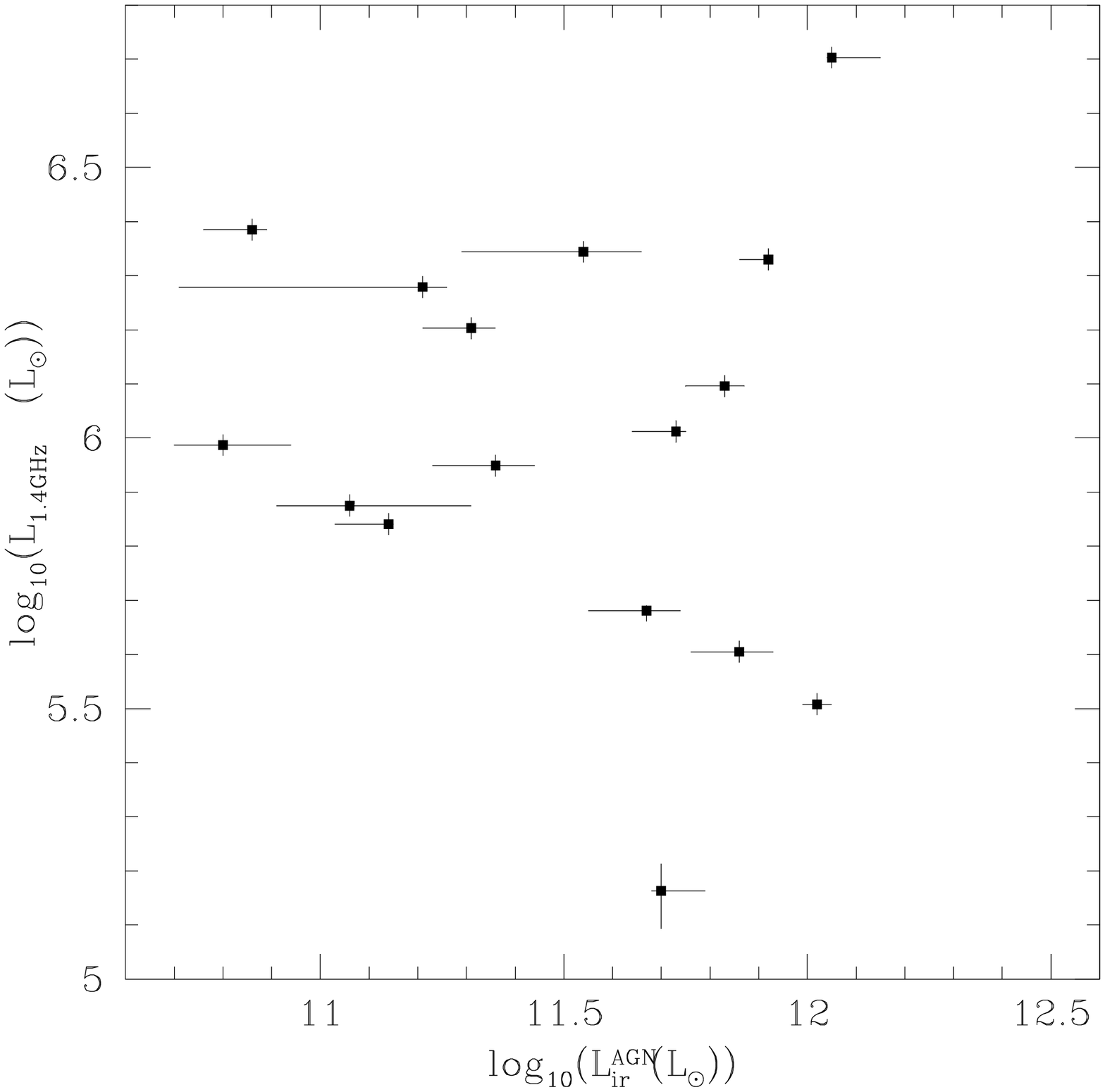,width=89mm}
\end{minipage}
\caption
{
(Left) Starburst luminosity plotted against 1.4GHz radio luminosity. (Right) AGN 
luminosity plotted against 1.4GHz radio luminosity. 
\label{belch}
}
\end{figure*}

Using the results from this paper we can examine the origin of the radio-IR correlation for ULIRGs, and 
the factors that may bias it. 1.4GHz fluxes were obtained for most of the objects in our sample from the
NVSS catalogues, these fluxes are presented in Table \ref{ulirgradio}. The left hand panel of Figure \ref{totvslradio}. 
shows total IR luminosity plotted against 1.4GHz luminosity, and a weak, though positive, correlation 
can be seen. Figure \ref{belch} shows the starburst and AGN luminosities plotted 
against 1.4GHz luminosity. The correlation between the radio luminosity and the starburst luminosity is 
stronger than for the total luminosity, whereas no correlation is observed with the AGN luminosity. We thus 
confirm that the radio-IR correlation in ULIRGs is due to star formation. 

We have further investigated this relation using the $q$ parameter. To allow future comparisons with 
galaxy samples which do not benefit from accurate estimates of IR luminosities from SED fits, we express 
$q$ as:

\begin{equation}
q = log \left(\frac{F_{60}}{F_{1.4}}\right)
\label{equnfircorr}
\end{equation}

\noindent where $F_{1.4}$ is the 1.4GHz flux and $F_{60}$ is the rest-frame $60\mu$m flux.  Rest-frame 
$60\mu$m fluxes were extracted from the best-fit total, starburst and AGN SEDs. In Table 
\ref{ulirgradio} we present values of $q$ calculated using these fluxes. We derive 
$\langle q_{total} \rangle = 2.36\pm0.06$,  $\langle q_{starburst} \rangle = 2.29\pm0.06$ and 
$\langle q_{AGN} \rangle = 1.29\pm0.25$, which are 
in agreement with previous estimates for ULIRGs. Individual 
$q$ values are plotted against total IR luminosity in the right hand panel of Figure \ref{totvslradio}. It 
is evident that $q_{total}$ and $q_{starburst}$ have only a small scatter about a mean value, but that the 
values of $q_{AGN}$ appear randomly distributed.

If the radio-IR correlation in ULIRGs is due to star formation, then what is the origin of the scatter 
about the mean values of $q_{starburst}$ and $q_{total}$? Broadly, there are four possibilities. Firstly, 
the IMF of the starburst may be skewed towards or away from producing high-mass stars, thus producing an 
over- or underabundance of radio supernova remnants, thus causing a scatter. This possibility is impossible 
to investigate within the context of this paper. Secondly, there may be contamination of the 1.4GHz 
luminosity from an obscured AGN, leading to an artificially suppressed $q$ value. We argue that this 
possibility, although undoubtedly present in the ULIRG population, is not the main cause of the observed 
scatter. If there were a significant number of obscured AGN causing the scatter, then we would expect to 
see some correlation between AGN luminosity and radio luminosity, and yet none is observed. Even if only 
the most IR-luminous AGN are plotted in Figure \ref{totvslradio}, no such correlation appears. The third 
possibility is that the starburst may be too young to have formed a significant population of radio supernova 
remnants, thus causing an artificial amplification of the $q$ parameter. We argue that this is a negligible 
effect in the ULIRG population for two reasons. Firstly, most of the starburst ages lie in the range 
10Myr - 37Myr, by which time a significant population of supernova remnants will certainly have formed. 
Secondly, if the youngest starbursts are removed from the left-hand plot of 
Figure \ref{belch} then the scatter becomes bigger rather than smaller. The fourth possibility 
is an old relativistic electron population from a previous starburst event. As described in section 
\ref{subsubsect:mult} it is likely that ULIRGs undergo multiple, discrete starburst events with 
lifetimes in the range 
$10^{6} - 10^{8}$ years. Although radio supernova remnants have lifetimes of only $10^{3}$ years at best, 
the relativistic electrons they produce have lifetimes of the order $10^{8}$ years \citep{con2}. This 
lifetime lies at the upper end of the lifetime range of a single starburst event, and is comparable 
to the total lifetime of a ULIRG. We thus conclude that most of the scatter around the radio-IR correlation 
for ULIRGs is due to skewed IMFs in the starburst, and relativistic electrons from a previous, separate 
starburst event.

\subsection{ULIRG evolution}\label{subsect:evo}

\begin{figure*}
\begin{minipage}{180mm}
\epsfig{figure=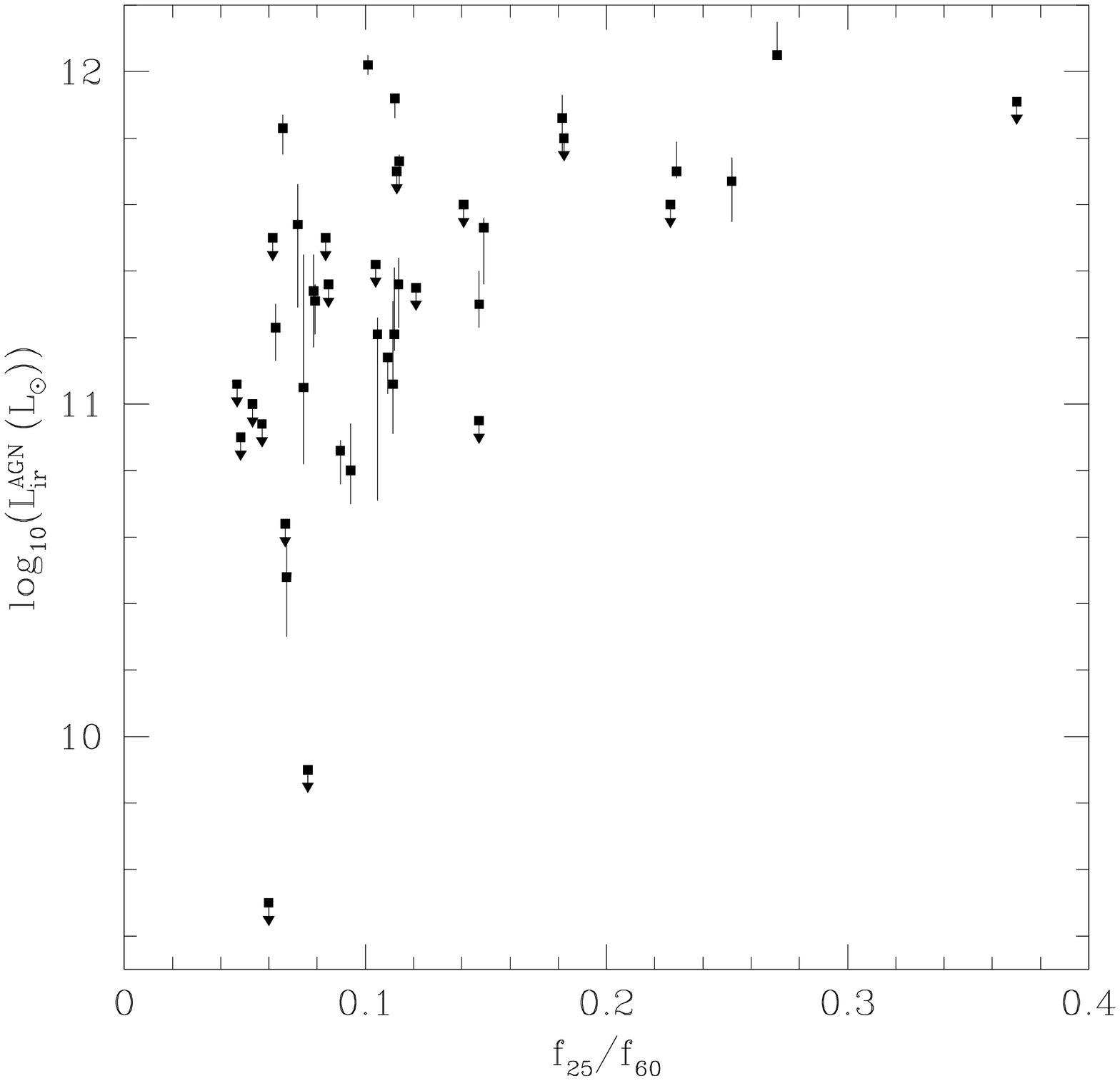,width=89mm}
\epsfig{figure=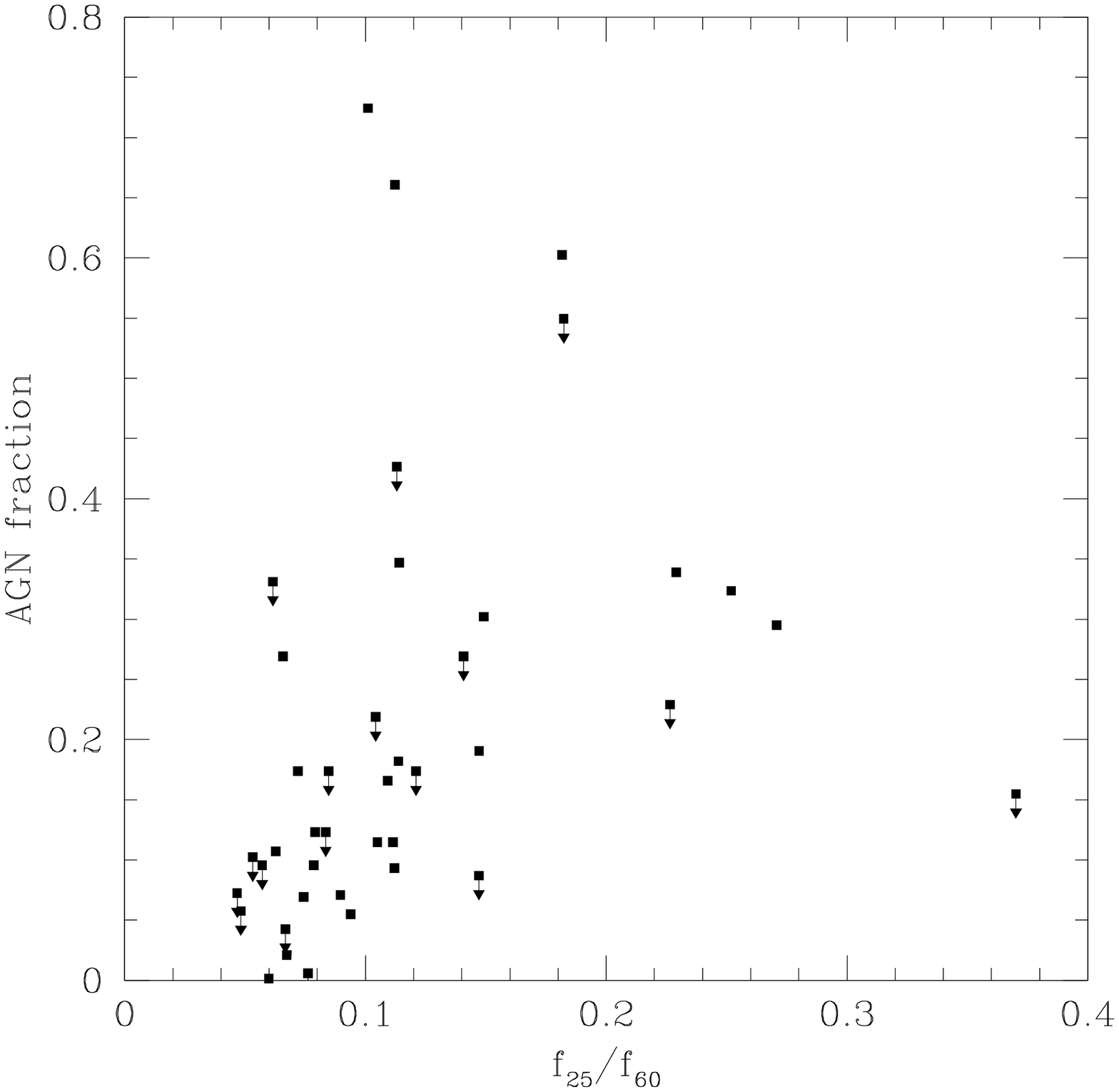,width=89mm}
\end{minipage}
\caption
{
$F_{25}/F_{60}$ colour plotted against (left) AGN luminosity, and (right) fractional AGN luminosity. 
\label{ircolplots}
}
\end{figure*}

\subsubsection{Do ULIRGs evolve into QSOs?}

The first picture of ULIRG evolution to be proposed was that of \citet{san2}, which asserted that ULIRGs as a class evolve into
optical QSOs.  According to this picture (hereafter referred to as the S88 picture), interactions and mergers between gas rich 
spirals transport gas to the central regions of the galaxies. This central gas concentration fuels starburst activity, 
and in the latter stages of the merger commences the fuelling of a central supermassive black hole, which rapidly comes to 
dominate the IR luminosity of the ULIRG. In the last stages the dust shrouding the black hole is blown away and the ULIRG 
evolves into an optical QSO. A qualitatively similar picture was also proposed by \citet{can}. A more recent evolutionary 
scenario for ULIRGs is that of \citet{far1}, also later suggested by \citet{tac}. In this scenario ULIRGs are not simply the 
dust shrouded precursors to optical QSOs but instead are a diverse population with a broad range of properties and 
evolutionary paths. Although some ULIRGs would still evolve into QSOs, this number would be a small and non-representative 
subset of the ULIRG population. 

Previous studies of the evolution of ULIRGs have highlighted some problems for the S88 picture. In the mid-IR, spectroscopic 
studies of ULIRGs \citep{rig,gen} find no evidence that the advanced merger systems are more AGN-like, based on the ULIRG 
nuclear separations. That AGN activity becomes more prevalent with merger advancement is a natural prediction of the S88 picture. 
Furthermore, a recent spectroscopic study of the stellar and gas kinematic in ULIRGs \citep{gen2} found that ULIRGs will likely 
evolve into only $\sim L_{*}$ ellipticals, and not into the most massive ellipticals seen locally. A corollary to this are 
the results from a large scale imaging and spectroscopy survey of ULIRGs \citep{vei3}, in which it was found that nearly half 
the sample may not evolve into an optical, post ULIRG AGN, and that merger induced QSO activity might only take place if both 
merger progenitors had $L^{*}$ luminosities or greater. 

Using the results from this paper, and from previous authors, we can examine these two evolutionary scenarios.
We first consider the growth of Supermassive Black Holes (SMBHs) in ULIRG mergers, and relate this to the SMBH masses 
seen in QSOs, and to the level of AGN activity seen in our sample. Nearly all optical QSOs with measured SMBH masses 
have $M_{BH}\geq10^{8}M_{\odot}$, in some cases reaching $M_{BH}\sim10^{10}M_{\odot}$ 
\citep{mcd,gcj}, hence it seems likely that a SMBH in this mass range is necessary for QSO activity. 
The growth of SMBHs in ULIRG mergers has been studied by \citet{ta}. They conclude that a SMBH of mass $\sim10^{8}M_{\sun}$ 
can form if any of the progenitor galaxies contains a `seed' SMBH of mass $\geq10^{7}M_{\sun}$ undergoing efficient 
Bondi type gas accretion over $\sim10^{8}$ years. 

Hence, if the S88 picture is correct and ULIRGs as a class evolve into QSOs, then the range of SMBH masses seen in spiral 
galaxies ($10^{6} < M_{BH}(M_{\odot}) < 10^{8}$) would have to be transformed by the merger into the SMBH mass range seen 
in QSOs ($10^{8} < M_{BH}(M_{\odot}) < 10^{10}$). This would require very efficient accretion onto the black hole for the 
entire duration of the merger. This itself implies that the lifetime of the IR-luminous AGN phase will be significantly 
greater than the lifetime of any single starburst event. Furthermore, in order to achieve the SMBH mass range seen in QSOs, 
then the AGN in a ULIRG observed at a random point during the merger would in many cases harbour a SMBH of mass
$M_{BH}\geq10^{8}M_{\odot}$. Overall therefore, if the S88 picture is correct, then most ULIRGs should contain a 
long-lived AGN which, by virtue of efficient accretion onto a SMBH with mass comparable to those seen in QSOs, should 
have an IR luminosity comparable to the bolometric luminosity seen in QSOs. Our results however do not support this, as 
we find all ULIRGs to contain a very luminous starburst, with only about half containing a luminous AGN, and that less than 
$10\%$ of ULIRGs are AGN dominated. Whilst we cannot rule out the S88 picture based on AGN activity alone, it seems unlikely 
that the level of AGN activity seen in our sample could transform the BH mass range in spirals into that seen in QSOs. 

We next consider the nature of the so-called `warm' ULIRGs, those with $F_{25}/F_{60}>0.25$, (where $F_{25}$ and $F_{60}$ 
are the IRAS $25\mu$m and $60\mu$m fluxes respectively). Objects with `warm' infrared colours have been previously shown 
to be more likely to contain IR-luminous AGN. Hence, if the S88 picture is true, then the greater prevalence of AGN in `warm' 
ULIRGs than in cool ones means that `warm' ULIRGs will be, on average, nearer the end of the ULIRG phase than `cool' ULIRGs, 
and closer to evolving into optical QSOs. Therefore, `warm' ULIRGs should have a higher {\it fractional} AGN luminosity on
average  
than `cool' ULIRGs as the AGN comes to dominate the bolometric emission and the starburst dies away. Figure \ref{ircolplots} 
shows $F_{25}/F_{60}$ colour plotted against AGN luminosity and fractional AGN luminosity for 
the objects in our sample. Values of $F_{25}/F_{60}$ are presented in Table \ref{ulirgsample}. As expected, those 
objects with a higher value of $F_{25}/F_{60}$ are more likely to contain a more luminous AGN. There is 
however no trend of increasing AGN fractional luminosity with increasing IR colour. 

Overall therefore, our results contradict the S88 picture, and imply that ULIRGs as a class do not evolve to become QSOs. 
Instead, the wide range of starburst and AGN luminosities derived for our sample imply multiple possible patterns of 
starburst and AGN activity over the lifetime of the ULIRG, in line with the \citet{far1} picture. The overall level of AGN 
activity is 
consistent with sufficient accretion to transform the range of black hole masses seen in spirals into the range of 
black hole masses seen in ellipticals, without recourse to an optical QSO phase. As a small number of our sample are AGN 
dominated it is likely that some ULIRGs do become optical QSOs, but the small number of such systems in our sample 
argues that such ULIRGs are rare, and thus unrepresentative of the ULIRG population as a whole.

\subsubsection{ULIRG evolution with redshift} \label{ulirglowhighz}

ULIRGs, despite being rare in the local Universe, are a cosmologically important galaxy population. At 
low redshifts the ULIRG luminosity function shows strong evolution with redshift \citep{vei2}. In addition, deep 
sub-mm surveys find that systems with ULIRG-like luminosities dominate the global energy density of the Universe at 
$z\geq1$ \citep{bar,lil,scot,fox}. Determining how systems with ULIRG-like luminosities evolve with redshift is therefore 
necessary to measure, for example, the overall star formation history of the Universe, currently a hotly debated topic 
(e.g. \citet{mad,rr0}). 

It is however not known whether the IR luminosity of these high redshift sources arises due to mergers between two or more 
large spiral galaxies, as in local ULIRGs, or via a different mechanism. If the former is true then the strong evolution 
seen in the ULIRG luminosity function will most likely be due to mergers between two or more large spiral galaxies becoming 
more common with increasing redshift. If the latter is true then this indicates that the strong evolution seen in the ULIRG 
luminosity function is probably due to different galaxy formation processes becoming more important at high redshifts. 
Currently, the only sample of IR luminous sources at high redshift with resolved starburst and AGN components is the 
sample of HLIRGs presented by \citet{far3}. On morphological grounds there is some evidence that the trigger for the IR 
emission in local ULIRGs and high-$z$ HLIRGs is the same; nearly all ULIRGs show signs of morphological disturbance 
\citep{bor,far1}, as do approximately half of HLIRGs \citep{far2}.

In section \ref{subsect:power} we showed that the broad correlation 
between starburst and AGN luminosities for a wide range of galaxy types, including ULIRGs and HLIRGs, was most plausibly 
explained by the starburst and AGN luminosities being governed by the gas and dust masses in the nuclear regions. 
We now examine whether the {\it evolution} of the starburst and AGN components are the same or different between ULIRGs 
and HLIRGs, to determine whether HLIRGs are the high redshift analogues of ULIRGs or a different galaxy population, by 
comparing the ULIRGs from our study with the HLIRGs from \citet{far3}. Both samples span a similar range in fractional 
AGN luminosity, ranging from almost pure starbursts to AGN dominated systems. Amongst the ULIRGs most systems are starburst 
dominated with a mean starburst fraction of $82\%$, whereas in the HLIRG population approximately half the systems are AGN 
dominated, with a mean starburst fractional luminosity of $\sim35\%$. This difference could be because a more luminous AGN 
is required to generate the higher total luminosities in HLIRGs, however starburst dominated HLIRGs are found amongst the 
most luminous members of the HLIRG population. This then constitutes a distinct difference between the two samples. Amongst 
local ULIRGs, starbursts are the dominant contributor to the total luminosity whereas amongst HLIRGs, although a starburst 
could in principle accomplish the same thing, this is not observed and AGN activity is much more prevalent than in ULIRGs. 
This can also be seen in Figures \ref{totvssbagn} and \ref{ltotvsagnfrac}, in which the starburst and AGN luminosities are 
compared to the total IR luminosities in ULIRGs, HLIRGs and PG QSOs. Furthermore, it can also be seen from these figures, 
particularly the right-hand panel of Figure \ref{totvssbagn}, that the pattern of starburst and AGN activity in HLIRGs is 
much more reminiscent of the pattern seen in QSOs than in ULIRGs. 

Based on this comparison, we conclude that IR-luminous galaxies at $z\sim0$ and $z\geq1$ are physically different galaxy 
populations. We speculate that this difference is due to different galaxy formation processes in the low and high 
redshift Universe, and that the trigger for starburst and AGN activity in high redshift IR-luminous sources is more 
similar to the trigger for QSO activity, rather than the trigger for ULIRG activity in the local Universe. At low-$z$, 
ULIRGs are formed via the merger of two or more gas rich spiral galaxies. At high redshift however the galaxy formation 
processes were likely more diverse, with hierarchical buildup from many small dwarf galaxies or monolithic collapse of 
a large disk of gas forming a primeval galaxy both leading possibilities.

\section{Conclusions}\label{sect:conc}

We have studied the properties of a sample of 41 local Ultraluminous Infrared Galaxies using archival 
optical and infrared photometry and advanced radiative transfer models for starbursts and AGN. Our 
conclusions are: \\

\noindent 1) All of the sample contain a luminous starburst, whereas about half contain a luminous AGN. 
The mean starburst fractional luminosity is $82\%$, and in $\sim90\%$ of the sample the starburst 
produces more than half the total IR emission.  We conclude that the fraction of purely AGN powered ULIRGs 
in the local Universe is less than $2\%$. By combining our objects with other galaxy samples we find 
that starburst and AGN luminosities correlate over 6 orders of magnitude in total IR luminosity and 
over a wide range of galaxy types suggesting that a common physical factor, most plausibly the gas 
masses in the nuclear regions, govern both the starburst and AGN luminosities. 

\noindent 2) The starburst luminosity shows a strong positive correlation with total luminosity, as does the AGN 
luminosity, albeit less strongly. We however find no trend for increasing fractional AGN luminosity with 
increasing total luminosity, contrary to previous claims. We find that these claims were based on finding 
generally more luminous AGN in more luminous ULIRGs, rather than increasing AGN dominance with increasing 
total luminosity. 

\noindent 3) We derive a mean starburst age range in ULIRGs of $1.0\times10^{7} - 4.0\times10^{7}$ years. 
Together with previous estimates for the lifetime of a ULIRG 
and an AGN, we find that most ULIRGs must undergo multiple starbursts during their lifetime. When 
combined with recent simulations of pair and multiple galaxy mergers we infer that mergers between 
more than two galaxies must be common in the ULIRG population. 

\noindent 4) By comparison with previous results we find that the mid-IR $F_{7.7}/C_{7.7}$ line-continuum ratio 
gives no indication of the luminosity of the starburst in ULIRGs. We also find that $F_{7.7}/C_{7.7}$ 
ratio gives no indication of the fractional AGN luminosity. We find therefore that a large value of 
$F_{7.7}/C_{7.7}$ only indicates the presence of a moderately obscured starburst, but gives no information 
on the presence or properties of heavily obscured starbursts and AGN. As such, the $F_{7.7}/C_{7.7}$ 
ratio on its own is not a reliable diagnostic of the power source in ULIRGs. 

\noindent 5) The total 1.4GHz radio flux for the objects in our sample correlates strongly with the starburst 
luminosity but shows no correlation with the AGN luminosity. From this we infer that the radio-IR 
correlation in ULIRGs is due to star formation, in line with previous results. Furthermore, we propose 
that the scatter in the correlation is due to a skewed IMF of the starburst and/or a relic relativistic 
electron population from a previous starburst event, rather than contamination from an obscured AGN. 

\noindent 6) The rarity of luminous AGN and AGN dominated systems argues against a simple 
evolutionary model for ULIRGs in which they all evolve to become optical QSOs. Furthermore, although 
`warm' ULIRGs generally contain more luminous AGN than do the `cool' ULIRGs, there is no difference in fractional 
AGN luminosity between the `warm' ULIRGs and the `cool' ULIRGs. We find therefore that ULIRGs as a class do not 
evolve to become QSOs, but instead follow multiple evolutionary paths in transforming merging spirals into emerging 
ellipticals, and that only a few ULIRGs become optical QSOs, as suggested by \citep{far1}. 

\noindent 7) By comparing our local sample to a sample of HLIRGs at $z\sim1$ we find that AGN activity is much 
higher in the $z\sim1$ sample. We infer that the two samples are distinct populations and postulate 
that different galaxy formation processes at high-$z$ are responsible for this difference.

\section*{Acknowledgments}
We thank Ben Tristem for helpful discussion, and the referee for a very helpful report. The work presented 
has made use of the NASA/IPAC Extragalactic Database (NED), which is operated by the Jet Propulsion Laboratory 
under contract with NASA, and the Digitized Sky Surveys, which were produced at the Space Telescope Science 
Institute under U.S. Government grant NAG W-2166. The images of these surveys are based on photographic data 
obtained using the Oschin Schmidt Telescope on Palomar Mountain and the UK Schmidt Telescope. This work was in
part supported by PPARC (grant number GR/K98728). D.F. was supported in part for this work by NASA grant 
NAG 5-3370 and by the Jet Propulsion Laboratory, California Institute of Technology, under contract 
with NASA. JA gratefully acknowledges the support from the Science and
Technology Foundation (FCT, Portugal) through the fellowship
BPD-5535-2001 and the research grant ESO-FNU-43805-2001.

 \label{lastpage}

\end{document}